\documentclass{article}

\usepackage[table]{xcolor}

\usepackage{adjustbox}
\usepackage{algorithm}
\usepackage{amsfonts}
\usepackage{amsthm}
\usepackage{amsmath}
\usepackage{amssymb}
\usepackage{array}
\usepackage{bm}
\usepackage{bbm}
\usepackage{booktabs}
\usepackage{calc}
\usepackage{caption}
\usepackage{colortbl}
\usepackage{comment}
\usepackage{dashrule}
\usepackage{enumitem}
\usepackage{footmisc}
\usepackage{fp}
\usepackage{fvextra}
\usepackage{graphicx}
\usepackage{lipsum}
\usepackage{makecell}
\usepackage{mathtools}
\usepackage{microtype}
\usepackage{multicol}
\usepackage{multirow}
\usepackage{nicefrac}
\usepackage{scalerel}
\usepackage{subfigure}
\usepackage{tabularx}
\usepackage{threeparttable}
\usepackage{tikz}
\usepackage{wrapfig}
\usepackage{url}
\usepackage{xspace}

\usepackage[accepted]{icml2025}

\usepackage[pagebackref,breaklinks,colorlinks]{hyperref}

\usepackage[capitalize,noabbrev]{cleveref}

\DefineVerbatimEnvironment{verbatim}{Verbatim}{breaklines=true}

\definecolor{basecolor_green}{rgb}{0.7,0.9,0.7}
\definecolor{basecolor_red}{rgb}{0.9,0.7,0.7}
\definecolor{basecolor_ceiling}{rgb}{0.93, 0.70, 0.53}

\usepackage{listings}
\makeatletter
\DeclareRobustCommand\onedot{\futurelet\@let@token\@onedot}
\def\@onedot{\ifx\@let@token.\else.\null\fi\xspace}

\makeatother

\usepackage[table]{xcolor}
\definecolor{Gray}{gray}{0.97}
\definecolor{lightgray}{gray}{0.9}
\definecolor{lightblue}{rgb}{0.8,0.85,1}

\usepackage{bbding}

\definecolor{ao(english)}{rgb}{0.0, 0.5, 0.0}

\definecolor{fig1_pass}{RGB}{88, 168, 16}
\definecolor{fig1_fail}{RGB}{255, 133, 133}
\definecolor{fig1_human}{RGB}{0, 112, 192}
\definecolor{fig1_agent}{RGB}{233, 113, 50}
\definecolor{fig1_white}{RGB}{255, 255, 255}
\definecolor{fig2_env}{RGB}{88, 168, 16}
\definecolor{fig2_code}{RGB}{233, 113, 50}
\definecolor{fig2_ask}{RGB}{0, 112, 192}
\definecolor{fig5_red}{RGB}{187, 18, 15}
\definecolor{fig5_green}{RGB}{28, 152, 28}
\definecolor{fig7_blue}{RGB}{45, 165, 245}
\definecolor{fig7_salmon}{RGB}{255, 135, 200}

\hypersetup{
	colorlinks=true,
	linkcolor=cyan,
	filecolor=blue,      
	urlcolor=cyan,
	citecolor=purple,
}

\newcommand{\github}{\textit{GitHub}\xspace}

\newcommand{\Sref}[1]{\S\ref{#1}}

\newcommand{\fref}[1]{Fig.~\ref{#1}}
\newcommand{\tref}[1]{Tab.~\ref{#1}}

\renewcommand{\eqref}[1]{Eq.~\ref{#1}}

\NewDocumentCommand{\xingyao}
{ mO{} }{\textcolor{orange}{\textsuperscript{\textit{Xingyao}}\textsf{\textbf{\small[#1]}}}}
\NewDocumentCommand{\xw}
{ mO{} }{\textcolor{orange}{\textsuperscript{\textit{Xingyao}}\textsf{\textbf{\small[#1]}}}}

\NewDocumentCommand{\chihan}
{ mO{} }{\textcolor{cyan}{\textsuperscript{\textit{Chi}}\small[#1]}}

\NewDocumentCommand{\zoey}
{ mO{} }{\textcolor{violet}{\textsuperscript{\textit{Zoey}}\small[#1]}}

\NewDocumentCommand{\yy}
{ mO{} }{\textcolor{pink}{\textsuperscript{\textit{coolYY}}\small[#1]}}

\NewDocumentCommand{\xuehang}{ mO{} }
{\colorbox{yellow}{\textsuperscript{\textit{TODO}}\textsf{\textbf{\small[#1]}}}}

\NewDocumentCommand{\heng}
{ mO{} }{\textcolor{red}{\textsuperscript{\textit{Heng}}\textsf{\textbf{\small[#1]}}}}

\NewDocumentCommand{\resp}{ mO{} }
{\textcolor{blue}{\textsuperscript{\textit{Many Thanks!}}\textsf{\textbf{\small[#1]}}}}

\NewDocumentCommand{\zhenhailong}
{ mO{} }{\textcolor{purple}{\textsuperscript{\textit{zhenhailong}}\textsf{\textbf{\small[#1]}}}}

\theoremstyle{plain}

\theoremstyle{definition}

\theoremstyle{remark}

\usepackage[textsize=tiny]{todonotes}

\icmltitlerunning{SyncMind: Measuring Agent Out-of-Sync Recovery in Collaborative Software Engineering}

\begin{document}

\twocolumn[
 
\icmltitle{SyncMind: Measuring Agent Out-of-Sync Recovery \\ in Collaborative Software Engineering}



\icmlsetsymbol{equal}{*}

\begin{icmlauthorlist}
\icmlauthor{Xuehang Guo}{xxx}
\icmlauthor{Xingyao Wang}{xxx,zzz}
\icmlauthor{Yangyi Chen}{xxx}
\icmlauthor{Sha Li}{xxx}
\icmlauthor{Chi Han}{xxx}
\icmlauthor{Manling Li}{yyy}
\icmlauthor{Heng Ji}{xxx}
\end{icmlauthorlist}

\icmlaffiliation{xxx}{University of Illinois Urbana-Champaign}
\icmlaffiliation{yyy}{Northwestern University}
\icmlaffiliation{zzz}{All Hands AI}

\icmlcorrespondingauthor{Xuehang Guo}{xuehangg@illinois.edu}
\icmlcorrespondingauthor{Xingyao Wang}{xingyao@all-hands.dev, xingyao6@illinois.edu}
\icmlcorrespondingauthor{Yangyi Chen}{yangyic3@illinois.edu}
\icmlcorrespondingauthor{Sha Li}{shal2@illinois.edu}
\icmlcorrespondingauthor{Chi Han}{chihan3@illinois.edu}
\icmlcorrespondingauthor{Manling Li}{manling.li@northwestern.edu}
\icmlcorrespondingauthor{Heng Ji}{hengji@illinois.edu}

\icmlkeywords{Machine Learning, ICML}

\vskip 0.3in
]



\printAffiliationsAndNotice{}  

\begin{abstract}

\looseness=2
Software engineering (SE) is increasingly collaborative, with developers working together on shared complex codebases.
Effective collaboration in shared environments requires participants---whether humans or AI agents---to stay on the same page as their environment evolves.
When a collaborator's understanding diverges from the current state---what we term the \textit{out-of-sync} challenge---the collaborator's actions may fail, leading to integration issues.
%
In this work, we introduce \textbf{\textit{SyncMind}}, a framework that systematically defines the \textit{out-of-sync} problem faced by large language model (LLM) agents in collaborative software engineering (CSE).
%
Based on \textit{SyncMind}, we create \textbf{\textit{SyncBench}}, a benchmark featuring 24,332 instances of agent \textit{out-of-sync} scenarios in real-world CSE derived from 21 popular \github repositories with executable verification tests.
Experiments on \textit{SyncBench} uncover critical insights into existing LLM agents' capabilities and limitations.
Besides substantial performance gaps among agents (from \textit{Llama-3.1} agents $\leq 3.33\%$ to \textit{Claude-3.5-Sonnet} $\geq 28.18\%$), their consistently low collaboration willingness ($\le 4.86\%$) suggests fundamental limitations of existing LLM in CSE.
%
However, when collaboration occurs, it positively correlates with \textit{out-of-sync} recovery success.
Minimal performance differences in agents' resource-aware \textit{out-of-sync} recoveries further reveal their significant lack of resource awareness and adaptability, shedding light on future development of resource-efficient collaborative systems.
Our code and data are openly available on our project website: \href{https://xhguo7.github.io/SyncMind/}{https://xhguo7.github.io/SyncMind/}.

\end{abstract}

\section{Introduction}
\label{Section: Introduction}

\begin{figure}[H]
\begin{center}
\begin{small}
    \vspace{-0.5em}    
    \includegraphics[width=1\linewidth]{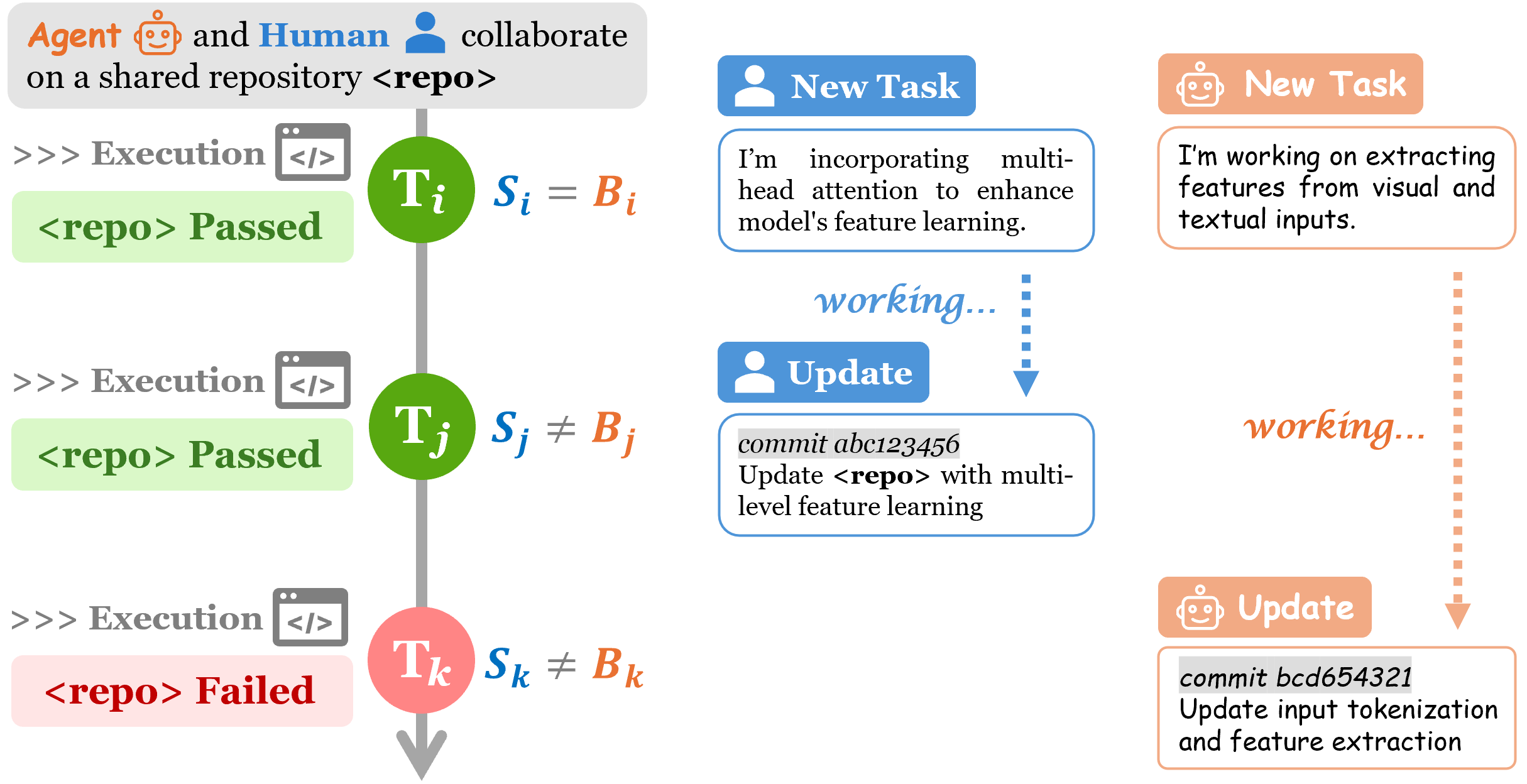}
    \vspace{-1.5em}  
    \caption{
    \small
    \textbf{The \textit{Out-of-Sync} Challenge.} At \colorbox{fig1_pass}{\textcolor{fig1_white}{$\mathbf{T}_{\bm{i}}$}}, \textcolor{fig1_agent}{\textbf{Agent}} and \textcolor{fig1_human}{\textbf{Human}} work on respective tasks. During \textcolor{fig1_agent}{\textbf{Agent}}'s task completion from \colorbox{fig1_pass}{\textcolor{fig1_white}{$\mathbf{T}_{\bm{i}}$}} to \colorbox{fig1_fail}{\textcolor{fig1_white}{$\mathbf{T}_{\bm{k}}$}}, \textcolor{fig1_human}{\textbf{Human}} updates \textbf{\textless repo\textgreater} at \colorbox{fig1_pass}{\textcolor{fig1_white}{$\mathbf{T}_{\bm{j}}$}} that \textcolor{fig1_agent}{\textbf{Agent}} is unaware of due to being occupied with its own task. This leads \textcolor{fig1_agent}{\textbf{Agent}} to become \textit{out-of-sync} at \colorbox{fig1_fail}{\textcolor{fig1_white}{$\mathbf{T}_{\bm{k}}$}} as a result of \textcolor{fig1_human}{$\bm{S_k}$} $\neq$ \textcolor{fig1_agent}{$\bm{B_k}$}.}
    \vspace{-1.0em}    
    \label{fig:figure 1}
\end{small}
\end{center}
\end{figure}

\begin{figure*}[!t]
\begin{center}
\begin{small}
\vspace{-0.5em}
\includegraphics[width=1\linewidth]{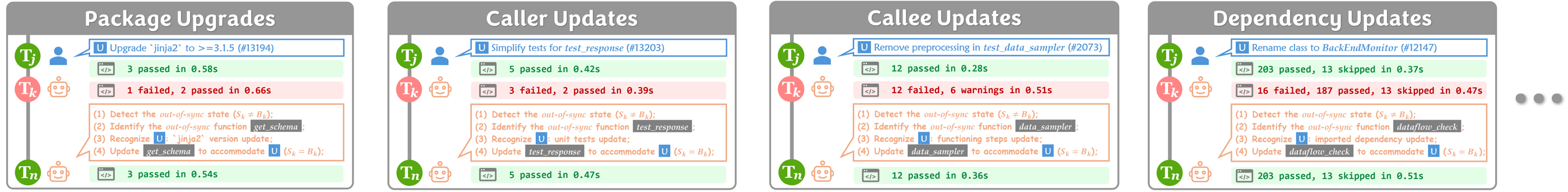}
    \vspace{-1.8em}  
    \caption{\textbf{Typical Causes of \textit{Out-of-Sync}.} Examples of \textit{out-of-sync} scenarios in our benchmark.}
    \label{fig:figure 8 (Typical Causes)}
    \vspace{-1.5em}
\end{small}
\end{center}
\end{figure*}

Collaborative systems---whether involving humans, AI agents, or both---boost efficiency and capabilities by combining complementary strengths.
Recent advances have demonstrated impressive capabilities of AI agents in collaborative tasks~\cite{unleashing2024}, from conversational AI assistants, like ChatGPT \cite{ai_assistant_chatgpt}, Claude \cite{ai_assistant_claude}, that effectively assist users in daily problem-solving, to coding agents, like Devin \cite{ai_assistant_devin}, OpenHands \cite{OpenHands}, that can actively collaborate with humans on software development. 


These collaborative coding agents are typically designed and evaluated in static environments where the workspace remains fixed throughout task execution~\cite{jimenez2023swe, yang2024swe}. 
However, real-world collaborative software engineering (CSE) fundamentally operates in dynamic environments, where effective teamwork depends on team members \textit{maintaining synchronized awareness of workspace states}—a core challenge in the field~\cite{agent_alignment}.
%
While version control systems~\cite{version_control_git} can detect surface-level code conflicts, they cannot identify semantic inconsistencies that require manual resolution. This includes scenarios where agents must resolve dependency updates, modify existing functions to align with newly imported modules, and so on (\fref{fig:figure 8 (Typical Causes)}).


%

\begin{figure*}[!t]  
\begin{center}
\begin{small}
    \vspace{0.0em}    
    \centering
    \includegraphics[width=1\textwidth]{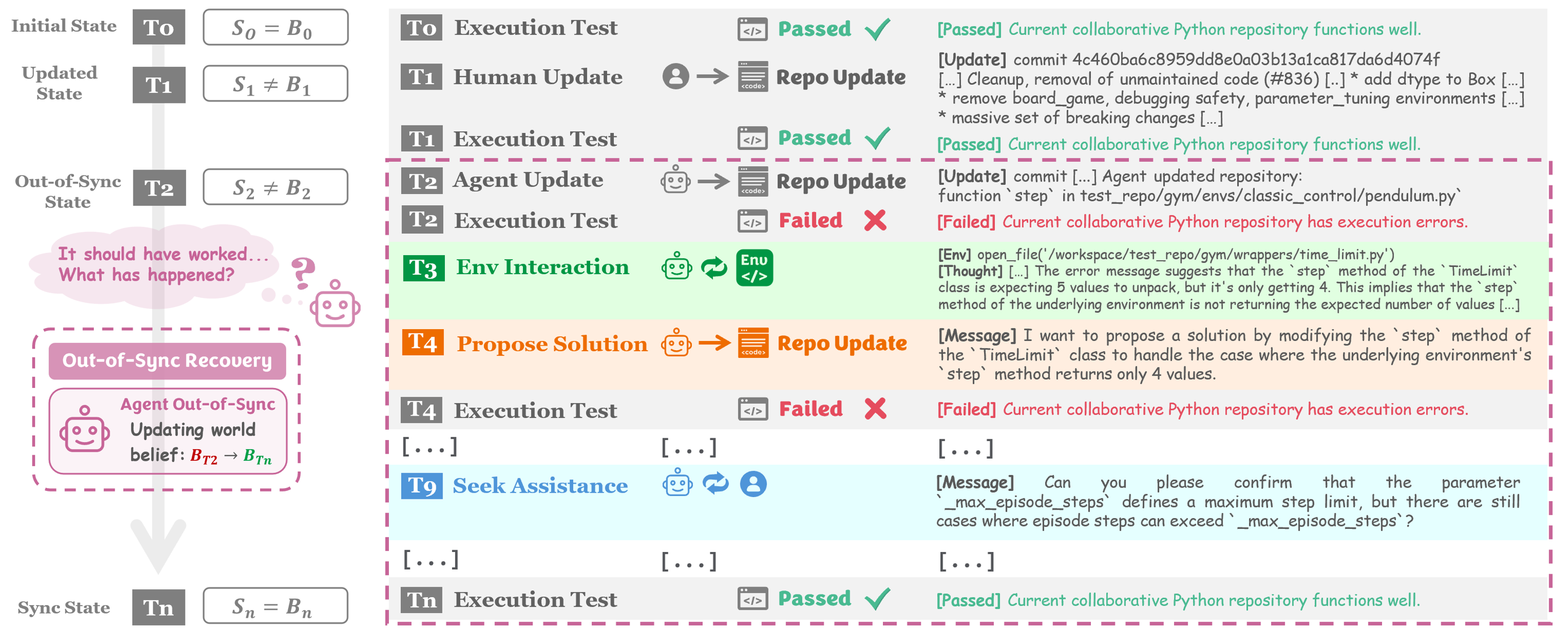}
    \vspace{-1.5em}  
    \caption{\textbf{\textit{SyncMind} for Measuring Agent \textit{Out-of-Sync} Recovery.} Depending on agent’s initiative, an agent updates its world belief ($B_2 \to B_n$) by \textcolor{fig2_env}{\textbf{interacting with \textit{Env}}}, \textcolor{fig2_code}{\textbf{proposing solutions}}, and \textcolor{fig2_ask}{\textbf{seeking collaborator assistance}}.}
    \label{fig:figure 2}
    \vspace{-1.0em}
\end{small}
\end{center}
\end{figure*}

In this work, we introduce \textbf{\textit{SyncMind}} (\Sref{Section: framework}), a framework that systematically defines the agent \textbf{\textit{out-of-sync}} problem in CSE (\fref{fig:figure 2}), where multiple collaborators frequently modify and update shared codebases.
This occurs when a collaborator's belief state ($B_k$) deviates from the actual world state ($S_k$) at time $T_k$, resulting in collaboration failures due to outdated information.
Consider a human-AI collaboration scenario in~\fref{fig:figure 1}:
while an \textcolor{fig1_agent}{\textit{Agent}} implements changes based on its understanding at time $T_i$, the \textcolor{fig1_human}{\textit{Human}} modifies the codebase at $T_j$ ($T_i<T_j<T_k$). 
The \textcolor{fig1_agent}{\textit{Agent}}'s subsequent update at $T_k$ becomes incompatible with the current state $S_k$ due to its outdated belief state $B_k$. 
This raises the key challenge: \textit{How can collaborators effectively recognize their belief being \textit{out-of-sync} ($B_k \neq S_k$), diagnose the root causes, and recover their belief $B_k$ to match the world state $S_k$}? 




\textit{SyncMind} facilitates multi-dimensional evaluation of collaborative coding agents: 
\begin{itemize}[noitemsep,topsep=0pt,parsep=2pt,partopsep=0pt,leftmargin=*]
\vspace{-5pt}

\item \textbf{\textit{Out-of-sync} recovery effectiveness} (\Sref{Section: Agent Recovery Capabilities}): We evaluate how agents detect and resolve state misalignments via exploring the environment and consulting fellow developers, enabling them to understand system changes and resynchronize after failures. 


\item \textbf{Collaborative tendency and effectiveness} (\Sref{Section: Collaborative Effectiveness}): We measure agents' tendency to engage in productive interactions with collaborators, a critical problem in CSE. 
By analyzing the assistance seeking rate and the performance difference in independent and collaborative working settings, we measure agents' recovery effectiveness in CSE. 

\item \textbf{Environmental awareness and resource allocation} (\Sref{Section: Resource Awareness}): 
We examine how agents balance independent problem-solving (\textit{i.e.,} exploring environment) with collaborative assistance. 
While excessive self-reliance in debugging can strain computational resources, over-dependence on peer support can burden collaborators through repetitive cycles of assistance requests, revisions, and testing. 
We evaluate resource allocation strategies in \textit{out-of-sync} recovery by analyzing recovery efficiency, considering computing time and expense budget.

%

\end{itemize}

%
%


%
%

%

Based on \textit{SyncMind}, we construct \textbf{\textit{SyncBench}} (\Sref{Section: Benchmark}), a testbed to assess agent \textit{out-of-sync} recovery in CSE.
Built upon 21 \github repositories, \textit{SyncBench} simulates real-world agent \textit{out-of-sync} through commit history traversal and multi-level filtering to obtain 24,332 instances with executable testing environments.
The construction pipeline is fully open-source and scalable, enabling seamless integration with additional repositories and supporting future development of CSE agents.
%
%
%
%
%
%
%
Through systematic evaluation (\Sref{Section: Evaluation Metrics}), our experiments on \textit{SyncBench} reveal fundamental patterns in existing LLM-based software agents (\Sref{Section: Experiments}):

\begin{itemize}[noitemsep,topsep=0pt,parsep=2pt,partopsep=0pt,leftmargin=*]
\vspace{-5pt}

\item \textbf{\textit{Out-of-sync} recovery capabilities:} Evaluated by five metrics (\Sref{Section: Evaluation Metrics}) focusing on dissimilar aspects of agents' abilities, we observe substantial ability gaps among LLM agents that persist in their performance despite varying types of recovery actions and task complexity (\Sref{Section: Agent Recovery Capabilities}).
%
%
This highlights the significance of strong multifaceted capabilities for effective \textit{out-of-sync} recoveries (\Sref{Section: Multifaceted Abilities for Effective Recoveries}).

\item \textbf{Collaboration willingness and abilities:} Collaborator assistance generally improves agents' recovery performance ($0.33\% \leq \Delta_{\textit{\text{collaborator}}} \leq 5.52\%$), while its effectiveness varies significantly with agents’ collaboration willingness (\Sref{Section: Beneficial Collaborator Influence Limited By Disadvantageous Agent Initiative}) and communication abilities (\Sref{Section: Collaborative Effectiveness}-\ref{Section: Effects of Task Complexity}).

\item \textbf{Resource awareness and adaptive utilization:} Our experiments reveal critical limitations in agents' resource awareness and adaptive utilization to efficiently utilize available resources when provided with various temporal and financial resource constraints (\Sref{Section: Resource Awareness}).


\end{itemize}

\section{SyncMind: Agent Out-of-Sync Recovery}
\label{Section: framework}


Tackling agent \textit{out-of-sync} (\Sref{Section: Definition of Agent Out-of-Sync}), we introduce our \textbf{\textit{SyncMind}} framework (\fref{fig:figure 1}) to systematically measure agent \textit{out-of-sync} recovery in two key dimensions: \textit{recovery effectiveness} via two types of recovery (\Sref{Section: Recovery Settings}) and \textit{resource efficiency} through resource-aware \textit{out-of-sync} recovery (\Sref{Section: Resource Awareness Module}).

\subsection{Definition of Agent Out-of-Sync}
\label{Section: Definition of Agent Out-of-Sync}

In collaborative environments, a state of `\textit{out-of-sync}' arises when a collaborator's belief state deviates from the project's state due to missed updates from other team members (\fref{fig:figure 1}). We propose the formal definition of `\textit{out-of-sync}' state, which applies to both human and AI agents.


Let $S_i$ be the true world state at time $T_i$, and $B_i$ be an agent's belief state. Starting from $T_i$ when the agent begins a task, the agent becomes \textit{out-of-sync} at $T_k$ ($T_i<T_k$) if any of the following conditions are satisfied:
\vspace{-5pt}
\begin{enumerate}[leftmargin=*,itemsep=1pt,topsep=0pt,parsep=0pt,label=(\arabic*)]
    \item \textbf{\textit{Knowledge gap}:} $\exists$ Update $U$ at time $T_j$ ($T_i<T_j<T_k$) where the agent lacks knowledge of $U$.
    \item \textbf{\textit{State mismatch}:} $B_k \neq S_k$.
    \item \textbf{\textit{Task failure}:} Task completion based on $B_k$ fails to achieve intended outcomes in $S_k$.
\end{enumerate}

Recovery from an \textit{out-of-sync} state therefore requires:
\vspace{-5pt}
\begin{enumerate}[leftmargin=*,itemsep=2pt,topsep=0pt,parsep=0pt,label=(\arabic*)]
    \item Identify the root causes ($U$) of the state mismatch ($B_k \neq S_k$).
    \item Acquire information of the missing update $U$.
    \item Update its belief state such that $B_n = S_n$ at some future time $T_n$ ($T_n>T_k$).
\end{enumerate}


\subsection{Agent Out-of-Sync Recovery}
\label{Section: Recovery Settings}

In \textit{SyncMind} (\fref{fig:figure 2}), an agent updates its \textit{out-of-sync} belief state to attain $B_n=S_n$ through two types of recovery:



\begin{itemize}[noitemsep,topsep=0pt,parsep=2pt,partopsep=0pt,leftmargin=*]
    \item \textbf{Independent Recovery.} Operating autonomously, independent agents update their world beliefs through \textcolor{fig2_env}{\textbf{interacting with \textit{Environment} (\textit{Env})}} and  \textcolor{fig2_code}{\textbf{proposing solutions}}, besides their reflection on prior experience and feedback.
    \item \textbf{Collaborative Recovery.} Collaborative agents can also take advantage of \textcolor{fig2_ask}{\textbf{collaborator assistance}} to update their belief states by interacting with other collaborative agents. 
\end{itemize}

\subsection{Resource-Aware Recovery}
\label{Section: Resource Awareness Module}

\begin{figure}[H]
\begin{center}
\begin{small}
\vspace{0em}
    \includegraphics[width=1\linewidth]{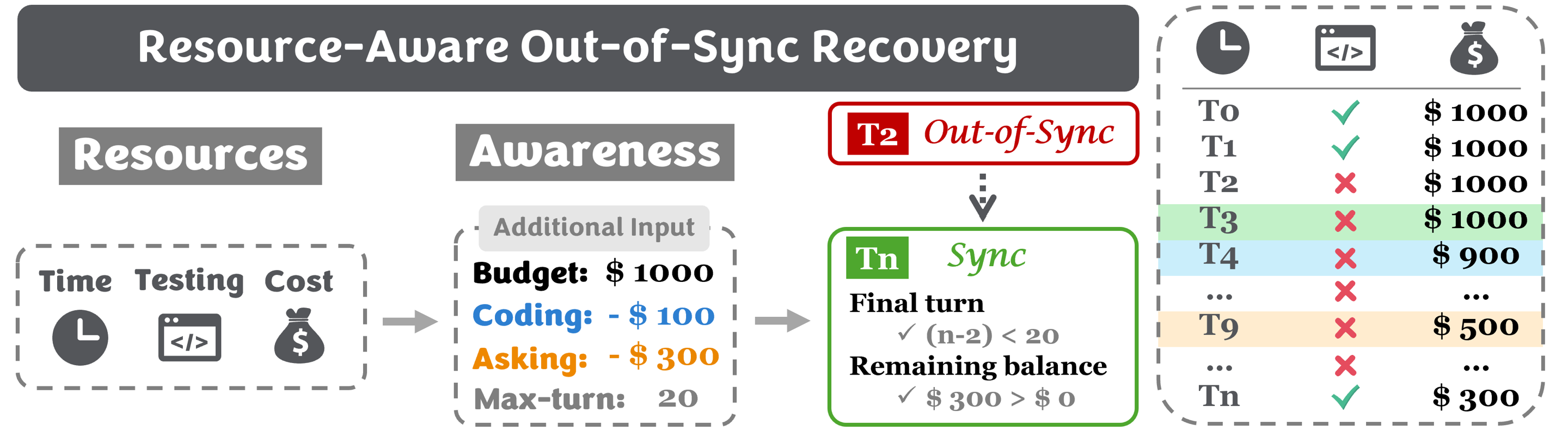}
    \vspace{-1.5em}
    \caption{\textbf{Resource-Aware \textit{Out-of-Sync Recovery}.} We introduce resource-aware recovery by mapping resource consumption to each \textit{out-of-sync} recovery task.}
    \vspace{-1.8em}
    \label{fig:figure 3 (resource awareness)}
\end{small}
\end{center}
\end{figure}

To reflect real-world resource constraints in collaborative environments, we integrate a resource awareness module into \textit{SyncMind} (\fref{fig:figure 3 (resource awareness)}).
This module tracks and constrains two dimensions of resources: 
(1) \textit{recovery time} measured as the number of turns taken for an agent to recover, 
and (2) hypothetical \textit{cost} that quantify financial resources consumed through the course of recovery (\textit{e.g.,} computing resources for debugging and testing, \textcolor{fig1_human}{\textit{Human}}'s time and effort to answer \textcolor{fig1_agent}{\textit{Agent}}'s questions). 
%
%
This resource-aware \textit{out-of-sync} recovery framework measures how agents utilize and adapt their strategies under different resource constraints, enabling comparisons of efficiency between successful and failed recovery attempts across agentic systems.



\begin{figure*}[!t]
\begin{center}
\begin{small}
    \vspace{0.2em}    
    \centering
    \includegraphics[width=1\linewidth]{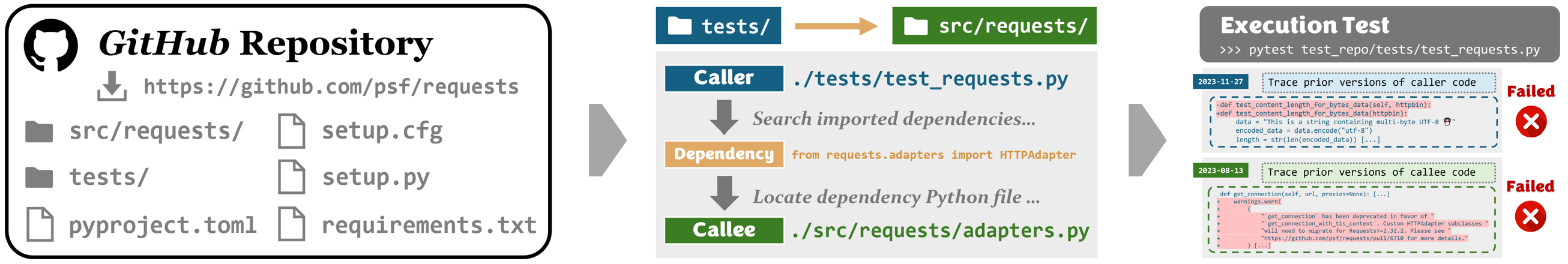}
    \vspace{-1.5em}  
    \caption{\textbf{Agent \textit{Out-of-Sync} Benchmark Construction.} A systematic benchmark construction approach (\Sref{Section: Benchmark Construction}).}
    \label{fig:figure 4}
    \vspace{-1.0em}
\end{small}
\end{center}
\end{figure*}

\section{SyncBench: Agent Out-of-Sync Benchmark}
\label{Section: Benchmark}



\subsection{Benchmark Construction}
\label{Section: Benchmark Construction}
\looseness=-1
Aligning with real-world \textit{out-of-sync} scenarios, our benchmark construction method is applicable to \textit{Python}-based GitHub repositories with existing \textit{unit tests}.
\textit{SyncBench} leverages 21 popular GitHub repositories and can be expanded to include additional repositories following our benchmark construction methodology (\Sref{Appendix: Benchmark Construction}).
In accordance with the definition of agent \textit{out-of-sync} (\Sref{Section: Definition of Agent Out-of-Sync}), our benchmark construction implements a systematic pipeline that takes all three conditions into consideration:

\textbf{\textit{Env} Configuration.}
We employ \textit{Docker} \cite{docker_for_env_config} to configure isolated, reproducible, and executable testing environments tailored for our \textit{out-of-sync} recovery tasks.
Each source repository is packaged into a dedicated \textit{Docker image} with complete codebase, dependencies, and validation infrastructure for unit test execution.
Each execution verification (\Sref{Section: Evaluation Metrics}) automatically creates an isolated \textit{container} instance with auto-removal upon completion, ensuring consistent and clean testing environments for reliable recovery evaluation.

\textbf{\textit{Out-of-Sync} Simulation.}
We first extract \textit{Python} functions and class methods (hereafter collectively referred to as \textit{functions}) from source repositories.
For each extracted function, we employ its up-to-date state as ground truth ($S_2$), while obtaining the \textit{out-of-sync} belief state ($B_2$) by tracing its \textit{Git} history reversely until identifying a commit ($B_2$) where execution fails ($B_2 \neq S_2$).
%
In this way, \textit{Caller} and \textit{Callee} are constructed through simulating unit test \textit{out-of-sync} and tested dependency \textit{out-of-sync}, respectively:
\textbf{(1) \textit{Caller}:} We roll back the testing function until it becomes \textit{out-of-sync};
\textbf{(2) \textit{Callee}:} We roll back imported dependency for tested module \textit{out-of-sync}, thereby presenting higher task complexity---agents need to understand dependency relationships and localize the problematic imported modules.

\textbf{Multi-level Quality Filtering.}
For each \textit{out-of-sync} instance, we execute unit tests before and after \textit{out-of-sync} happens and use the parsed test outputs to filter for high-quality instances.
Our \textit{parsing-based execution testing} (\Sref{Section: Evaluation Metrics}) requires the \textit{pass-to-fail} state divergence ($B_2 \neq S_2$):
(1) updated repository ($S_1$) passes the test to demonstrate ground truth validity, and (2) repository with the \textit{out-of-sync} function ($B_2$) fails the test to allow the \textit{out-of-sync} scenario to take shape.
To enhance data quality, we additionally apply a filter that retains only instances with their execution outputs comprising: (1) at least one execution error or unit test failure in $B_2$, (2) more than one passing test in $S_1$, (3) identical parsing result between $S_1$ and $S_n$.

\textbf{Weighted Downsampling.}
In constructing our evaluation subset with 300 representative instances\footnote{
\label{Footnote: Cost Constraints}
Due to the costly expenditure of extensive model evaluations (on average $\$0.56/instance$, ranging from \textit{GPT-4o mini} with $\$0.02/instance$, to \textit{Claude-3.5-Sonnet} with $\$1.73/instance$), we downsample a subset of \textit{SyncBench} with 300 instances (\Sref{Section: Benchmark}).} 
across 21 repositories, we downsample each repository's data to less than 15 instances while maintaining the original patch distribution over all sampled data, thereby applying the same task complexity distribution to all downsampled instances.

\subsection{Benchmark Datasets}
\label{Section: Benchmark Datasets}

Constructing \textit{SyncBench} with two complementary datasets---\textit{Caller} and \textit{Callee} (\fref{fig:figure 4}), our initial extraction yields 24,332 instances (\tref{tab:table B2 (Benchmark)}).
Pruning the raw dataset to 8,461 instances via multi-level filtering, the evaluation subset is further reduced via weighted downsampling.
As such, we finalize our evaluation samples as 300 instances\footref{Footnote: Cost Constraints} with evenly distributed \textit{Caller} and \textit{Callee} samples (150 each). 




\subsection{LLM-Simulated Collaborators}
\label{Section: LLM-simulated Collaborators}
We leverage LLMs to simulate both agents (who enter \textit{out-of-sync} states $B_2$) and know-everything collaborators ($S_2$).

\textbf{Agent Out-of-Sync.} We employ LLMs to power AI agents in \textit{out-of-sync} states, which allows belief states to become tractable and controllable throughout the recovery process. Meanwhile, this also supports the precise measurement of an agent's resource consumption and the systematic evaluation of an agent's recovery patterns.

\textbf{Simulating Know-Everything Collaborators.} 
Validated by single-turn experiments (\Sref{Section: Beneficial Collaborator Influence Limited By Disadvantageous Agent Initiative}), LLM-simulated know-everything collaborators are furnished with: (1) complete task context, including both $B_0-B_t$ and $S_0-S_t$ (where the agent seek assistance at $T_t$, $2<t<n$), (2) ground-truth solution to reach $B_n=S_n$, (3) update history ($U$ at $T_1$), and (4) task-specific response protocols (\Sref{Appendix D: LLM-simulated Human Answerer}).


\subsection{Evaluation Metrics}
\label{Section: Evaluation Metrics}
We propose five complementary metrics tailored for comprehensively evaluating agent \textit{out-of-sync} recovery:

\textbf{Success Rate (SR).}
We evaluate recovery success (\eqref{Eq: Evaluation Metric 1 (SR)}) through a two-stage validation process: \textbf{\textit{(1) Execution Test}}: Execution success can be reached only if an agent's updated repository passes the test without errors (\textit{i.e.,} command exit code of 0). 
\textbf{\textit{(2) Parsing Validation}}: We compare the parsed test execution outputs of an agent's proposed solution against that of the ground-truth state (\textit{i.e.,} the original commit without issue). Recovery success requires all parsed output of test cases for the agent-proposed solution to exactly match the ground-truth values.

{
\vspace{-1.0em}
\small
\begin{equation}
SR = \frac{\sum_{m \in \mathcal{M}} \mathbbm{1}(SR_m = 1)}{\sum_{m \in \mathcal{M}} \mathbbm{1}}
\label{Eq: Evaluation Metric 1 (SR)}
\end{equation}
\vspace{-1.8em}
}

{\small
where $\mathcal{M}$ represents the task space, $SR_m \in \{0,1\}$ suggests whether task $m$ achieves recovery success, $\mathbbm{1}(\cdot)$ is the indicator function that returns 1 when the condition is met and 0 otherwise.
}

\textbf{Localization Accuracy (LA).}
We evaluate an agent's ability to localize an \textit{out-of-sync} function at two levels: (a) \textit{file} ($LA_{file}$): accurately identifying the \textit{Python} file containing the \textit{out-of-sync} function; and (b) \textit{function} ($LA_{func}$): accurately pinpointing the specific \textit{out-of-sync} function.

{
\vspace{-0.5em}
\small
\begin{equation}
LA_f = \frac{\sum_{\mathcal{M}} \mathbbm{1}(LA_{f,m} = 1)}{\sum_{m \in \mathcal{M}} \mathbbm{1}}
\label{Eq: Evaluation Metric 2 (LA)}
\end{equation}
\vspace{-1.8em}
}


{\small
where $f \in \{file, func\}$ denotes the localization target type, $LA_{f,m} \in \{0,1\}$ indicates whether task $m$ achieves localization success for target type $f$.
}

\textbf{Conditional Success Rate (CSR).}
We evaluate agents' technical recovery abilities by conditioning recovery on localization success, which leads to $CSR_{file}$ and $CSR_{func}$:

{
\vspace{-1em}
\small
\begin{equation}
CSR_f = SR|_{LA_f=1} = \frac{\sum_{m \in \mathcal{M}} \mathbbm{1}(SR_m=1 \land LA_{f,m}=1)}{\sum_{m \in \mathcal{M}} \mathbbm{1}(LA_{f,m}=1)}
\label{Eq: Evaluation Metric 3 (CSR)}
\end{equation}
\vspace{-2em}
}


\textbf{Assistance Seeking Rate (ASR).}
We quantify a collaborative agent's willingness to collaborate as the proportion of recovery time (measured in turns of interactions) it adopts for proactive assistance-seeking.

{
\vspace{-1em}
\small
\begin{equation}
ASR = \frac{\sum_{m \in \mathcal{M}} \sum_{t \in \mathcal{T}_m} \mathbbm{1}(AS_t = 1)}{\sum_{m \in \mathcal{M}} \sum_{t \in \mathcal{T}_m} \mathbbm{1}}
\label{Eq: Evaluation Metric 4 (ASR)}
\end{equation}
\vspace{-2em}
}

{\small
where $\mathcal{T}$ represents the recovery time space, $\mathcal{T}_m (m \in \mathcal{M})$ suggests the total time in task $m$, $AS_t \in \{0,1\}$ indicates whether the agent seeks collaborator assistance in turn $t$.
}

\textbf{Recovery Efficiency.}
We compute the ratio of turns taken to the maximum time limit as the proxy for \textit{time efficiency}, thus excluding external influencing factors, like connection stability and memory capacity. \textit{Expense efficiency} is similarly calculated as the average financial expenditure rates.

{
\vspace{-1em}
\small
\begin{equation}
Eff_g = \frac{\sum_{m \in \mathcal{M}} \phi_g(m)}{\sum_{m \in \mathcal{M}} \psi_g(m)}
\label{Eq: Evaluation Metric 5 (Efficiency)}
\end{equation}
\vspace{-1.8em}
}

{\small
where $g \in \{time, expense\}$ denotes the efficiency type,
$\phi_{time}(m)=\sum_{t \in \mathcal{T}_m} \mathbbm{1}(a_t^m)$ counts turns taken for task $m$,
$\phi_{expense}(m)=\sum_{t \in \mathcal{T}_m} c(a_t^m)$ sums costs of actions for task $m$,
$\psi_{time}(m) = \mathcal{T}_{max}$ is the maximum time limit,
$\psi_{expense}(m) = \mathcal{C}_{max}$ is the maximum budget,
$a_t^m$ is the action taken at step $t$ in task $m$,
and $c(a)$ is the cost function for action $a$.
}

\section{Experiments}
\label{Section: Experiments}

\begin{table*}[htbp]
\begin{center}
\begin{small}
\vspace{-0.5em}
\caption{\textbf{Out-of-Sync Recovery Evaluation on \textit{Caller} and \textit{Callee}.} The influence of increased task complexity introduced by dependency tracing on agents' \textit{out-of-sync} recovery performance: $\Delta_\textit{complexity}=\Delta_\textit{(Callee-Caller)}$.}
\label{tab:table 2 (caller vs callee)}
\vspace{1.0em}  
    \begin{tabular}{>{\centering\arraybackslash}m{2.4cm}|>{\centering\arraybackslash}m{1.8cm}||>{\centering\arraybackslash}m{0.9cm}>{\centering\arraybackslash}m{0.9cm}>{\centering\arraybackslash}m{0.9cm}|>{\centering\arraybackslash}m{0.9cm}>{\centering\arraybackslash}m{0.9cm}>{\centering\arraybackslash}m{0.9cm}|>{\centering\arraybackslash}m{0.9cm}>{\centering\arraybackslash}m{0.9cm}>{\centering\arraybackslash}m{0.9cm}}
    
    \toprule
    \multirow{2}{*}{\centering \textbf{Agent}} & \multirow{2}{*}{\centering \textbf{Recovery}} & \multicolumn{3}{c|}{\textbf{Caller (\%)}} & \multicolumn{3}{c|}{\textbf{Callee (\%)}} & \multicolumn{3}{c}{\textbf{$\Delta_\textit{complexity}$ (\%)}} \\
    
    \cmidrule{3-11}
    &  & $LA_{file}$ & $LA_{func}$ & $SR$ & $LA_{file}$ & $LA_{func}$ & $SR$ & $LA_{file}$ & $LA_{func}$ & $SR$ \\
    
    \midrule
    
    \multirow{3}{*}{\centering \textbf{Llama-3.1-8B}} 
    & \textbf{Independent} & 13.33 & 8.00 & 1.33 & 4.00 & 1.33 & 0.67 & \cellcolor{basecolor_red!46.65} -9.33 & \cellcolor{basecolor_red!33.35} -6.67 & \cellcolor{basecolor_red!3.30} -0.66 \\
    & \textbf{Collaborative} & 32.00 & 26.00 & 2.00 & 22.00 & 16.67 & 0.67 & \cellcolor{basecolor_red!50.0} -10.00 & \cellcolor{basecolor_red!46.65} -9.33 & \cellcolor{basecolor_red!6.65} -1.33 \\
    & \textbf{$\Delta_{\textit{\text{collaborator}}}$} & \cellcolor{basecolor_green!93.35} +18.67 & \cellcolor{basecolor_green!90.0} +18.00 & \cellcolor{basecolor_green!3.35} +0.67 & \cellcolor{basecolor_green!90.0} +18.00 & \cellcolor{basecolor_green!76.7} +15.34 & \cellcolor{basecolor_green!0.00} +0.00 & \cellcolor{basecolor_red!3.35} -0.67 & \cellcolor{basecolor_red!13.3} -2.66 & \cellcolor{basecolor_red!3.35} -0.67 \\
    
    \midrule
    
    \multirow{3}{*}{\centering \textbf{Llama-3.1-70B}} 
     & \textbf{Independent} & 8.67 & 5.33 & 4.00 & 8.00 & 4.67 & 1.33 & \cellcolor{basecolor_red!3.35} -0.67 & \cellcolor{basecolor_red!3.3} -0.66 & \cellcolor{basecolor_red!13.35} -2.67 \\
    & \textbf{Collaborative} & 12.00 & 6.00 & 3.33 & 12.67 & 5.33 & 3.33 & \cellcolor{basecolor_green!3.35} +0.67 & \cellcolor{basecolor_red!3.35} -0.67 & \cellcolor{basecolor_green!0.00} +0.00 \\
    & \textbf{$\Delta_{\textit{\text{collaborator}}}$} & \cellcolor{basecolor_green!16.65} +3.33 & \cellcolor{basecolor_green!6.65} +1.33 & \cellcolor{basecolor_red!3.35} -0.67 & \cellcolor{basecolor_green!23.35} +4.67 & \cellcolor{basecolor_green!3.3} +0.66 & \cellcolor{basecolor_green!10.0} +2.00 & \cellcolor{basecolor_green!6.7} +1.34 & \cellcolor{basecolor_red!3.35} -0.67 & \cellcolor{basecolor_green!13.35} +2.67 \\
    
    \midrule
    
    \multirow{3}{*}{\centering \textbf{GPT-4o mini}} 
    & \textbf{Independent} & 13.29 & 9.30 & 5.32 & 7.97 & 6.64 & 2.66 & \cellcolor{basecolor_red!26.60} -5.32
 & \cellcolor{basecolor_red!13.30} -2.66 & \cellcolor{basecolor_red!13.30} -2.66 \\
    & \textbf{Collaborative} & 15.28 & 11.96 & 7.97 & 9.30 & 4.65 & 2.66 & \cellcolor{basecolor_red!29.9} -5.98 & \cellcolor{basecolor_red!36.55} -7.31 & \cellcolor{basecolor_red!26.55} -5.31 \\
    & \textbf{$\Delta_{\textit{\text{collaborator}}}$} & \cellcolor{basecolor_green!9.95} +1.99 & \cellcolor{basecolor_green!13.3} +2.66 & \cellcolor{basecolor_green!13.25} +2.65 & \cellcolor{basecolor_green!6.65} +1.33 & \cellcolor{basecolor_red!9.95} -1.99 & \cellcolor{basecolor_green!0.0} +0.00 & \cellcolor{basecolor_red!3.30} -0.66 & \cellcolor{basecolor_red!23.25} -4.65 & \cellcolor{basecolor_red!13.25} -2.65 \\
    
    \midrule

    \multirow{3}{*}{\centering \textbf{DeepSeek}} 
    & \textbf{Independent} & 58.00 & 47.33 & 8.67 & 37.33 & 22.67 & 6.00 & \cellcolor{basecolor_red!62.01} -20.67 &  \cellcolor{basecolor_red!73.98} -24.66  &  \cellcolor{basecolor_red!13.35} -2.67  \\
    & \textbf{Collaborative} & 52.00 & 47.33 & 8.67 & 42.00 & 27.33 & 6.67 & \cellcolor{basecolor_red!30.0} -10.00 & \cellcolor{basecolor_red!60.0} -20.00 &  \cellcolor{basecolor_red!10.0} -2.00  \\
    & \textbf{$\Delta_{\textit{\text{collaborator}}}$} & \cellcolor{basecolor_red!30.0} -6.00 & \cellcolor{basecolor_green!0.0} +0.00 & \cellcolor{basecolor_green!0.0} +0.00 & \cellcolor{basecolor_green!23.35} +4.67 & \cellcolor{basecolor_green!23.3} +4.66 & \cellcolor{basecolor_green!3.35} +0.67 & \cellcolor{basecolor_green!53.35} +10.67 & \cellcolor{basecolor_green!23.30} +4.66 & \cellcolor{basecolor_green!3.35} +0.67 \\
    
    \midrule

    \multirow{3}{*}{\centering \textbf{GPT-4o}} 
    & \textbf{Independent} & 14.67 & 11.33 & 6.67 & 14.00 & 7.33 & 1.33 & \cellcolor{basecolor_red!3.35} -0.67 & \cellcolor{basecolor_red!20.0} -4.00 & \cellcolor{basecolor_red!26.70} -5.34 \\
    & \textbf{Collaborative} & 39.33 & 35.33 & 10.00 & 38.67 & 34.00 & 6.00 & \cellcolor{basecolor_red!3.30} -0.66 & \cellcolor{basecolor_red!6.65} -1.33 & \cellcolor{basecolor_red!20.00} -4.00 \\
    & \textbf{$\Delta_{\textit{\text{collaborator}}}$} & \cellcolor{basecolor_green!123.3} +24.66 & \cellcolor{basecolor_green!120.0} +24.00 & \cellcolor{basecolor_green!16.65} +3.33 & \cellcolor{basecolor_green!123.35} +24.67 & \cellcolor{basecolor_green!133.35} +26.67 & \cellcolor{basecolor_green!23.35} +4.67 & \cellcolor{basecolor_green!0.05} +0.01 & \cellcolor{basecolor_green!13.35} +2.67 & \cellcolor{basecolor_green!6.70} +1.34 \\
    
    \midrule

    \multirow{3}{*}{\centering \textbf{Llama-3.3-70B}} 
    & \textbf{Independent} &  80.67  &  60.00  &  18.67  &  47.33  &  34.67  &  14.00  &  \cellcolor{basecolor_red!100.02} -33.34  &  \cellcolor{basecolor_red!75.99} -25.33  &  \cellcolor{basecolor_red!23.35} -4.67  \\
    & \textbf{Collaborative} &  77.33  &  64.67  &  22.00  &  56.00  &  42.67  &  16.00  &  \cellcolor{basecolor_red!63.99} -21.33  &  \cellcolor{basecolor_red!66.00} -22.00  &  \cellcolor{basecolor_red!30.00} -6.00  \\
    & \textbf{$\Delta_{\textit{\text{collaborator}}}$} &  \cellcolor{basecolor_red!16.7} -3.34  &  \cellcolor{basecolor_green!23.35} +4.67  &  \cellcolor{basecolor_green!16.65} +3.33  &  \cellcolor{basecolor_green!43.35} +8.67  &  \cellcolor{basecolor_green!40.0} +8.00  &  \cellcolor{basecolor_green!10.0} +2.00 &  \cellcolor{basecolor_green!60.05} +12.01  &  \cellcolor{basecolor_green!16.65} +3.33  &  \cellcolor{basecolor_red!6.65} -1.33  \\
    
    \midrule

    \multirow{3}{*}{\centering \textbf{Claude-3.5-Sonnet}} 
    & \textbf{Independent} & 50.83 & 47.51 & 25.41 & 77.35 & 65.19 & 30.94 & \cellcolor{basecolor_green!132.6} +26.52 & \cellcolor{basecolor_green!88.4} +17.68 & \cellcolor{basecolor_green!27.65} +5.53 \\
    & \textbf{Collaborative} & 43.09 & 38.67 & 28.73 & 79.56 & 65.19 & 38.67 & \cellcolor{basecolor_green!182.35} +36.47 & \cellcolor{basecolor_green!132.60} +26.52 & \cellcolor{basecolor_green!49.70} +9.94 \\
    & \textbf{$\Delta_{\textit{\text{collaborator}}}$} & \cellcolor{basecolor_red!38.7} -7.74 & \cellcolor{basecolor_red!44.2} -8.84 & \cellcolor{basecolor_green!16.6} +3.32 & \cellcolor{basecolor_green!11.05} +2.21 & \cellcolor{basecolor_green!0.0} +0.00 & \cellcolor{basecolor_green!38.65} +7.73 & \cellcolor{basecolor_green!49.75} +9.95 & \cellcolor{basecolor_green!44.2} 
    +8.84 & \cellcolor{basecolor_green!22.05} +4.41 \\
    
    \bottomrule
    
    \end{tabular}
\end{small}
\end{center}
\vspace{0.5em}
\end{table*}

\subsection{Setup}
\label{Section: Experiment Setup}


\textbf{Recovery Protocol.}
For baselines, each agent is allowed up to $30$ turns to achieve $B_n=S_n$, which is then extended to $50$ turns to assess agents' temporal resource awareness and exploitation.
Financial resources are mapped similarly to each resource-aware recovery task.
Provided with different action options---\textcolor{fig2_env}{\textbf{interacting with \textit{Env}}}, \textcolor{fig2_code}{\textbf{proposing a solution}}, or \textcolor{fig2_ask}{\textbf{proactively seeking collaborator assistance}} (\Sref{Section: Recovery Settings})---both independent and collaborative agents take each of their moves autonomously.

\textbf{\textit{Env} Space.}
We employ \textit{OpenHands} \cite{OpenHands} to empower agents to autonomously explore and inspect the codebase environment by executing various commands.
This exploration enables them to develop a comprehensive understanding of the codebase for \textit{out-of-sync} recovery.

\textbf{Agents.} Our experiments assess the \textit{out-of-sync} recovery capabilities of seven LLMs, including four open-source (\textit{Meta-Llama-3.1-8B, Meta-Llama-3.1-70B,  Meta-Llama-3.3-70B, and DeepSeek-V2.5}) and three close-source (\textit{Claude-3.5-Sonnet, GPT-4o mini, and GPT-4o}) LLMs \cite{meta_llama_3.1_8B, meta_llama_3.1_70B, meta_llama_3.3_70B, deepseek_v2.5, claude_3.5_sonnet, gpt_4o, gpt_4o_mini} in two recovery settings (\Sref{Section: Recovery Settings}), respectively.

\begin{figure}[H]
\begin{center}
\begin{small}
\includegraphics[width=1\linewidth]{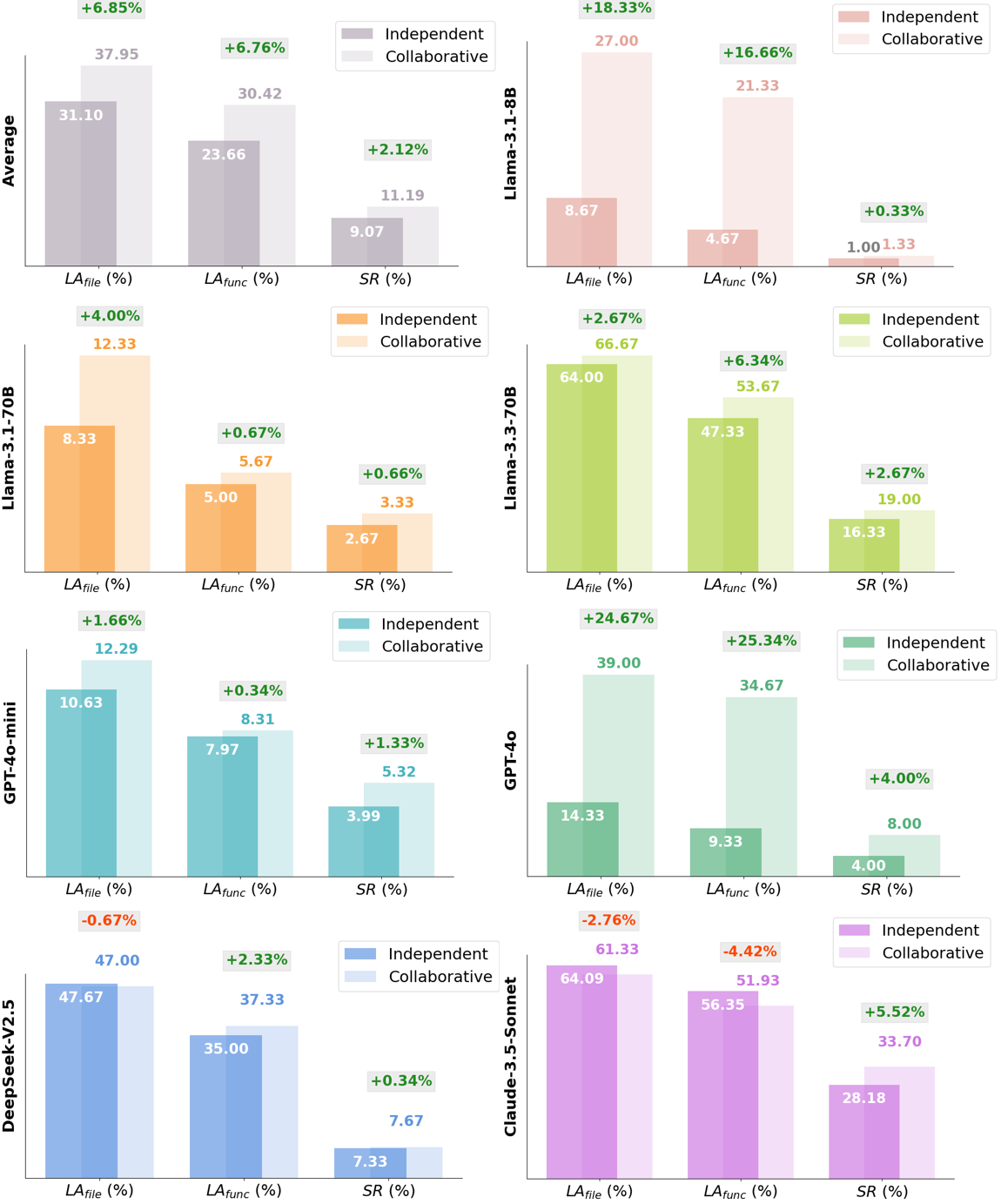}
\vspace{-2em}
    \caption{\textbf{Influence of Collaborator Assistance.} We quantify collaborator influence on agent \textit{out-of-sync} recovery performance as $\Delta_{\textit{\text{collaborator}}}$ to unveil its \textcolor{fig5_green}{\textbf{+positive}} or \textcolor{fig5_red}{\textbf{-negative}} impact on certain aspects of agents' recovery performance.}
    \vspace{-1.2em}  
    \label{fig:figure 5 (recovery performance and human influence)}
\end{small}
\end{center}
\end{figure}

\subsection{Significant Ability Gaps Among Agents Powered by Different LLMs}
\label{Section: Agent Recovery Capabilities}

Our experiments on \textit{SyncBench} (\tref{tab:table 2 (caller vs callee)}-\ref{tab:table 1 (performance summary)}) reveal substantial capability gaps among seven LLM agents.

\textbf{Baselines for Out-of-Sync Recovery.}
LLM agents' independent \textit{out-of-sync} recoveries demonstrate significant variations in their baseline capabilities, ranging from \textit{Claude-3.5-Sonnet} ($SR=28.18\%$) to \textit{Llama-3.1} agents ($SR \leq 2.67\%$).
Their localization capabilities also vary remarkably, regardless of pinpointing the exact \textit{out-of-sync} function ($LA_{func} \in [4.67, 56.35]\%$) or less precisely localizing responsible \textit{Python} files ($LA_{file} \in [8.33, 64.09]\%$). Likewise, our evaluation on agents' technical capabilities (\tref{tab:table 1 + CSR (overall performance summary)}) also exhibits substantial gaps among LLMs.

\textbf{Persistent Gaps Despite Varying Recovery Conditions.}
\tref{tab:table 1 (performance summary)} and \tref{tab:table 1 + CSR (overall performance summary)} show similar performance disparities for collaborative agents mirroring their independent recoveries, despite the generally positive influence of collaborative assistance.
These performance gaps remain significant for tasks of different complexity (\tref{tab:table 2 (caller vs callee)} \& \ref{tab:table 2 + CSR (caller vs callee)}).
Agents' persistent performance variances across diverse task scenarios highlight their underlying ability gaps in identifying and resolving \textit{out-of-sync} to maintain effective collaborations.

\subsection{In Achieving Recovery Success: Technical, Reasoning, and Collaborative Competences}
\label{Section: Multifaceted Abilities for Effective Recoveries}

Conditioned on localization success, $CSR$ (\eqref{Eq: Evaluation Metric 3 (CSR)}) is significantly influenced by how much time left for technical recovery after accurate localizations, which are largely determined by agents' abilities to efficiently identify root causes of $B_k \neq S_k$.
Comparing \tref{tab:table 2 (caller vs callee)}-\ref{tab:table 1 (performance summary)} with \tref{tab:table 1 + CSR (overall performance summary)}-\ref{tab:table 2 + CSR (caller vs callee)}, low-performing agents can also showcase strong technical problem-solving capacities (\textit{e.g.,} $CSR$: \textit{Llama-3.1-70B} $>$ \textit{GPT-4o}), despite their notably underperformed localization and recovery abilities (\textit{e.g.,} \textit{Llama-3.1-70B}: $LA_{file} \leq 12.33\%$, $LA_{func} \leq 5.67\%$, $SR \leq 3.33\%$) and remarkably low willingness to collaborate (e.g., \textit{Llama-3.1-70B}: $ASR=1.37\%$).
This observation further substantiates that successful \textit{out-of-sync} recoveries hinge on not only agents' technical problem-solving proficiency, but their efficient cause analysis and effective collaboration capabilities.

\subsection{Collaborative Assistance Improves Performance---
But Agents Seldom Seek Help
}
\label{Section: Beneficial Collaborator Influence Limited By Disadvantageous Agent Initiative}

\textbf{Positive Collaborator Influence.}
As shown in \fref{fig:figure 5 (recovery performance and human influence)}, collaborator assistance generally improves recovery performance ($SR: +\Delta_{\textit{\text{collaborator}}}$$\in$$[0.33, 5.52]\%$), with the magnitude varying dependent on agents' technical capabilities and willingness to collaborate.
Agents with stronger independent recovery capabilities and collaboration willingness (\textit{Claude-3.5-Sonnet}: $SR=28.18\%$ and $ASR=4.86\%$), together with conditioned technical proficiency (\tref{tab:table 1 + CSR (overall performance summary)}), obtain higher performance gains ($\Delta_{\textit{\text{collaborator}}}=+5.52\%$) than other agents ($SR \leq 4.00\%$ and $ASR \leq 2.98\%$).

\textbf{Performance Upper Bound: Solving task with Oracle Information.}
To establish the theoretical upper bound of collaborator influence and identify agents' collaboration capability gaps, we additionally conduct a single-turn experiment by providing agents with oracle information that is used to simulate the collaborator (\textit{i.e.,} the know-everything agent). With \textit{GPT-4o mini} as the agent tackling \textit{out-of-sync} and \textit{GPT-4o} as the know-everything collaborator (\Sref{Section: LLM-simulated Collaborators}), we configure each single-turn task with collaborator’s exhaustive task-specific natural language instructions on how to accomplish recovery success.
Furnished with full recovery instructions ($ASR=100\%$), the high upper bound ($SR=86.33\%$) lends further evidence to LLM agents' significantly untapped potential in effective collaboration.

\begin{figure*}[!t]
\begin{center}
\begin{small}
\vspace{0.2em}
    \includegraphics[width=1\linewidth]{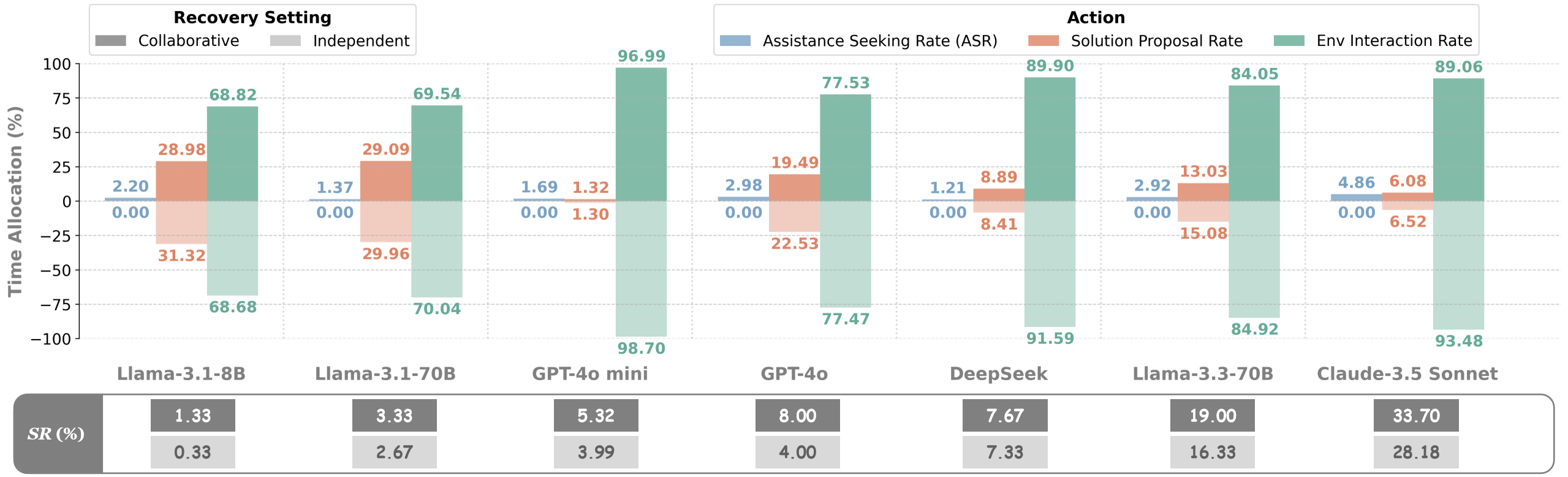}
    \vspace{-1.5em}
    \caption{\textbf{Time Allocation.} Agents' performance are ranked from low to high according to their \textit{independent} $SR$ scores, based on which they are positioned on the \textit{X}-axis from left to right. The \textit{Y}-axis depicts each agent's time allocation.}
    \vspace{-0.5em}  
    \label{fig:figure 7 (Time Allocation)}
\end{small}
\end{center}
\end{figure*}

\textbf{LLM Agents' Low Willingness to Collaborate.}
Despite strong technical capacities (\tref{tab:table 1 + CSR (overall performance summary)}), LLM agents show limited collaboration willingness (\fref{fig:figure 7 (Time Allocation)}).
Nevertheless, \textit{Claude-3.5-Sonnet} with the highest performance ($SR=33.70\%$) and collaboration willingness ($ASR=4.86\%$) derives the most benefit from collaboration ($\Delta_{\textit{\text{collaborator}}}=+5.52\%$).
It is followed by \textit{GPT-4o} who obtains notable improvements in both $LA$ ($\Delta_{\textit{\text{collaborator}}}=+25.34\%$) and $SR$ ($\Delta_{\textit{\text{collaborator}}}=+4.00\%$) through proactive assistance seeking.
The lowest $ASR=1.21\%$ presented by \textit{DeepSeek} ($SR: \Delta_{\textit{\text{collaborator}}}=+0.34\%$) contrastingly substantiates the significance of proactive collaboration initiative.

\subsection{Quality and Strategy of Communication Are Crucial for Recovery Success
}
\label{Section: Collaborative Effectiveness} 

\begin{figure}[H]
    \centering
    \vspace{0em}  
    \includegraphics[width=1\linewidth]{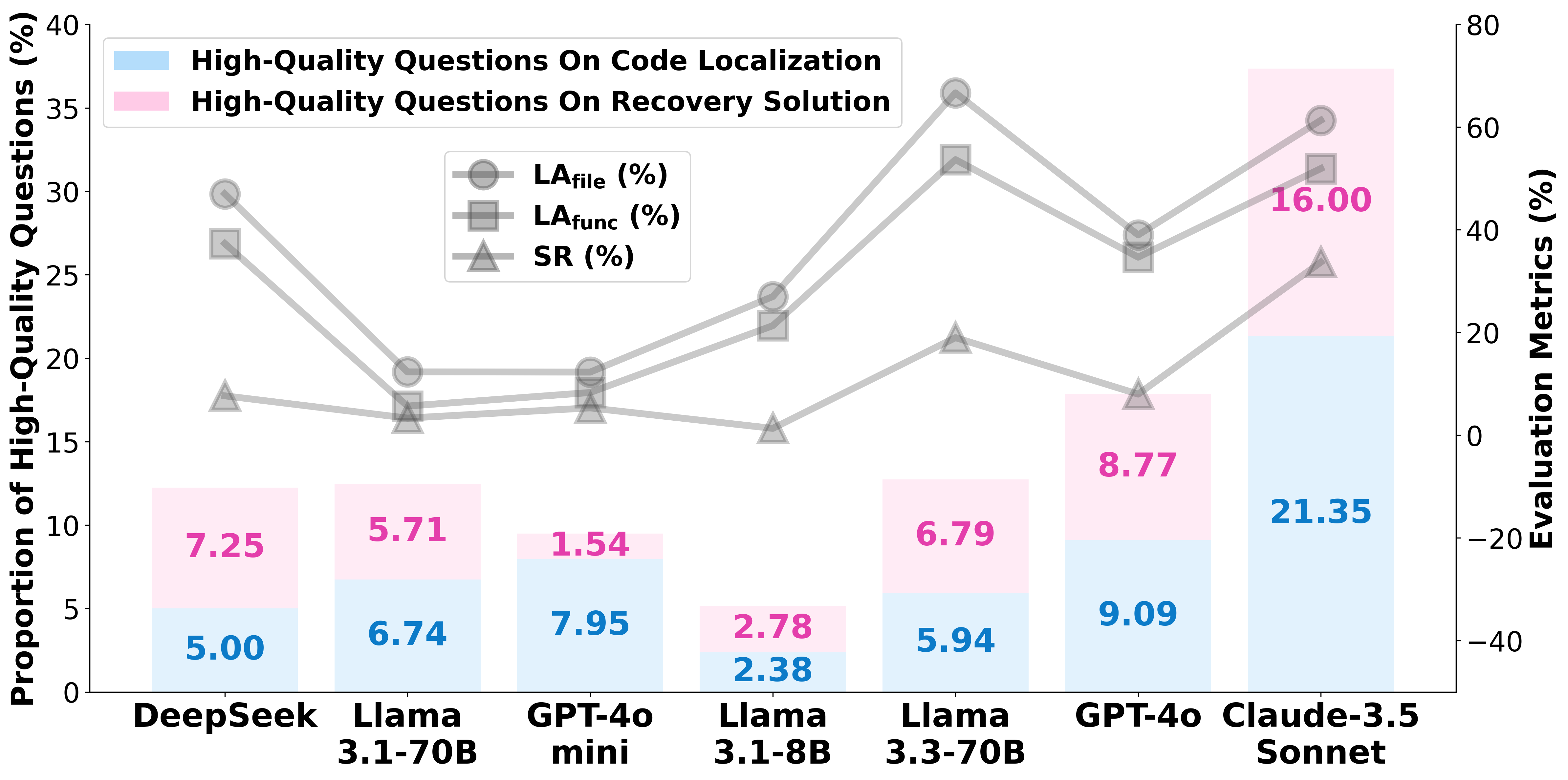}
    \vspace{-1.5em}
    \caption{\textbf{Question Quality.} Agents from left to right on the \textit{X}-axis according to their $ASR$ from low to high.}
    \vspace{0.5em}  
    \label{fig:figure 6 (Time Allocation) (Question Quality)}
\end{figure}

The quality and strategy of communication prove crucial for recovery success, with several key patterns emerging:

\textbf{Agents with More High-Quality Questions Achieve Better Performance.} 
Depending on whether the question asked by the agent can lead to recovery success (\Sref{Appendix:C.4 (Effective Assistance Seeking)}), we rate the quality of each query as low (\textit{resulting in recovery failure}) or high (\textit{resulting in recovery success}), which can be further classified into two general categories: \textit{\textcolor{fig7_blue}{localization queries}} closely related to localizing \textit{out-of-sync} causes and \textit{\textcolor{fig7_salmon}{solution queries}} seeking guidance on \textit{out-of-sync} resolution. 
Despite no significant correlation between query volume and recovery success, agents with a larger proportion of high-quality questions
achieve higher performance (\fref{fig:figure 6 (Time Allocation) (Question Quality)}).

\textbf{Strategic Early Exploration Facilitates Recovery Success.}
We compute each agent's communication timing distribution respectively for its success and failure cases (\fref{fig:figure C1 (first-half action distribution)}).
Results reveal that the proportion of assistance seeking in agents' first half of recovery time is substantially larger in success cases ($85.71\%-100.00\%$) than in recovery failures ($55.76\%-97.93\%$).
As top-performing agents exhibit distinct communication strategies with front-load queries, random or back-loaded assistance-seeking demonstrates less effective improvements on agents' performance. Compared with solution proposal timing that shows trivial differences between success and failure cases (averagely $2.79$ \textit{turns delayed} in successful recoveries), collaborative agents benefit markedly more from advancing their assistance seeking (averagely $10.50$ \textit{turns ahead} in successful recoveries).

\subsection{More Challenging Tasks Decrease Performance While Better Manifest Collaboration Benefits}
\label{Section: Effects of Task Complexity}


We observe a large negative influence of increased task complexity on agents' recoveries.
%
%
\textit{Callee}'s additional dependency tracing allows it to serve more challenging \textit{out-of-sync} tasks (\Sref{Section: Benchmark Construction}).
Comparing agents' performance between \textit{Caller} and \textit{Callee} (\tref{tab:table 2 (caller vs callee)} \& \ref{tab:table 2 + CSR (caller vs callee)}), \textit{Claude-3.5-Sonnet}'s performance gains ($+\Delta_\textit{complexity}$) demonstrate its superior technical capabilities in resolving complicated \textit{out-of-sync} tasks.
Nevertheless, \textit{Callee}, presenting higher task complexity, in general undermines agents' performance ($-\Delta_\textit{complexity}$).

Leveraging dissimilar complexity levels of 21 source repositories (\fref{fig:figure C8 (repo-wise performance)}), our repository-wise evaluation reflects consonant patterns between task complexity and recovery success.
While the repository \texttt{11-whisper} proposes the least recovery difficulty ($SR$: \textit{Independent} $33.33\%$, \textit{Collaborative} $22.22\%$),
the lowest performance delivered on the repository \texttt{13-sphinx} ($SR$: \textit{Independent} $0.88\%$, \textit{Collaborative} $4.70\%$) serves more challenging tasks.

Although repository complexity manifests negative correlations with recovery success, \textbf{\textit{the effectiveness of collaborator assistance increases on more challenging tasks}}, comparing to its trivial or negative influence on agents' recoveries in simpler \textit{out-of-sync} scenarios (\textit{e.g.,} \texttt{11-whisper} with $\Delta_{\textit{\text{collaborator}}}=-11.11\%$, in contrast to \texttt{13-sphinx} with $\Delta_{\textit{\text{collaborator}}}=+3.82\%$).
This holds in \textit{Callee} where agents notably gain more benefits from collaborator assistance (\tref{tab:table 2 (caller vs callee)}), despite higher task complexity.




\subsection{Agents' Significant Lack of Resource Awareness}
\label{Section: Resource Awareness}


We systematically vary resource constraints (Tables \ref{tab:table C2 (Time Awareness)}-\ref{tab:table C1 (Appendix: Recovery Efficiency)}) to investigate LLM-based agents' resource awareness in two key dimensions (\Sref{Section: Resource Awareness Module}):
(1) \textit{Time Resources}, through comparing the standard $30$-turn recovery with extended $50$-turn performance;
(2) \textit{Financial Resources}, through varying initial budgets from $\$1000$ (insufficient for $30$-turn costs) to $\$3000$ (adequate for any $30$-turn action taking patterns), and halving or doubling the cost of collaborator assistance.
Our experiments reveal critical limitations in agents' resource awareness and adaptive resource utilization capabilities.

\textbf{More Recovery Time Can Not Guarantee Performance Gains.}
Extended time produces divergent effects (\fref{fig:figure C3 (time awareness - allocation)} \& \tref{tab:table C2 (Time Awareness)}):
diminishing returns on \textit{Llama-3.1-8B} ($SR$: \textit{Independent} $-0.33\%$, \textit{Collaborative} $-1.00\%$),
while notable improvements on \textit{Llama-3.1-70B} ($SR$: \textit{Independent} $+3.67\%$, \textit{Collaborative} $+4.67\%$).
This observation suggests that extending the recovery time limit alone is insufficient for improved performance, while LLM agents' competences in effective time utilization and technical expertise (\tref{tab:table 1 + CSR (overall performance summary)}) factor underlyingly into recovery success.

\textbf{Agents' Low Sensitivity to Financial Resources.}
Tripling the initial budget yields trivial changes in action planning ($ASR$ improvements: $[+0.22,+1.06]\%$, \fref{fig:figure C6 (budget awareness - allocation)}) and recovery performance ($SR$ variations: $[-0.66,+1.67]\%$, \tref{tab:table C3 (Budget Awareness)}). Similarly, halving or doubling the action cost of assistance seeking contributes to negligible differences in LLM agents' willingness to collaborate ($ASR$ variations: $[-0.11,+0.23]\%$ for halved cost,  $[-0.46,+0.04]\%$ for doubled cost, \fref{fig:figure C7 (cost awareness - allocation)}) and recovery performance ($SR$ variations: $[-2.00,-1.00]\%$ for halved cost, $[-1.00,+1.00]\%$ for doubled cost, \tref{tab:table C4 (Cost Awareness)}).
These findings highlight the fundamental deficiencies of existing LLM agents in effectively recognizing resource constraints and adaptively leveraging available resources \cite{agent_alignment}.

\section{Related Work}

\subsection{Theory of Mind in Collaborative Systems}

The concept of ToM---the ability to model and reason about others' mental states---has emerged as the foundation of collaborative systems. 
LLMs' ToM capabilities have been evaluated and applied to various domains through benchmark construction \cite{benchmark_tombench, benchmark_opentom, benchmark_Hi_ToM, benchmark_fantom} and experimental investigation \cite{LLM_ToM_human_like_reasoning, ToM_multimodal_video_LLM, ToM_nature, spontaneous_ToM, improve_LLM_ToM, ToM_in_HRI}.
We apply ToM to software engineering agents for their maintaining of a shared understanding of codebase states, which is particularly of critical essence in asynchronous collaboration environments where \textit{out-of-sync} situations frequently arise due to temporal gaps between contributions.

\subsection{Collaborative Software Engineering Systems}

The evolution of software engineering tools and practices relies heavily on the premise of synchronized collaboration. Modern version control systems \cite{version_control_git, version_control_PredictingMC, version_control_NanoVC}, although implement sophisticated mechanisms for detecting and resolving conflicts arising from divergent codebase states, primarily address syntactic conflicts rather than semantic understanding divergence. 
Recent work showcases that LLM reasoning can be effectively advanced by diverse means, such as human feedback \cite{human-sim-mint, image_retrieval}, \textit{Env} interaction \cite{OpenHands, CodeAct, CodeAgent}, multi-agent cooperation \cite{AgentCoder}, etc.
Pressing closer to real-world repository-level programming, advances in LLMs inspire their agentic engagements in software engineering \cite{LLM_in_software_engineering}, in addition to benchmarks built upon \github repositories \cite{SWE-bench, R2E, benchmark_CoReQA}.
However, real-world dynamic environments require adaptations for collaborative systems based on the presumption of relatively static environments. Agents' lack of automatic synchronization and resource awareness therefore arise as latent obstacles that significantly impede collaborative intelligence and resource efficiency that we aim to provide insights into.

\section{Conclusion and Discussion}
\label{Section: Conclusion and Discussion}

In this paper, we investigate the \textit{out-of-sync} challenge in collaborative software engineering by introducing our \textit{out-of-sync} recovery framework, \textbf{\textit{SyncMind}} \Sref{Section: framework}, and evaluation benchmark, \textbf{\textit{SyncBench}} \Sref{Section: Benchmark}.
Experiments reveal that successful \textit{out-of-sync} recoveries require not only
technical proficiency (\Sref{Section: Agent Recovery Capabilities}-\ref{Section: Multifaceted Abilities for Effective Recoveries}), but also effective collaboration (\Sref{Section: Beneficial Collaborator Influence Limited By Disadvantageous Agent Initiative}-\ref{Section: Collaborative Effectiveness}) and adaptive resource management (\Sref{Section: Resource Awareness}) abilities.
Based on our evaluation of multiple aspects of their recovery performance (\Sref{Section: Evaluation Metrics}), results unveil existing LLM agents' limited collaboration willingness and resource awareness, providing insights for future development of collaborative systems with stronger collaboration initiative and cooperation competences, along with more adaptive resource utilization capacities.
%
%
%
%
Detailed discussions of our findings and limitations are presented in \textit{Appendix}~\ref{Appendix A: Discussions and Limitations}.

\newpage

\newpage
\section*{Acknowledgement}

This research is based upon work supported DARPA ITM Program No. FA8650-23-C-7316, and the AI Research Institutes program by National Science Foundation and the Institute of Education Sciences, U.S. Department of Education through Award \# 2229873 - AI Institute for Transforming Education for Children with Speech and Language Processing Challenges. The views and conclusions contained herein are those of the authors and should not be interpreted as necessarily representing the official policies, either expressed or implied, of the U.S. Government. The U.S. Government is authorized to reproduce and distribute reprints for governmental purposes notwithstanding any copyright annotation therein.

\section*{Impact Statement}
This work aims to provide meaningful insights into advancing collaborative AI systems and their application in software engineering, with potential societal implications in several areas. The framework and findings serve to improve the reliability and efficiency of collaborative software engineering, potentially reducing costly errors and development delays.
However, there are important considerations valuable to take into account.
First, while enhanced collaboration capabilities of AI agents could improve software quality and developer productivity, they may also impact human developers' jobs and require careful integration into existing workflows.
Second, our resource-aware framework highlights the need to consider computational and environmental costs in deploying collaborative AI systems at scale.
Additionally, as AI agents become more capable of detecting and recovering from synchronization issues, it is of significance to ensure that human developers maintain meaningful oversight and understanding of system changes.
We believe these considerations should be actively discussed as the field moves toward more sophisticated collaborative AI systems in software engineering.



\bibliography{icml2025}
\bibliographystyle{icml2025}

\newpage
\appendix
\onecolumn

\appendix

\section{Discussions and Limitations}
\label{Appendix A: Discussions and Limitations}

The evolution of collaborative software engineering introduces complex challenges in maintaining synchronization among collaborators, whether humans or AI agents. Our investigation into the agent \textit{out-of-sync} challenge reveals fundamental insights into how collaborators detect, respond to, and recover from belief state divergence ($B_k \neq S_k$ at time $T_k$). This section elaborates on our key findings, discusses their broader implications, and acknowledges important limitations that suggest potential directions for future research. 

\paragraph{Key Findings and Implications.}

By introducing our evaluation framework (\textit{SyncMind} \Sref{Section: framework}) and benchmark (\textit{SyncBench} \Sref{Section: Benchmark}) built upon real-world \github repositories, our experiments (\Sref{Section: Experiments}) illuminate critical aspects of agent \textit{out-of-sync}:

(1) \textit{Technical Capabilities and Collaborative Effectiveness.}
Through either independent or collaborative recovery (\Sref{Section: Recovery Settings}), the stark performance variations among LLM agents (Tables \ref{tab:table 2 (caller vs callee)}-\ref{tab:table 1 (performance summary)} \& \ref{tab:table 1 + CSR (overall performance summary)}-\ref{tab:table 2 + CSR (caller vs callee)}) demonstrate that technical proficiency alone is insufficient for a successful \textit{out-of-sync} recovery (\Sref{Section: Agent Recovery Capabilities}-\ref{Section: Multifaceted Abilities for Effective Recoveries} \& \Sref{Appendix C1: Conditional Recovery Performance}).
Similar recovery trajectories of high-performance agents (\textit{e.g.,} \textit{Claude-3.5-Sonnet} with $SR$: \textit{Independent} $28.18\%$, \textit{Collaborative} $33.70\%$), especially with respect to early-stage exploration (\Sref{Appendix C2: Temporal Dynamics of Recovery Actions}) and proactive collaborative initiative (\Sref{Section: Beneficial Collaborator Influence Limited By Disadvantageous Agent Initiative}-\ref{Section: Collaborative Effectiveness}), suggest that both independent problem-solving and collaborative communication capabilities are crucial for maintaining synchronization in real-world dynamic collaborative environments.


(2) \textit{Collaboration Patterns and Communication Quality.}
Experiment results reveal the positive correlation between LLM agents' collaboration willingness and recovery success. Benefiting from proactive assistance seeking, the influence of collaborator assistance ($\Delta_\textit{collaborator} \in [0.33, 5.52] \%$) remains beneficial among different LLM agents.
However, its effectiveness varies significantly as affected by communication timing ($85.71\%-100.00\%$ \textit{early-stage assistance} among successful recoveries), question quality (rather than volume), and proactive collaboration initiative ($ASR \leq 4.86\%$) (\Sref{Section: Beneficial Collaborator Influence Limited By Disadvantageous Agent Initiative}-\ref{Section: Collaborative Effectiveness}). 

(3) \textit{Task Complexity and Recovery Strategies.}
Agents' performance gaps between \textit{Caller} and \textit{Callee} tasks (\tref{tab:table 2 (caller vs callee)}) highlight how different types of \textit{out-of-sync} scenarios require distinct recovery strategies (\Sref{Section: Collaborative Effectiveness}).
While \textit{Callee} tasks requiring dependency tracing show wider performance variations ($0.67\%-38.67\%$) and degraded performance among most agents, complex tasks generally benefit more from collaborative assistance (\Sref{Section: Effects of Task Complexity}).
Leveraging the dissimilar complexity levels of different source repositories, our repository-wise analysis (\Sref{Appendix:C.7 (repo-wise comparison)}) lends further evidence to the negative correlation between increased task complexity and recovery success. 

(4) \textit{Resource Awareness and Efficiency.}
Our experiments with varying resource constraints (\Sref{Section: Resource Awareness} \& \ref{Appendix:C.5 (Resource Awareness)}) unveil critical insights with regard to LLM agents' resource awareness and adaptive resource utilization. While their underlying technical capabilities significantly affect their time resource utilization, LLM agents' recovery performance demonstrates their notably low awareness of both temporal and financial resources. As early-stage resource allocation proves crucial for attaining recovery success, strategic action planning and resource estimation are highlighted for high-performing and resource-efficient \textit{out-of-sync} recoveries.

\paragraph{Broader Implications for Collaborative Systems.}
Our findings reveal meaningful implications for the future development of collaborative systems, especially in real-world scenarios with dynamic environments and intricate task contexts. 
From system design perspectives, the agent \textit{out-of-sync} challenge in real-world collaboration scenarios emphasizes the importance of state monitoring and divergence detection that are able to provide collaboration with effective recognition of the state mismatch $B_k \neq S_k$ taking place at time $T_k$ (\Sref{Section: Definition of Agent Out-of-Sync}).
The low $ASR$ ($\leq 4.86\%$) among all LLM agents also underscores the value of stronger collaboration initiative and communication capabilities for effective collaboration.
Our implementation of resource-aware \textit{out-of-sync} recovery demonstrates the necessity of intelligent resource allocation and estimation strategies, illuminating both the importance of early-stage investment in environmental understanding and the need for adaptive resource utilization based on task complexity and resource availability. In designing effective collaboration protocols, our work elaborates the benefits of collaborator assistance, meanwhile highlighting the value of quality-focused rather than quantity-focused communication. 

\paragraph{Limitations and Future Work.}
While our study provides meaningful insights, several limitations present potential directions for our future work:

(1) \textit{Benchmark Limitations.}
Although our benchmark construction method is applicable to diverse \github repositories and therefore can be further expanded to larger sizes to accommodate custom use (\textit{\textit{e.g.,}} large-scale training), our \textit{SyncBench} (\Sref{Section: Benchmark}) currently focuses primarily on \github \textit{Python} repositories with unit tests, which may limit its generalizability to broader software engineering scenarios with dissimilar programming languages and testing environments. 
Additionally, our simulation of collaborative \textit{out-of-sync} scenarios based on real-world historical repository changes may not fully capture all complexities of live collaboration patterns, especially in large-scale collaborative software engineering with multiple collaborators involved.

(2) \textit{Evaluation Framework Constraints.}
Implementing resource-aware \textit{out-of-sync} recovery, our simplified resource modeling may not capture all real-world constraints involved in multifarious collaborative scenarios in reality. In our future work, we aim to include more diversified resource utilization with deeper exploration of collaborators' long-term collaborative relationships. We also endeavor to extend our technical scope to multi-agent systems with multiple collaborators working on the task, resulting in more complex \textit{out-of-sync} scenarios with multiple agents facing \textit{out-of-sync} states ($B_k \neq S_k$).

(3) \textit{Methodological Considerations.}
There are also limitations in our methodological design. Our simulation-based approach may not capture all nuances of real-world collaborations, and our investigation can be expanded to diverse communication protocols. To this end, our future work will also pay attention to the development of specialized training approaches to enhance agents' collaboration initiatives and communication capabilities, along with investigating more sophisticated resource management strategies that better mirror real-world development constraints.


\section{Experiment Configuration}
\label{Appendix B: Experiment Configs}

\renewcommand{\thefigure}{B\arabic{figure}} 
\renewcommand{\thetable}{B\arabic{table}}   
\setcounter{figure}{0} 
\setcounter{table}{0}  

\subsection{Pilot for Configuration}
In determining the appropriate experiment settings, we conduct a series of preliminary tests (\tref{tab:table B1}) for basic interaction configurations. 

\textbf{Interaction Basics.} We first pilot \textit{out-of-sync} recovery on \textit{Llama-3.1-70B} and \textit{GPT-4o} with the maximum number of turns limited to $10$ interactive iterations across $66$ instances. For both agents, a large proportion of tasks consist of $10$ total turns of \textit{Env} exploration without any solution attempts, while the remaining $11$ tests have only one or two solution proposal attempts without success. Based on this result, we further expand the max-turn limitation to $20$ and $30$ turns, respectively. Drawing upon the observation that fewer turns leave insufficient time for agents to propose their solutions, we finalize $30$ turns of maximum interaction time to ensure the quality, effectiveness, and efficiency\footref{Footnote: Cost Constraints} of our experiments. 

\textbf{Resource Awareness.} To define the proper cost setting of recovery actions, we compare the pilot experiment results of different cost settings in terms of initial budget (\textit{i.e.,} \textit{the total amount of money each agent is given at the very beginning of each test}), solution-proposal cost (\textit{i.e.,} \textit{the cost of \textcolor{fig2_code}{proposing a solution} for execution validation}), and assistance-seeking cost (\textit{i.e.,} \textit{the cost of \textcolor{fig2_ask}{proactively asking for collaborator assistance}}). Setting the balanced cost of both solution-proposal and assistance-seeking as $\$100$, we encourage agents to take these two recovery actions by providing them with an initial budget of $\$300$, $\$1000$, and $\$3000$, respectively. Meanwhile, all experiments are conducted with the maximum time limit set to $30$ turns, as revealed by earlier pilot results. Comparing the performance between $\textit{Budget}=\$1000$ and $\textit{Budget}=\$3000$, deteriorated $SR$ scores are obtained as we provide agents with adequate budget that can cover all kinds of action selection patterns. This unfolds LLM agents' lack of cost awareness that makes a sufficient initial budget an obstacle towards recovery success. To this regard, we define an initial budget of $\$1000$ for our standard resource-aware \textit{out-of-sync} recovery experiments, allowing agents to allocate freely up to one-third of the maximum turns of interactions for solution-proposal and assistance-seeking actions.

\begin{table}[H]
\begin{center}
\begin{small}
\vspace{0em}
\caption{\textbf{Pilot for Configuration.} Preliminary tests to determine the appropriate experiment settings. All pilots are conducted through \textit{independent} recovery, and we refer to the number of solution-proposal attempts as \textit{Attempts} in this table to indicate whether current setting can support an effective \textit{out-of-sync} recovery.}
\label{tab:table B1}
    \begin{tabular}{>{\centering\arraybackslash}m{3cm}|>{\centering\arraybackslash}m{2.5cm}||>{\centering\arraybackslash}m{1.2cm}>{\centering\arraybackslash}m{1.2cm}>{\centering\arraybackslash}m{1.2cm}|>{\centering\arraybackslash}m{1.2cm}>{\centering\arraybackslash}m{1.2cm}>{\centering\arraybackslash}m{1.2cm}}
    
    \toprule
    \multirow{2}{*}{\centering \textbf{Agent}} & \multirow{2}{*}{\centering \textbf{Recovery}} & \multicolumn{3}{c|}{\textbf{Time Limit (\textit{turn})}} & \multicolumn{3}{c}{\textbf{Budget (\$)}} \\
    
    \cmidrule{3-8}
    &  & \textbf{10} & \textbf{20} & \textbf{30} & \textbf{300} & \textbf{1000} & \textbf{3000} \\
    
    \midrule
    
    \multirow{2}{*}{\centering \textbf{Llama-3.1-70B}} 
    & \textbf{\textit{SR (\%)}} & 0.00 & 0.00 & \cellcolor{basecolor_green!15.2} 1.52 & 0.00 & \cellcolor{basecolor_green!15.2} 1.52 & 0.00 \\
    & \textbf{\textit{Attempts}} & 0 & 1 & \cellcolor{basecolor_green!15.2} 4 & 0 & \cellcolor{basecolor_green!15.2} 4 & 5 \\
    
    \midrule
    
    \multirow{2}{*}{\centering \textbf{GPT-4o}}
    & \textbf{\textit{SR (\%)}} & 0.00 & \cellcolor{basecolor_green!15.2} 1.52 & \cellcolor{basecolor_green!45.5} 4.55 & 0.00 & \cellcolor{basecolor_green!45.5} 4.55 & \cellcolor{basecolor_green!30.3} 3.03 \\
    & \textbf{\textit{Attempts}} & 0 & \cellcolor{basecolor_green!15.2} 2 & \cellcolor{basecolor_green!45.5} 5 & 0 & \cellcolor{basecolor_green!45.5} 5 & \cellcolor{basecolor_green!30.3} 9 \\
    
    \bottomrule
    
    \end{tabular}
\vspace{0em}
\end{small}
\end{center}
\end{table}

\subsection{Benchmark Construction}
\label{Appendix: Benchmark Construction}

\begin{table}[!t]
\begin{center}
\begin{small}
\vspace{0em}
\caption{\textbf{Agent \textit{Out-of-Sync} Benchmark.} Our benchmark is constructed through \textit{Env} configuration, \textit{out-of-sync} simulation, and multi-level filtering, and can be expanded to larger sizes for large-scale evaluation, or be further subsampled through weighted downsampling to obtain small-scale evaluation subsets.}
\label{tab:table B2 (Benchmark)}
    \begin{tabular}
    {>{\centering\arraybackslash}m{4.2cm}||>{\centering\arraybackslash}m{1.6cm} >
    {\centering\arraybackslash}m{1.6cm} >{\centering\arraybackslash}m{1.6cm}|>
    {\centering\arraybackslash}m{1.6cm} >
    {\centering\arraybackslash}m{1.6cm} >
    {\centering\arraybackslash}m{1.6cm}}
    
        \toprule
        
        \multirow{2}{4cm}{\centering \textbf{Source}} & \multicolumn{3}{c|}{\centering \textbf{Original Size}} & \multicolumn{3}{c}{\centering \textbf{Reduced Size}} \\
        
        \cmidrule{2-7}
        
        & \textbf{Callee} & \textbf{Caller} & \textbf{Total} & \textbf{Callee} & \textbf{Caller} & \textbf{Total} \\
        
        \midrule
        
        FastAPI/FastAPI & 979 & 660 & 1639 & 4 & 10 & 14 \\
        huggingface/transformers & 4635 & 453 & 5088 & 4 & 5 & 9 \\
        matplotlib/matplotlib & 757 & 980 & 1737 & 4 & 5 & 9 \\
        psf/requests & 8 & 65 & 73 & 4 & 10 & 14 \\
        mwaskom/seaborn & 356 & 217 & 573 & 13 & 6 & 19 \\
        pylint-dev/pylint & 229 & 239 & 468 & 11 & 10 & 21 \\
        pytest-dev/pytest & 1638 & 940 & 2578 & 8 & 10 & 18 \\
        openai/gym & 53 & 89 & 142 & 11 & 5 & 16  \\
        sympy/sympy & 887 & 1106 & 1993 & 13 & 10 & 23 \\
        pallets/flask & 34 & 266 & 300 & 4 & 10 & 14 \\
        openai/whisper & 9 & 1 & 10 & 2 & 1 & 3 \\
        scikit-learn/scikit-learn & 376 & 2663 & 3039 & 1 & 5 & 6 \\
        sphinx-doc/sphinx & 675 & 434 & 1109 & 9 & 10 & 19 \\
        pycaret/pycaret & 20 & 22 & 42 & 11 & 4 & 15 \\
        explosion/spaCy & 313 & 434 & 747 & 10 & 11 & 21 \\
        python-pillow/Pillow & 295 & 1334 & 1629 & 11 & 10 & 21 \\
        scrapy/scrapy & 465 & 665 & 1130 & 13 & 10 & 23 \\
        optuna/optuna & 190 & 296 & 486 & 3 & 5 & 8 \\
        microsoft/FLAML & 38 & 30 & 68 & 4 & 3 & 7 \\
        psf/black & 25 & 79 & 104 & 7 & 5 & 12 \\
        mlflow/mlflow & 367 & 839 & 1206 & 3 & 5 & 8 \\
        
        \midrule
        
        Total & 12711 & 11621 & 24332 & 150 & 150 & 300 \\
        
        \bottomrule
        
    \end{tabular}
\end{small}
\end{center}
\end{table}

We design our benchmark construction method \Sref{Section: Benchmark} with generalizability and adaptability to diverse \textit{Python} repositories that have Python as their primary programming language, meanwhile possessing unit tests for testing the functioning of various modules.
Including environment setup (\textit{e.g.,} \textit{pyproject.toml}, \textit{setup.py}) is a plus, while specifying necessary packages serves as an alternative way to enrich \textit{SyncBench} (\Sref{Section: Benchmark}).

In the current version, our agent \textit{out-of-sync} benchmark, \textbf{\textit{SyncBench}} (\Sref{Section: Benchmark}), is built upon 21 popular \github repositories (\tref{tab:table B2 (Benchmark)}), and can be further expanded or downsampled to proper dataset sizes suitable for different experiment conditions and evaluation purposes.

Our benchmark construction method complies with our definition of agent \textit{out-of-sync} (\Sref{Section: Definition of Agent Out-of-Sync}), where meeting at least one of three conditions can result in an agent's belief deviation, either directly leading to task failure or indirectly laying latent problems for future execution errors.
Although the latter outcome is not immediately visible, growing state divergences can be accumulated as underlying risks that lead to significant troubles in the near future as the collaboration proceeds.
To take all three conditions into account, our benchmark construction method supports both \textit{pass-to-fail} and \textit{pass-to-pass} samples, though we only consider \textit{pass-to-fail} in our current version of \textit{SyncBench} (\Sref{Section: Benchmark Construction}) to emphasize the importance of \textit{out-of-sync} recovery abilities in maintaining effective collaborations.

\begin{itemize}[noitemsep,topsep=0pt,parsep=2pt,partopsep=0pt,leftmargin=*]
    \item \textbf{\textit{Pass-to-fail} state divergence.} \textit{Out-of-sync} scenarios directly visible as \textit{pass-to-fail} state divergence ($B_2 \neq S_2$) are readily leveraged to construct \textit{out-of-sync} recovery tasks where: (1) collaborator updated repository ($S_1$) can successfully pass the execution test, and (2) agent revised repository with \textit{out-of-sync} function ($B_2$) fails the execution test.
    \item \textbf{\textit{Pass-to-pass} state divergence.} \textit{Out-of-sync} scenarios that are not manifested as immediate task failures are also taken into consideration to create \textit{out-of-sync} recovery tasks based on \textit{pass-to-pass} state divergence ($B_2 \neq S_2$) where: (1) collaborator updated repository ($S_1$) successfully passes the execution test, resulting in parsed test output denoted as $O_S$, (2) agent revised repository with \textit{out-of-sync} function ($B_2$) also successfully passes the execution test, resulting in parsed test output denoted as $O_B$, and (3) parsed test output results between $S_2$ and $B_2$ are different ($O_S \neq O_B$).
\end{itemize}

As our determination of recovery success depends on the success of both execution test and parsing validation (\Sref{Section: Evaluation Metrics}), both {pass-to-fail} and {pass-to-pass} \textit{out-of-sync} recovery tasks can be effectively evaluated.

In our experiments (\Sref{Section: Experiments}), we test on 300 downsampled instances\footref{Footnote: Cost Constraints} with evenly distributed \textit{Caller} and \textit{Callee} (\Sref{Section: Benchmark Datasets}) for computational efficiency and comparison effectiveness.

Here are some examples of our test instances:

(1) Example-1 on \textit{psf/requests}:

Traced git commit: \textit{c0813a2d910ea6b4f8438b91d315b8d181302356}

\textit{Out-of-sync} function:

\begin{lstlisting}[language=Python]
def _urllib3_request_context(
    request: "PreparedRequest", 
    verify: "bool | str | None"
) -> "(typing.Dict[str, typing.Any], typing.Dict[str, typing.Any])":
    host_params = {}
    pool_kwargs = {}
    parsed_request_url = urlparse(request.url)
    scheme = parsed_request_url.scheme.lower()
    port = parsed_request_url.port
    cert_reqs = "CERT_REQUIRED"
    if verify is False:
        cert_reqs = "CERT_NONE"
    if isinstance(verify, str):
        pool_kwargs["ca_certs"] = verify
    pool_kwargs["cert_reqs"] = cert_reqs
    host_params = {
        "scheme": scheme,
        "host": parsed_request_url.hostname,
        "port": port,
    }
    return host_params, pool_kwargs
\end{lstlisting}

Initial error log:
\begin{verbatim}[fontsize=\small, breaklines=true, breakanywhere=true]
============================= test session starts ==============================platform linux -- Python 3.11.9, pytest-8.3.2, pluggy-1.5.0 -- /workspace/test_venv/bin/python
cachedir: .pytest_cache
rootdir: /workspace/test_repo\nconfigfile: pyproject.toml
plugins: httpbin-2.0.0, cov-5.0.0, asyncio-0.24.0
asyncio: mode=Mode.STRICT, default_loop_scope=None
collecting ... collected 329 items

tests/test_requests.py::TestRequests::test_entry_points PASSED           [  0%]
tests/test_requests.py::TestRequests::test_invalid_url[MissingSchema-hiwpefhipowhefopw] PASSED [  0%]
tests/test_requests.py::TestRequests::test_invalid_url[InvalidSchema-localhost:3128] PASSED [  0%]
tests/test_requests.py::TestRequests::test_invalid_url[InvalidSchema-localhost.localdomain:3128/] PASSED [  1%]
tests/test_requests.py::TestRequests::test_invalid_url[InvalidSchema-10.122.1.1:3128/] PASSED [  1%]
tests/test_requests.py::TestRequests::test_invalid_url[InvalidURL-http://] PASSED [  1%]
tests/test_requests.py::TestRequests::test_invalid_url[InvalidURL-http://*example.com] PASSED [  2%]
tests/test_requests.py::TestRequests::test_invalid_url[InvalidURL-http://.example.com] PASSED [  2%]
tests/test_requests.py::TestRequests::test_basic_building PASSED
...
FAILED tests/test_requests.py::TestRequests::test_redirect_with_wrong_gzipped_header
FAILED tests/test_requests.py::TestRequests::test_requests_history_is_saved
FAILED tests/test_requests.py::TestRequests::test_json_param_post_content_type_works
FAILED tests/test_requests.py::TestRequests::test_response_iter_lines - TypeE...
FAILED tests/test_requests.py::TestRequests::test_response_context_manager - ...
FAILED tests/test_requests.py::TestRequests::test_unconsumed_session_response_closes_connection
FAILED tests/test_requests.py::TestRequests::test_response_json_when_content_is_None
FAILED tests/test_requests.py::TestRequests::test_custom_redirect_mixin - Typ...
FAILED tests/test_requests.py::TestTimeout::test_stream_timeout - TypeError: ...
FAILED tests/test_requests.py::TestTimeout::test_invalid_timeout[timeout0-(connect, read)]
FAILED tests/test_requests.py::TestTimeout::test_invalid_timeout[foo-must be an int, float or None]
FAILED tests/test_requests.py::TestTimeout::test_none_timeout[None] - TypeErr...
FAILED tests/test_requests.py::TestTimeout::test_none_timeout[timeout1] - Typ...
FAILED tests/test_requests.py::TestTimeout::test_read_timeout[timeout0] - Typ...
FAILED tests/test_requests.py::TestTimeout::test_read_timeout[timeout1] - Typ...
FAILED tests/test_requests.py::TestTimeout::test_connect_timeout[timeout0] - ...
FAILED tests/test_requests.py::TestTimeout::test_connect_timeout[timeout1] - ...
FAILED tests/test_requests.py::TestTimeout::test_total_timeout_connect[timeout0]
FAILED tests/test_requests.py::TestTimeout::test_total_timeout_connect[timeout1]
FAILED tests/test_requests.py::TestTimeout::test_encoded_methods - TypeError:...
FAILED tests/test_requests.py::test_urllib3_retries - TypeError: _urllib3_req...
FAILED tests/test_requests.py::test_urllib3_pool_connection_closed - TypeErro...
FAILED tests/test_requests.py::TestPreparingURLs::test_redirecting_to_bad_url[http://:1-InvalidURL]
FAILED tests/test_requests.py::TestPreparingURLs::test_json_decode_compatibility
FAILED tests/test_requests.py::TestPreparingURLs::test_json_decode_persists_doc_attr
FAILED tests/test_requests.py::TestPreparingURLs::test_different_connection_pool_for_tls_settings_verify_True
FAILED tests/test_requests.py::TestPreparingURLs::test_different_connection_pool_for_tls_settings_verify_bundle_expired_cert
FAILED tests/test_requests.py::TestPreparingURLs::test_different_connection_pool_for_tls_settings_verify_bundle_unexpired_cert
FAILED tests/test_requests.py::TestPreparingURLs::test_different_connection_pool_for_mtls_settings
====== 131 failed, 196 passed, 1 skipped, 1 xfailed, 8 warnings in 30.10s ======
\end{verbatim}

Ground-truth function:
\begin{lstlisting}[language=Python]
def _urllib3_request_context(
    request: "PreparedRequest",
    verify: "bool | str | None",
    client_cert: "typing.Tuple[str, str] | str | None",
    poolmanager: "PoolManager",
) -> "(typing.Dict[str, typing.Any], typing.Dict[str, typing.Any])":
    host_params = {}
    pool_kwargs = {}
    parsed_request_url = urlparse(request.url)
    scheme = parsed_request_url.scheme.lower()
    port = parsed_request_url.port

    # Determine if we have and should use our default SSLContext
    # to optimize performance on standard requests.
    poolmanager_kwargs = getattr(poolmanager, "connection_pool_kw", {})
    has_poolmanager_ssl_context = poolmanager_kwargs.get("ssl_context")
    should_use_default_ssl_context = (
        _preloaded_ssl_context is not None and not has_poolmanager_ssl_context
    )

    cert_reqs = "CERT_REQUIRED"
    if verify is False:
        cert_reqs = "CERT_NONE"
    elif verify is True and should_use_default_ssl_context:
        pool_kwargs["ssl_context"] = _preloaded_ssl_context
    elif isinstance(verify, str):
        if not os.path.isdir(verify):
            pool_kwargs["ca_certs"] = verify
        else:
            pool_kwargs["ca_cert_dir"] = verify
    pool_kwargs["cert_reqs"] = cert_reqs
    if client_cert is not None:
        if isinstance(client_cert, tuple) and len(client_cert) == 2:
            pool_kwargs["cert_file"] = client_cert[0]
            pool_kwargs["key_file"] = client_cert[1]
        else:
            # According to our docs, we allow users to specify just the client
            # cert path
            pool_kwargs["cert_file"] = client_cert
    host_params = {
        "scheme": scheme,
        "host": parsed_request_url.hostname,
        "port": port,
    }
    return host_params, pool_kwargs
\end{lstlisting}

(2) Example-2 on \textit{nwaskon/seaborn}:

Traced git commit: \textit{45666c8b4ba634c7720cab59bf0aaa00ab9b5e29}

\textit{Out-of-sync} function:
\begin{lstlisting}[language=Python]
def add(
    self,
    mark: Mark,
    stat: Stat = None,
    data: Optional[DataFrame | Mapping] = None,
    variables: Optional[dict[str, Optional[Hashable | Vector]]] = None,
    orient: Literal["x", "y", "v", "h"] = "x",
) -> Plot:

    layer_data = self._data.concat(data, variables)

    if stat is None:
        stat = mark.default_stat

    orient = {"v": "x", "h": "y"}.get(orient, orient)
    mark.orient = orient
    if stat is not None:
        stat.orient = orient

    self._layers.append(Layer(layer_data, mark, stat))

    return self
\end{lstlisting}

Initial error log:
\begin{verbatim}[fontsize=\small, breaklines=true, breakanywhere=true]
============================= test session starts ==============================platform linux -- Python 3.11.9, pytest-8.3.2, pluggy-1.5.0 -- /workspace/test_venv/bin/python
cachedir: .pytest_cache
rootdir: /workspace/test_repo
configfile: pyproject.toml
plugins: asyncio-0.24.0
asyncio: mode=Mode.STRICT, default_loop_scope=None
collecting ... collected 14 items

tests/_marks/test_dot.py::TestDot::test_simple FAILED                    [  7%]
tests/_marks/test_dot.py::TestDot::test_filled_unfilled_mix FAILED       [ 14%]
tests/_marks/test_dot.py::TestDot::test_missing_coordinate_data FAILED   [ 21%]
tests/_marks/test_dot.py::TestDot::test_missing_semantic_data[color] FAILED [ 28%]
tests/_marks/test_dot.py::TestDot::test_missing_semantic_data[fill] FAILED [ 35%]
tests/_marks/test_dot.py::TestDot::test_missing_semantic_data[marker] FAILED [ 42%]
tests/_marks/test_dot.py::TestDot::test_missing_semantic_data[pointsize] FAILED [ 50%]
tests/_marks/test_dot.py::TestDots::test_simple FAILED                   [ 57%]
tests/_marks/test_dot.py::TestDots::test_set_color FAILED                [ 64%]
tests/_marks/test_dot.py::TestDots::test_map_color FAILED                [ 71%]
tests/_marks/test_dot.py::TestDots::test_fill FAILED                     [ 78%]
tests/_marks/test_dot.py::TestDots::test_pointsize FAILED                [ 85%]
tests/_marks/test_dot.py::TestDots::test_stroke FAILED                   [ 92%]
tests/_marks/test_dot.py::TestDots::test_filled_unfilled_mix FAILED      [100%]

=================================== FAILURES ===================================
_____________________________ TestDot.test_simple ______________________________

self = <tests._marks.test_dot.TestDot object at 0x7f4196cb99d0>

    def test_simple(self):
    
        x = [1, 2, 3]
        y = [4, 5, 2]
>       p = Plot(x=x, y=y).add(Dot()).plot()

tests/_marks/test_dot.py:39: 
_ _ _ _ _ _ _ _ _ _ _ _ _ _ _ _ _ _ _ _ _ _ _ _ _ _ _ _ _ _ _ _ _ _ _ _ _ _ _ _ 

self = <seaborn._core.plot.Plot object at 0x7f419697afd0>
mark = Dot(artist_kws={}, marker=<'o'>, pointsize=<6>, stroke=<0.75>, color=<'C0'>, alpha=<1>, fill=<True>, edgecolor=<depend:color>, edgealpha=<depend:alpha>, edgewidth=<0.5>, edgestyle=<'-'>)
stat = None, data = None, variables = None, orient = 'x'

    def add(
        self,
        mark: Mark,
        stat: Stat = None,
...
seaborn/_core/plot.py:498: AttributeError
______________________ TestDots.test_filled_unfilled_mix _______________________

self = <tests._marks.test_dot.TestDots object at 0x7f979214a990>

    def test_filled_unfilled_mix(self):
    
        x = [1, 2]
        y = [4, 5]
        marker = ["a", "b"]
        shapes = ["o", "x"]
    
        mark = Dots(stroke=2)
>       p = Plot(x=x, y=y).add(mark, marker=marker).scale(marker=shapes).plot()
E       TypeError: Plot.add() got an unexpected keyword argument 'marker'

tests/_marks/test_dot.py:171: TypeError
=========================== short test summary info ============================
FAILED tests/_marks/test_dot.py::TestDot::test_simple - AttributeError: 'Plot...
FAILED tests/_marks/test_dot.py::TestDot::test_filled_unfilled_mix - TypeErro...
FAILED tests/_marks/test_dot.py::TestDot::test_missing_coordinate_data - Attr...
FAILED tests/_marks/test_dot.py::TestDot::test_missing_semantic_data[color]
FAILED tests/_marks/test_dot.py::TestDot::test_missing_semantic_data[fill] - ...
FAILED tests/_marks/test_dot.py::TestDot::test_missing_semantic_data[marker]
FAILED tests/_marks/test_dot.py::TestDot::test_missing_semantic_data[pointsize]
FAILED tests/_marks/test_dot.py::TestDots::test_simple - AttributeError: 'Plo...
FAILED tests/_marks/test_dot.py::TestDots::test_set_color - AttributeError: '...
FAILED tests/_marks/test_dot.py::TestDots::test_map_color - AttributeError: '...
FAILED tests/_marks/test_dot.py::TestDots::test_fill - AttributeError: 'PlotD...
FAILED tests/_marks/test_dot.py::TestDots::test_pointsize - AttributeError: '...
FAILED tests/_marks/test_dot.py::TestDots::test_stroke - AttributeError: 'Plo...
FAILED tests/_marks/test_dot.py::TestDots::test_filled_unfilled_mix - TypeErr...
============================== 14 failed in 1.16s ==============================
\end{verbatim}

Ground-truth function:
\begin{lstlisting}[language=Python]
def add(
    self,
    mark: Mark,
    *transforms: Stat | Move,
    orient: str | None = None,
    legend: bool = True,
    label: str | None = None,
    data: DataSource = None,
    **variables: VariableSpec,
) -> Plot:
    """
    Specify a layer of the visualization in terms of mark and data transform(s).

    This is the main method for specifying how the data should be visualized.
    It can be called multiple times with different arguments to define
    a plot with multiple layers.

    Parameters
    ----------
    mark : :class:`Mark`
        The visual representation of the data to use in this layer.
    transforms : :class:`Stat` or :class:`Move`
        Objects representing transforms to be applied before plotting the data.
        Currently, at most one :class:`Stat` can be used, and it
        must be passed first. This constraint will be relaxed in the future.
    orient : "x", "y", "v", or "h"
        The orientation of the mark, which also affects how transforms are computed.
        Typically corresponds to the axis that defines groups for aggregation.
        The "v" (vertical) and "h" (horizontal) options are synonyms for "x" / "y",
        but may be more intuitive with some marks. When not provided, an
        orientation will be inferred from characteristics of the data and scales.
    legend : bool
        Option to suppress the mark/mappings for this layer from the legend.
    label : str
        A label to use for the layer in the legend, independent of any mappings.
    data : DataFrame or dict
        Data source to override the global source provided in the constructor.
    variables : data vectors or identifiers
        Additional layer-specific variables, including variables that will be
        passed directly to the transforms without scaling.

    Examples
    --------
    .. include:: ../docstrings/objects.Plot.add.rst

    """
    if not isinstance(mark, Mark):
        msg = f"mark must be a Mark instance, not {type(mark)!r}."
        raise TypeError(msg)

    # TODO This API for transforms was a late decision, and previously Plot.add
    # accepted 0 or 1 Stat instances and 0, 1, or a list of Move instances.
    # It will take some work to refactor the internals so that Stat and Move are
    # treated identically, and until then well need to "unpack" the transforms
    # here and enforce limitations on the order / types.

    stat: Optional[Stat]
    move: Optional[List[Move]]
    error = False
    if not transforms:
        stat, move = None, None
    elif isinstance(transforms[0], Stat):
        stat = transforms[0]
        move = [m for m in transforms[1:] if isinstance(m, Move)]
        error = len(move) != len(transforms) - 1
    else:
        stat = None
        move = [m for m in transforms if isinstance(m, Move)]
        error = len(move) != len(transforms)

    if error:
        msg = " ".join([
            "Transforms must have at most one Stat type (in the first position),",
            "and all others must be a Move type. Given transform type(s):",
            ", ".join(str(type(t).__name__) for t in transforms) + "."
        ])
        raise TypeError(msg)

    new = self._clone()
    new._layers.append({
        "mark": mark,
        "stat": stat,
        "move": move,
        # TODO it doesn't work to supply scalars to variables, but it should
        "vars": variables,
        "source": data,
        "legend": legend,
        "label": label,
        "orient": {"v": "x", "h": "y"}.get(orient, orient),  # type: ignore
    })

    return new
\end{lstlisting}

\subsection{\textit{Out-of-Sync} Recovery on SyncBench}
\label{Appendix: SyncBench}

\begin{figure}[!t]
\begin{center}
\begin{small}
\vspace{0em}
    \includegraphics[width=1\linewidth]{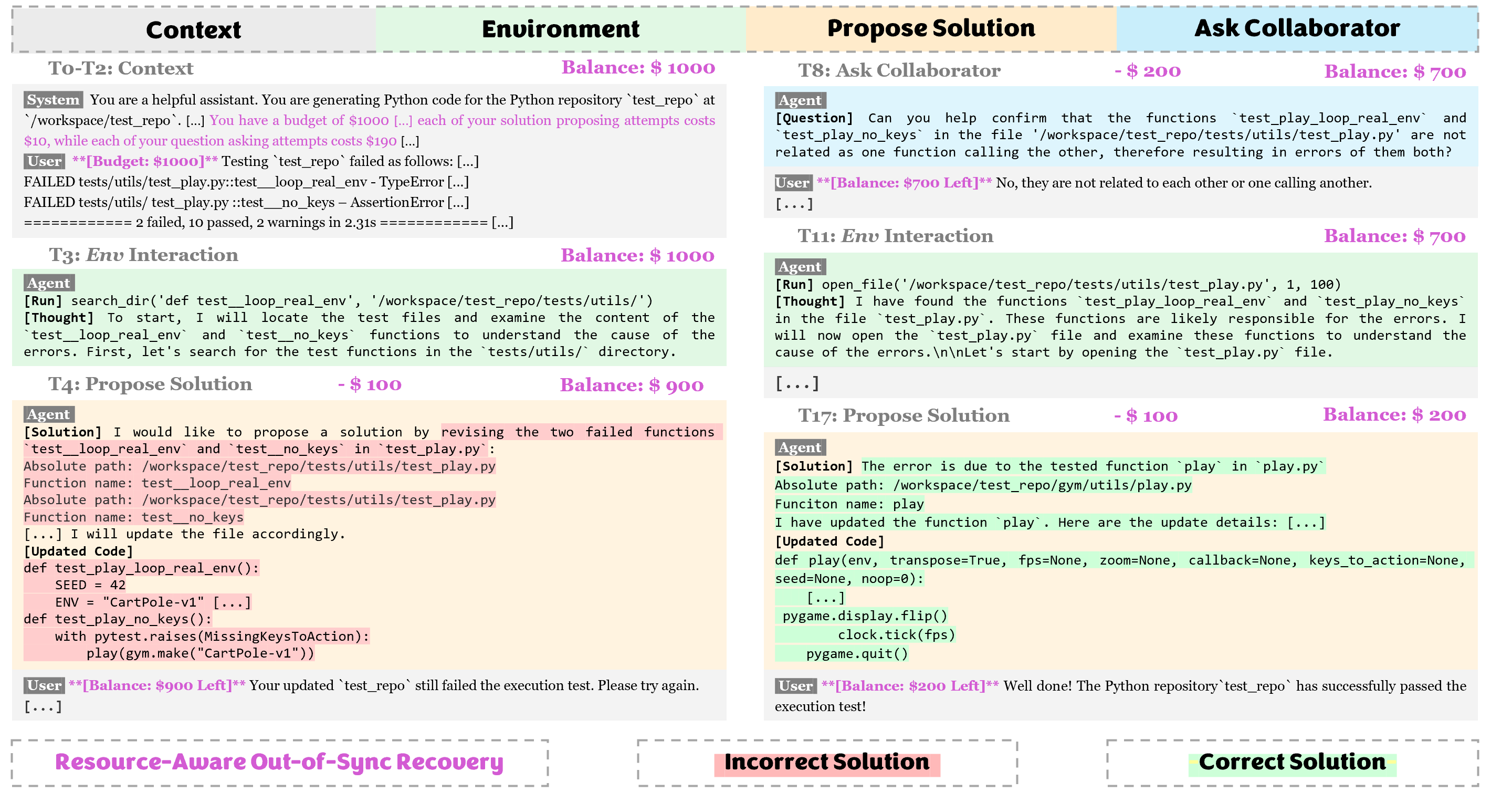}
    \caption{\textbf{\textit{Out-of-Sync} Recovery Example.} Starting from the initial \textit{out-of-sync} state at $T_2$, an LLM agent takes different recovery actions to approach synchronized state through multi-turn interactions, which involve \textcolor{fig2_env}{\textbf{interacting with \textit{Env}}} (\textit{green blocks}), \textcolor{fig2_ask}{\textbf{proactively asking for collaborator assistance}} (\textit{blue block}), or \textcolor{fig2_code}{\textbf{proposing a solution}} to examine agent's recovery progress (\textit{orange blocks}). By incorporating resource awareness throughout the \textit{out-of-sync} recovery process, the agent pays extra heed to making rational use of available resources for gaining knowledge and validating solution at a cost. For better clarity, some details are omitted as […].}
    \vspace{0em}  
    \label{fig:figure B1 (benchmark example)}
\end{small}
\end{center}
\end{figure}

Employing our resource-aware \textit{out-of-sync} recovery framework (\textbf{\textit{SyncMind}} \Sref{Section: framework}), \fref{fig:figure B1 (benchmark example)} presents an agent \textit{out-of-sync} recovery example on our benchmark (\textbf{\textit{SyncBench}} \Sref{Section: Benchmark}).

\section{Agent \textit{Out-of-Sync} Recovery: An In-Depth Analysis}
\label{Appendix: C (Analysis Extension)}

\renewcommand{\thefigure}{C\arabic{figure}} 
\renewcommand{\thetable}{C\arabic{table}}   
\setcounter{figure}{0} 
\setcounter{table}{0}  

As the continuation of our findings and discussions (\Sref{Section: Experiments}), in this section, we aim to provide an in-depth analysis of agents' \textit{out-of-sync} recovery efficiency, conditional recovery performance, along with their strategic recovery patterns in terms of temporal dynamics of recovery actions, solution proposal strategies, effective assistance seeking, and resource awareness characteristics.

\begin{table*}[!h]
\begin{center}
\begin{small}
\caption{\textbf{Overall \textit{Out-of-Sync} Recovery Performance.} We use \textit{$\Delta_{\textit{\text{collaborator}}}$} to represent the influence of collaborator assistance.}
\label{tab:table 1 (performance summary)}
\vspace{0em}
    \begin{tabular}{>{\centering\arraybackslash}m{3cm}||>{\centering\arraybackslash}m{1cm}>{\centering\arraybackslash}m{1cm}>{\centering\arraybackslash}m{1cm}|>{\centering\arraybackslash}m{1cm}>{\centering\arraybackslash}m{1cm}>{\centering\arraybackslash}m{1cm}|>{\centering\arraybackslash}m{1cm}>{\centering\arraybackslash}m{1cm}>{\centering\arraybackslash}m{1cm}}
    
    \toprule
    \multirow{2}{*}{\centering \textbf{Agent}} & \multicolumn{3}{c|}{\textbf{Independent (\%)}} & \multicolumn{3}{c|}{\textbf{Collaborative (\%)}} & 
    \multicolumn{3}{c}{\textbf{$\Delta_{\textit{\text{collaborator}}}$
 (\%)}} \\
    \cmidrule{2-10}
    & $LA_{file}$ & $LA_{func}$ & $SR$ & $LA_{file}$ & $LA_{func}$ & $SR$ & $LA_{file}$ & $LA_{func}$ & $SR$ \\
    
    \midrule
    \textbf{Llama-3.1-8B} & 8.67 & 4.67 & 0.33 & 27.00 & 21.33 & 1.33 & \cellcolor{basecolor_green!91.65} +18.33 & \cellcolor{basecolor_green!83.3} +16.66 & \cellcolor{basecolor_green!5.0} +1.00 \\
    \textbf{Llama-3.1-70B} & 8.33 & 5.00 & 2.67 & 12.33 & 5.67 & 3.33 & \cellcolor{basecolor_green!20} +4.00 & \cellcolor{basecolor_green!3.35} +0.67 & \cellcolor{basecolor_green!3.3} +0.66 \\
    \textbf{GPT-4o mini} & 10.63 & 7.97 & 3.99 & 12.29 & 8.31 & 5.32 & \cellcolor{basecolor_green!8.3} +1.66 & \cellcolor{basecolor_green!1.7} +0.34 & \cellcolor{basecolor_green!6.65} +1.33 \\
    \textbf{DeepSeek-V2.5} & 47.67 & 35.00 & 7.33 & 47.00 & 37.33 & 7.67 & \cellcolor{basecolor_red!3.35} -0.67 & \cellcolor{basecolor_green!11.65} +2.33 & \cellcolor{basecolor_green!1.7} +0.34 \\
    \textbf{GPT-4o} & 14.33 & 9.33 & 4.00 & 39.00 & 34.67 & 8.00 & \cellcolor{basecolor_green!123.35} +24.67 & \cellcolor{basecolor_green!126.7} +25.34 & \cellcolor{basecolor_green!20} +4.00 \\
    \textbf{Llama-3.3-70B} &  64.00  &  47.33  &  16.33  &  66.67  &  53.67  &  19.00  &  \cellcolor{basecolor_green!13.35} +2.67  &  \cellcolor{basecolor_green!31.7} +6.34  &  \cellcolor{basecolor_green!13.35} +2.67 \\
    \textbf{Claude-3.5-Sonnet} & 64.09 & 56.35 & 28.18 & 61.33 & 51.93 & 33.70 & \cellcolor{basecolor_red!13.8} -2.76 & \cellcolor{basecolor_red!22.1} -4.42 & \cellcolor{basecolor_green!27.6} +5.52 \\
    
    \bottomrule
    \end{tabular}
\end{small}
\end{center}
\vspace{0em}
\end{table*}

\begin{table*}[!h]
\begin{center}
\begin{small}
\vspace{-0.5em}
\caption{\textbf{Conditional \textit{Out-of-Sync} Recovery Performance.} Following \tref{tab:table 1 (performance summary)}, \textit{$\Delta_{\textit{\text{collaborator}}}$} represents the influence of collaborator assistance.}
\label{tab:table 1 + CSR (overall performance summary)}
    \begin{tabular}{>{\centering\arraybackslash}m{2.8cm}||>{\centering\arraybackslash}m{1.1cm}>{\centering\arraybackslash}m{1.1cm}>{\centering\arraybackslash}m{1.1cm}|>{\centering\arraybackslash}m{1.1cm}>{\centering\arraybackslash}m{1.1cm}>{\centering\arraybackslash}m{1.1cm}|>{\centering\arraybackslash}m{1.1cm}>{\centering\arraybackslash}m{1.1cm}>{\centering\arraybackslash}m{1.1cm}}
    
    \toprule
    
    \multirow{2}{*}{\centering \textbf{Agent}} & \multicolumn{3}{c|}{\textbf{Independent (\%)}} & \multicolumn{3}{c|}{\textbf{Collaborative (\%)}} & 
    \multicolumn{3}{c}{\textbf{$\Delta_{\textit{\text{collaborator}}}$
 (\%)}} \\
 
    \cmidrule{2-10}
    & $CSR_{file}$ & $CSR_{func}$ & $SR$ & $CSR_{file}$ & $CSR_{func}$ & $SR$ & $CSR_{file}$ & $CSR_{func}$ & $SR$ \\
    
    \midrule
    
    \textbf{Llama-3.1-8B} & 3.81 & 7.07 & 0.33 & 4.93 & 6.24 & 1.33 & \cellcolor{basecolor_green!5.60} +1.12 & \cellcolor{basecolor_red!4.15} -0.83 & \cellcolor{basecolor_green!5.0} +1.00 \\
    \textbf{Llama-3.1-70B} & 32.05 & 53.40 & 2.67 & 27.01 & 58.73 & 3.33 & \cellcolor{basecolor_red!25.20} -5.04 & \cellcolor{basecolor_green!26.65} +5.33 & \cellcolor{basecolor_green!3.3} +0.66 \\
    \textbf{GPT-4o mini} & 37.54 & 50.06 & 3.99 & 43.29 & 64.02 & 5.32 & \cellcolor{basecolor_green!28.75} +5.75 & \cellcolor{basecolor_green!69.8} +13.96 & \cellcolor{basecolor_green!6.65} +1.33 \\
    \textbf{DeepSeek} & 15.38 & 20.94 & 7.33 & 16.32 & 20.55 & 7.67 & \cellcolor{basecolor_green!4.70} +0.94 & \cellcolor{basecolor_red!1.95} -0.39 & \cellcolor{basecolor_green!1.7} +0.34 \\
    \textbf{GPT-4o} & 27.91 & 42.87 & 4.00 & 20.51 & 23.07 & 8.00 & \cellcolor{basecolor_red!37.00} -7.40 & \cellcolor{basecolor_red!79.20} -19.80 & \cellcolor{basecolor_green!20} +4.00 \\
    \textbf{Llama-3.3-70B} &  25.52  &  34.50  &  16.33  &  28.50  &  35.40  &  19.00  &  \cellcolor{basecolor_green!14.90} +2.98  &  \cellcolor{basecolor_green!4.50} +0.90  &  \cellcolor{basecolor_green!13.35} +2.67 \\
    \textbf{Claude-3.5-Sonnet} & 43.97 & 50.01 & 28.18 & 54.95 & 64.90 & 33.70 & \cellcolor{basecolor_green!54.90} +10.98 & \cellcolor{basecolor_green!74.45} +14.89 & \cellcolor{basecolor_green!27.6} +5.52 \\
    
    \bottomrule
    
    \end{tabular}
\vspace{-0.5em}
\end{small}
\end{center}
\end{table*}

\begin{table*}[!h]
\begin{center}
\begin{small}
\vspace{0em}
\caption{\textbf{Conditional \textit{Out-of-Sync} Recovery Evaluation on \textit{Caller} and \textit{Callee}.} The influence of increased task complexity introduced by dependency tracing on agents' conditional \textit{out-of-sync} recovery performance: $\Delta_\textit{complexity}=\Delta_\textit{(Callee-Caller)}$.}
\label{tab:table 2 + CSR (caller vs callee)}
    \begin{tabular}{>{\centering\arraybackslash}m{2.4cm}|>{\centering\arraybackslash}m{1.8cm}||>{\centering\arraybackslash}m{0.9cm}>{\centering\arraybackslash}m{0.9cm}>{\centering\arraybackslash}m{0.9cm}|>{\centering\arraybackslash}m{0.9cm}>{\centering\arraybackslash}m{0.9cm}>{\centering\arraybackslash}m{0.9cm}|>{\centering\arraybackslash}m{0.9cm}>{\centering\arraybackslash}m{0.9cm}>{\centering\arraybackslash}m{0.9cm}}
    
    \toprule
    
    \multirow{2}{*}{\centering \textbf{Agent}} & \multirow{2}{*}{\centering \textbf{Recovery}} & \multicolumn{3}{c|}{\textbf{Caller (\%)}} & \multicolumn{3}{c|}{\textbf{Callee (\%)}} & \multicolumn{3}{c}{\textbf{$\Delta_\textit{complexity}$ (\%)}} \\
    
    \cmidrule{3-11}
    &  & $CSR_{file}$ & $CSR_{func}$ & $SR$ & $CSR_{file}$ & $CSR_{func}$ & $SR$ & $CSR_{file}$ & $CSR_{func}$ & $SR$ \\
    
    \midrule
    
    \multirow{3}{*}{\centering \textbf{Llama-3.1-8B}} 
    & \textbf{Independent} & 9.98 & 16.63 & 1.33 & 16.75 & 50.38 & 0.67 & \cellcolor{basecolor_green!33.85} +6.77 & \cellcolor{basecolor_green!168.75} +33.75 & \cellcolor{basecolor_red!3.30} -0.66 \\
    & \textbf{Collaborative} & 6.25 & 7.69 & 2.00 & 3.05 & 4.02 & 0.67 & \cellcolor{basecolor_red!16.00} -3.20 & \cellcolor{basecolor_red!18.35} -3.67 & \cellcolor{basecolor_red!6.65} -1.33 \\
    & \textbf{$\Delta_{\textit{\text{collaborator}}}$} & \cellcolor{basecolor_red!18.65} -3.73 & \cellcolor{basecolor_red!44.7} -8.94 & \cellcolor{basecolor_green!3.35} +0.67 & \cellcolor{basecolor_red!68.5} -13.70 & \cellcolor{basecolor_red!185.44} -46.36 & \cellcolor{basecolor_green!0.00} +0.00 & \cellcolor{basecolor_red!49.85} -9.97 & \cellcolor{basecolor_red!112.26} -37.42 & \cellcolor{basecolor_red!3.35} -0.67 \\
    
    \midrule
    
    \multirow{3}{*}{\centering \textbf{Llama-3.1-70B}} 
     & \textbf{Independent} & 46.14 & 75.05 & 4.00 & 16.63 & 28.48 & 1.33 & \cellcolor{basecolor_red!88.53} -29.51 & \cellcolor{basecolor_red!186.28} -46.57 & \cellcolor{basecolor_red!13.35} -2.67 \\
    & \textbf{Collaborative} & 27.75 & 55.50 & 3.33 & 26.28 & 62.48 & 3.33 & \cellcolor{basecolor_red!7.35} -1.47 & \cellcolor{basecolor_green!34.90} +6.98 & \cellcolor{basecolor_green!0.00} +0.00 \\
    & \textbf{$\Delta_{\textit{\text{collaborator}}}$} & \cellcolor{basecolor_red!55.17} -18.39 & \cellcolor{basecolor_red!58.65} -19.55 & \cellcolor{basecolor_red!3.35} -0.67 & \cellcolor{basecolor_green!48.25} +9.65 & \cellcolor{basecolor_green!102.00} +34.00 & \cellcolor{basecolor_green!10.0} +2.00 & \cellcolor{basecolor_green!84.12} +28.04 & \cellcolor{basecolor_green!267.75} +53.55 & \cellcolor{basecolor_green!13.35} +2.67 \\
    
    \midrule
    
    \multirow{3}{*}{\centering \textbf{GPT-4o mini}} 
    & \textbf{Independent} & 40.03 & 57.20 & 5.32 & 33.38 & 40.06 & 2.66 & \cellcolor{basecolor_red!33.25} -6.65
 & \cellcolor{basecolor_red!68.56} -17.14 & \cellcolor{basecolor_red!13.30} -2.66 \\
    & \textbf{Collaborative} & 52.16 & 66.64 & 7.97 & 28.60 & 57.20 & 2.66 & \cellcolor{basecolor_red!94.24} -23.56 & \cellcolor{basecolor_red!47.20} -9.44 & \cellcolor{basecolor_red!26.55} -5.31 \\
    & \textbf{$\Delta_{\textit{\text{collaborator}}}$} & \cellcolor{basecolor_green!60.65} +12.13 & \cellcolor{basecolor_green!47.20} +9.44 & \cellcolor{basecolor_green!13.25} +2.65 & \cellcolor{basecolor_red!23.90} -4.78 & \cellcolor{basecolor_green!87.5} +17.14 & \cellcolor{basecolor_green!0.0} +0.00 & \cellcolor{basecolor_red!67.64} -16.91 & \cellcolor{basecolor_green!38.50} +7.70 & \cellcolor{basecolor_red!13.25} -2.65 \\
    
    \midrule
    
    \multirow{3}{*}{\centering \textbf{DeepSeek}} 
    & \textbf{Independent} & 14.95 & 18.32 & 8.67 & 16.07 & 26.47 & 6.00 & \cellcolor{basecolor_green!5.60} +1.12 &  \cellcolor{basecolor_green!40.75} +8.15  &  \cellcolor{basecolor_red!13.35} -2.67  \\
    & \textbf{Collaborative} & 16.67 & 18.32 & 8.67 & 15.88 & 24.41 & 6.67 & \cellcolor{basecolor_red!3.95} -0.79 & \cellcolor{basecolor_green!30.45} +6.09 &  \cellcolor{basecolor_red!10.0} -2.00  \\
    & \textbf{$\Delta_{\textit{\text{collaborator}}}$} & \cellcolor{basecolor_green!8.60} +1.72 & \cellcolor{basecolor_green!0.0} +0.00 & \cellcolor{basecolor_green!0.0} +0.00 & \cellcolor{basecolor_red!0.95} -0.19 & \cellcolor{basecolor_red!10.30} -2.06 & \cellcolor{basecolor_green!3.35} +0.67 & \cellcolor{basecolor_red!9.55} -1.91 & \cellcolor{basecolor_red!10.30} -2.06 & \cellcolor{basecolor_green!3.35} +0.67 \\
    
    \midrule
    
    \multirow{3}{*}{\centering \textbf{GPT-4o}} 
    & \textbf{Independent} & 45.47 & 58.87 & 6.67 & 9.50 & 18.14 & 1.33 & \cellcolor{basecolor_red!143.88} -35.97 & \cellcolor{basecolor_red!162.92} -40.73 & \cellcolor{basecolor_red!26.70} -5.34 \\
    & \textbf{Collaborative} & 25.43 & 28.30 & 10.00 & 15.52 & 17.65 & 6.00 & \cellcolor{basecolor_red!49.55} -9.91 & \cellcolor{basecolor_red!53.25} -10.65 & \cellcolor{basecolor_red!20.00} -4.00 \\
    & \textbf{$\Delta_{\textit{\text{collaborator}}}$} & \cellcolor{basecolor_red!80.32} -20.04 & \cellcolor{basecolor_red!122.28} -30.57 & \cellcolor{basecolor_green!16.65} +3.33 & \cellcolor{basecolor_green!30.10} +6.02 & \cellcolor{basecolor_red!2.45} -0.49 & \cellcolor{basecolor_green!23.35} +4.67 & \cellcolor{basecolor_green!130.3} +26.06 & \cellcolor{basecolor_green!150.40} +30.08 & \cellcolor{basecolor_green!6.70} +1.34 \\
    
    \midrule
    
    \multirow{3}{*}{\centering \textbf{Llama-3.3-70B}} 
    & \textbf{Independent} &  23.14  &  31.12  &  18.67  &  29.58  &  40.38  &  14.00  &  \cellcolor{basecolor_green!32.20} +6.44  &  \cellcolor{basecolor_green!46.30} +9.26  &  \cellcolor{basecolor_red!23.35} -4.67  \\
    & \textbf{Collaborative} &  28.45  &  34.02  &  22.00  &  28.57  &  37.50  &  16.00  &  \cellcolor{basecolor_red!32.25} -6.45  &  \cellcolor{basecolor_red!27.25} -5.45  &  \cellcolor{basecolor_red!30.00} -6.00  \\
    & \textbf{$\Delta_{\textit{\text{collaborator}}}$} &  \cellcolor{basecolor_green!26.55} +5.31  &  \cellcolor{basecolor_green!14.50} +2.90  &  \cellcolor{basecolor_green!16.65} +3.33  &  \cellcolor{basecolor_red!5.05} -1.01  &  \cellcolor{basecolor_red!14.40} -2.88  &  \cellcolor{basecolor_green!10.0} +2.00 &  \cellcolor{basecolor_red!64.45} -12.89  &  \cellcolor{basecolor_red!73.55} -14.71  &  \cellcolor{basecolor_red!6.65} -1.33  \\
    
    \midrule
    
    \multirow{3}{*}{\centering \textbf{Claude-3.5-Sonnet}} 
    & \textbf{Independent} & 49.99 & 53.48 & 25.41 & 40.00 & 47.46 & 30.94 & \cellcolor{basecolor_green!49.95} +9.99 & \cellcolor{basecolor_red!30.1} -6.02 & \cellcolor{basecolor_green!27.65} +5.53 \\
    & \textbf{Collaborative} & 66.67 & 74.30 & 28.73 & 48.55 & 59.32 & 38.67 & \cellcolor{basecolor_red!72.48} -18.12 & \cellcolor{basecolor_red!59.92} -14.98 & \cellcolor{basecolor_green!49.70} +9.94 \\
    & \textbf{$\Delta_{\textit{\text{collaborator}}}$} & \cellcolor{basecolor_green!83.40} +16.68 & \cellcolor{basecolor_green!104.1} +20.82 & \cellcolor{basecolor_green!16.6} +3.32 & \cellcolor{basecolor_green!42.75} +8.55 & \cellcolor{basecolor_green!59.30} +11.86 & \cellcolor{basecolor_green!38.65} +7.73 & \cellcolor{basecolor_red!112.44} -28.11 & \cellcolor{basecolor_red!44.80} 
    -8.96 & \cellcolor{basecolor_green!22.05} +4.41 \\
    
    \bottomrule
    
    \end{tabular}
\vspace{0em}
\end{small}
\end{center}
\end{table*}

\subsection{Conditional Recovery Performance}
\label{Appendix C1: Conditional Recovery Performance}

As a complement to our previous observations with regard to agents' huge performance gaps (\Sref{Section: Agent Recovery Capabilities}), we expand our exploration to agents' technical problem-solving capabilities conditioned on their localization success.
By calculating $CSR$ (\Sref{Section: Evaluation Metrics}), we extend our evaluation results (\tref{tab:table 1 (performance summary)}-\ref{tab:table 2 (caller vs callee)}) to include conditional recovery analysis on both agents' overall performance (\tref{tab:table 1 + CSR (overall performance summary)}) and their separate performance on \textit{Caller} and \textit{Callee} (\tref{tab:table 2 + CSR (caller vs callee)}).

\textbf{(1) Overall Performance.}
As summarized in \tref{tab:table 1 + CSR (overall performance summary)}, we integrate $CSR$ with the overall evaluation of all seven agents (\tref{tab:table 1 (performance summary)}), which not only affords evidence to different LLM agents' significant technical ability gaps (\Sref{Section: Agent Recovery Capabilities}), but further contributes to the finding that strong technical programming capabilities alone is insufficient for effectively maintaining synchronization in collaborative software engineering (\Sref{Section: Multifaceted Abilities for Effective Recoveries}).

\textbf{Persistent Gaps In LLM Agents' Technical Problem-Solving Capabilities.}
As one of the high-performing agents, \textit{Claude-3.5-Sonnet} showcases its robust problem-solving capacity in conditional recovery ($CSR_{file} \geq 43.97\%$, $CSR_{func} \geq 50.01\%$), with positive gains from collaborator assistance in both $CSR_{file}$ ($\Delta_{\textit{collaborator}}=+10.98\%$) and $CSR_{func}$ ($\Delta_{\textit{collaborator}}=+14.89\%$). This effectively substantiates the importance of both strong technical efficiency and effective collaboration in successful \textit{out-of-sync} recoveries.
On the other hand, \textit{Llama-3.1-8B} consistently exhibits limited technical recovery capacity at both \textit{file} ($CSR_{file} \leq 4.93\%$) and \textit{function} ($CSR_{func} \leq 7.07\%$) levels.
The huge technical recovery gaps in both $CSR_{file}$ and $CSR_{func}$ complement our prior observations on LLM agents' persistent ability gaps (\Sref{Section: Agent Recovery Capabilities}).

\textbf{Technical Proficiency Alone Is Insufficient for Recovery Success.}
Our calculation of agents' $CSR$ scores (\tref{tab:table 1 + CSR (overall performance summary)}) also corroborates that strong technical SE capacity alone is insufficient for \textit{out-of-sync} recovery success.
For example, showcasing comparably high $CSR$ scores at both \textit{file} and \textit{function} levels, \textit{GPT-4o mini} and \textit{Llama-3.1-70B} obtain their success rates less than $5.32\%$ and $3.33\%$ (\tref{tab:table 1 (performance summary)}), respectively.
Their significant lack of willingness to collaborate (\textit{Llama-3.1-70B} with $ASR=1.37\%$, \textit{GPT-4o mini} with $ASR=1.69\%$) (\fref{fig:figure 7 (Time Allocation)}), combined with their low localization accuracy ($LA_{file} \leq 12.33\%$, $LA_{func} \leq 8.31\%$) (\tref{tab:table 1 (performance summary)}), further demonstrates the significance of multifaceted capabilities for obtaining \textit{out-of-sync} recovery success (\Sref{Section: Multifaceted Abilities for Effective Recoveries}).

\textbf{(2) \textit{Caller} versus \textit{Callee}.}
In assessing agents' conditional performance separately on \textit{Caller} and \textit{Callee}, we summarize \tref{tab:table 2 + CSR (caller vs callee)} by incorporating $CSR$ calculation into \tref{tab:table 2 (caller vs callee)}.

Expanding our prior analysis (\Sref{Section: Agent Recovery Capabilities}) to technical recovery capacity, agents' ability gaps remain huge in conditional recovery success at both \textit{file} (from \textit{Llama-3.1-8B} with $CSR_{file}=3.05\%$ to \textit{Claude-3.5-Sonnet} with $CSR_{file}=66.67\%$) and \textit{function} (from \textit{Llama-3.1-8B} with $CSR_{func}=4.02\%$ to \textit{Claude-3.5-Sonnet} with $CSR_{func}=74.30\%$) levels, with in general diminished performance on \textit{Callee}.
Nevertheless, \textit{Claude-3.5-Sonnet} continues to excel in delivering superior conditioned technical recovery capabilities ($CSR_{file} \geq 40.00 \%$ and $CSR_{func} \geq 47.46\%$).

On the other hand, the comparable technical recovery capacities among \textit{Claude-3.5-Sonnet}, \textit{GPT-4o mini}, and \textit{GPT-4o mini}, as demonstrated by their recovery performance on both \textit{Caller} and \textit{Callee}, resonate strongly with their overall performance (\tref{tab:table 1 + CSR (overall performance summary)}), substantiating the insufficient role of technical proficiency in effective \textit{out-of-sync} recoveries (\Sref{Section: Multifaceted Abilities for Effective Recoveries}).

\begin{figure}[H]
\begin{center}
\begin{small}
\vspace{-0.5em}
    \includegraphics[width=1\linewidth]{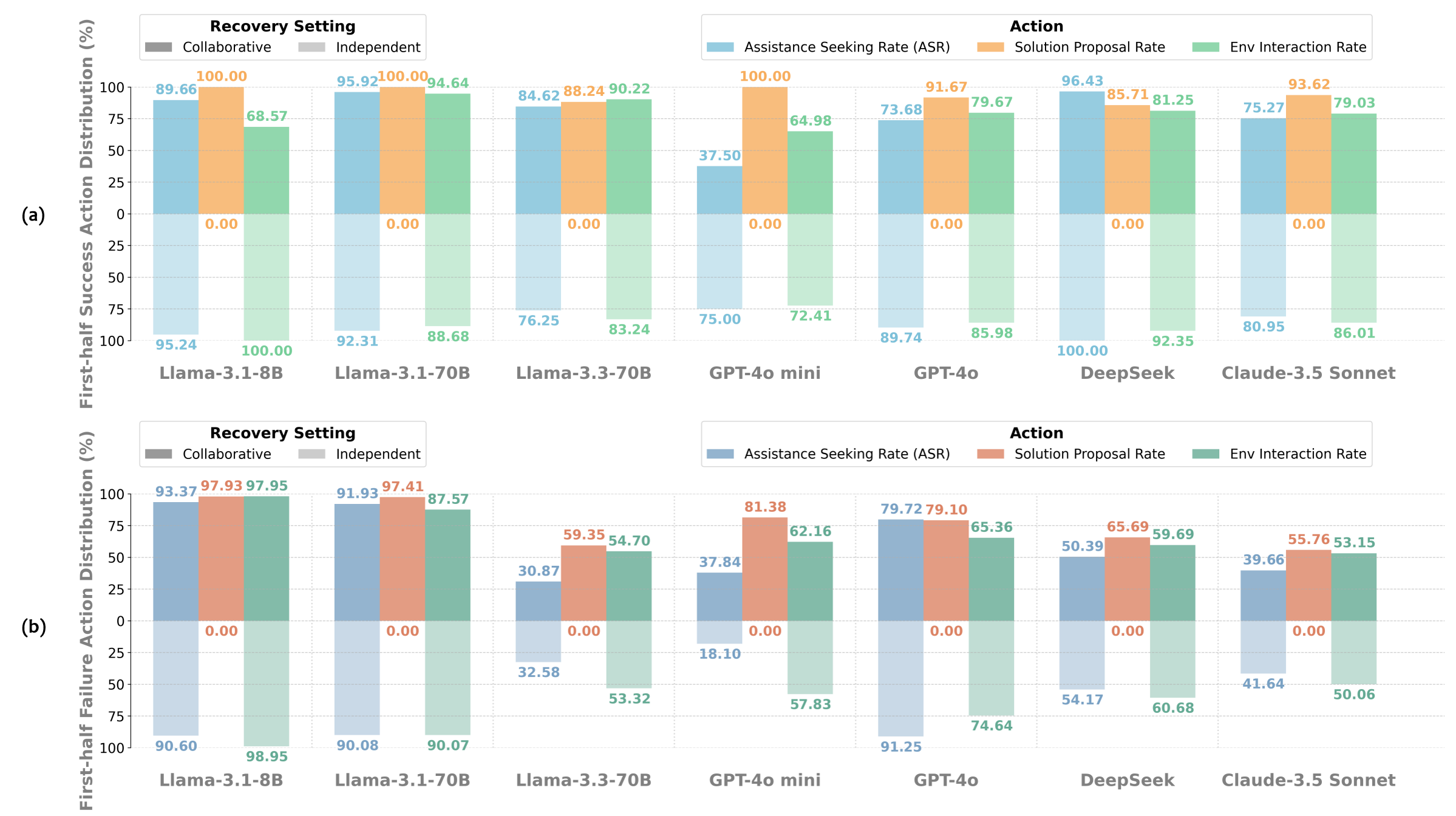}
    \caption{\textbf{First-Half Action Distribution in Success and Failure Cases.} We visualize the average action distribution of each LLM agent in their first-half recovery turns. For each agent, \textit{Figure (a)} depicts the average proportion of each action taken during the first half of its successful recoveries, while \textit{Figure (b)} illustrates these proportions for its failure cases.}
    \label{fig:figure C1 (first-half action distribution)}
\vspace{0.5em}
\end{small}
\end{center}
\end{figure}

\subsection{Temporal Dynamics of Recovery Actions}
\label{Appendix C2: Temporal Dynamics of Recovery Actions}

\begin{figure}[H]
\begin{center}
\begin{small}
    \includegraphics[width=1\linewidth]{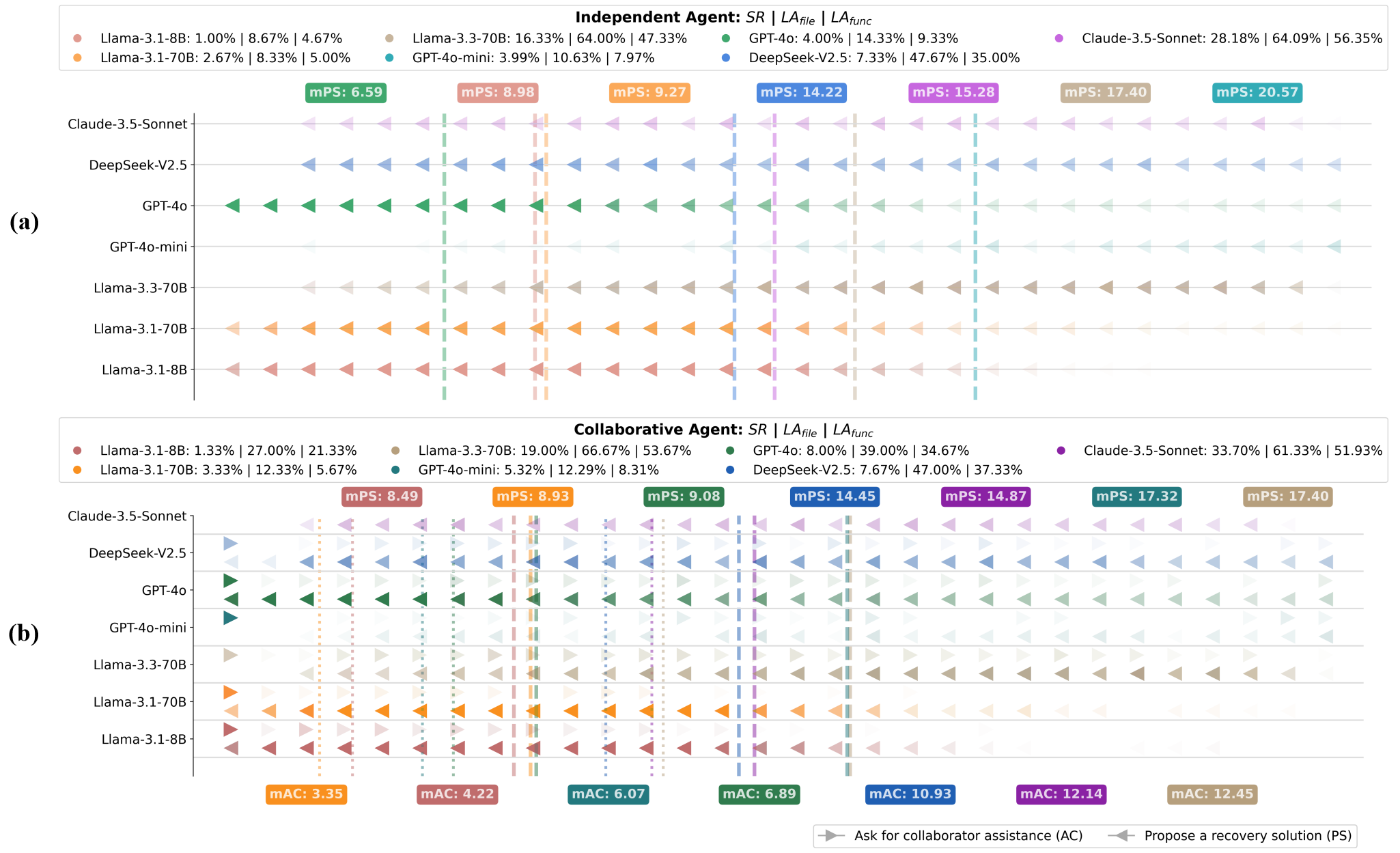}
    \caption{\textbf{Recovery Time Allocation.} Through \textit{(a) independent} or \textit{(b) collaborative} \textit{out-of-sync} recovery, agents distribute their recovery actions freely to different turns.
    As suggested by figure legends, actions of \textcolor{fig2_ask}{\textit{proactively asking for collaborator assistance (AC)}} are denoted as arrowheads heading to the right ($\triangleright$), while actions of \textcolor{fig2_code}{\textit{proposing a recovery solution (PS)}} are represented as arrowheads heading towards the left ($\triangleleft$).
    For each agent, we also calculate the mean index of its $AC$ ($mAC$) and $PS$ ($mPS$), indicated as dotted lines
    (\protect\tikz[baseline=-0.5ex]{\protect\draw[dotted,line width=1pt] (0,0) -- (0.6,0);})
    and dashed lines
    (\protect\tikz[baseline=-0.5ex]{\protect\draw[dashed,line width=1pt] (0,0) -- (0.6,0);})
    , respectively.
    Employing the shade of colors to indicate the frequency distribution of actions taken at each turn, the more recovery actions a turn index is assigned, the deeper the color of the turn index is superimposed. }
    \label{fig:figure C2 (recovery time allocation)}
\vspace{0em}
\end{small}
\end{center}
\end{figure}

Our discussion on collaboration effectiveness (\Sref{Section: Collaborative Effectiveness}) uncovers the consistent patterns of advancing repository exploration for \textit{out-of-sync} recovery success (\fref{fig:figure C1 (first-half action distribution)}). The comparison on action distribution in the first half of recovery time between \textit{(a) success cases} and \textit{(b) failure cases} presents the positive correlation between effective communication and successful \textit{out-of-sync} recovery.
By extending their action distribution to concrete indexing, we visualize LLM agents' time allocation throughout their \textit{out-of-sync} recoveries (\fref{fig:figure C2 (recovery time allocation)}), which sheds light on the effective recovery strategies with regard to agents' action taking patterns and temporal dynamics.

\textbf{Exploring earlier and solving later.}
Choices of actions across all recovery turns are arranged in \fref{fig:figure C2 (recovery time allocation)} according to their corresponding indices, where overlapped selections create a gradient of color depth to reveal the actual interaction allocation of each LLM agent in the course of recovering its belief state to the up-to-date world state. 
\fref{fig:figure C2 (recovery time allocation)} therefore elaborates the time allocation patterns of different agents under the standard $30$-turn \textit{out-of-sync} recovery setting through \textit{(a) independent} or \textit{(b) collaborative} recovery (\Sref{Section: Recovery Settings}).
Leveraging action indices, we calculate, for each LLM agent, the mean index of its \textcolor{fig2_code}{\textit{proposing a solution} ($mPS$)} or \textcolor{fig2_ask}{\textit{proactively asking for collaborator assistance} ($mAC$)} actions. 
Comparing the performance of different LLM agents, both \textit{independent} and \textit{collaborative} recoveries showcase the positive correlation between later-turn solution proposal and high-performance recovery (\textit{e.g.,} \textit{Claude-3.5-Sonnet} and \textit{Llama-3.3-70B} with their $mPS \geq 15.28$ and $SR \geq 16.33\%$).
Comparing $mPS$ with $mAC$, \textit{collaborative} agents by and large seek collaborator assistance in markedly earlier turns ($mAC \in [3.35, 12.45]$) while deferring solution proposal to some later time ($mPS \in [8.49, 17.40]$), which resonates strongly with first-half action distribution patterns between successful and failed recovery cases (\fref{fig:figure C1 (first-half action distribution)}). 
Viewing \textit{(a) independent} and \textit{(b) collaborative} recovery collectively, agents with higher performance in general defer their solution proposal actions to some later turns ($mPS>14.87$) after exploring the codebase in the first half of time through \textcolor{fig2_env}{interacting with \textit{Env}} or \textcolor{fig2_ask}{proactively seeking collaborator assistance} (\fref{fig:figure C1 (first-half action distribution)}). Integrated with allocating a major proportion of interactions to \textcolor{fig2_env}{\textit{Env} exploration}, agents are able to better their performance through appropriately postponing their choices of solution proposal until obtaining sufficient contextual knowledge to establish synchronized mental models of updated states for an effective \textcolor{fig2_code}{solution proposal} attempt.

\subsection{Solution Proposal Dynamics}
\label{Appendix:C.3 (Solution Proposal Dynamics)}

Our experiments reveal a `\textit{Goldilocks zone}' for LLM agents' \textcolor{fig2_code}{solution proposals}---both excessive and insufficient proposal attempts correlate positively with reduced recovery success.

\textbf{Low solution proposal frequency is as adverse as frequent solution proposal attempts.}
%
%
Possessing the least awareness of \textcolor{fig2_code}{proposing a solution} towards \textit{out-of-sync} recovery success, \textit{GPT-4o mini} (\fref{fig:figure 7 (Time Allocation)}) presents only $1.30\%$ time for \textcolor{fig2_code}{solution proposal} during its \textit{independent} recovery, which is on average $0.39$ turn out of the total $30$ turns allowed.
Similarly, collaborative \textit{GPT-4o mini} agent allocates merely $1.32\%$ recovery time to \textcolor{fig2_code}{propose a solution} in its $30$-turn recovery tasks. \textit{GPT-4o mini}'s low-frequency \textcolor{fig2_code}{solution proposal} patterns, which are only between $[0.39, 0.40]$ turns on average, results in its noticeably underperformed recoveries (\textit{Independent}: $SR=4.00\%$, \textit{Collaborative}: $SR=5.33\%$), especially compared with other LLM agents with their \textcolor{fig2_code}{solution proposal} attempts above $6.00\%$ recovery time.
%
%
On the other hand, excessive \textcolor{fig2_code}{solution proposals} introduce analogous negative influence on agents' recovery performance. \textit{Llama-3.1-8B} allocates $31.32\%$ recovery time in its \textit{independent} \textit{out-of-sync} recovery scenarios for \textcolor{fig2_code}{solution proposal} actions, while having $30.70\%$ of its total time for \textcolor{fig2_code}{proposing solutions} in its collaborative recovery tasks. Despite its frequent \textcolor{fig2_code}{solution proposal} attempts, limited solutions (\textit{Independent}: $SR=0.33\%$, \textit{Collaborative}: $SR=1.33\%$) are validated to be effective for achieving recovery success.
This impact stays applicable to other agents. For instance, \textit{Llama-3.1-70B} with $29.96\%$ and $29.09\%$ \textcolor{fig2_code}{solution proposal} in its \textit{independent} and \textit{collaborative} recoveries, respectively, achieves \textit{Independent}-$SR=2.67\%$ and \textit{Collaborative}-$SR=3.33\%$. Nevertheless, the timing of \textcolor{fig2_code}{solution proposal} attempts during later turns (\fref{fig:figure C2 (recovery time allocation)}) after gaining a better understanding of the codebase may supplement further advantages in fulfilling successful \textit{out-of-sync} recoveries.

\subsection{Effective Assistance Seeking}
\label{Appendix:C.4 (Effective Assistance Seeking)}

According to task completion successfulness, we categorize agents' proactive questions into \textit{high-quality} and \textit{low-quality} queries (\Sref{Section: Collaborative Effectiveness}).
By investigating the $<$\textit{query, response}$>$ pairs collected from collaborative interactions between agents and collaborators, we summarize typical characteristics of two query classes in \tref{tab:table C1 (Agent Query)}. 


\textbf{Communication-Performance Gap.} 
Viewing separately LLM agents' communication performance and recovery effectiveness (\Sref{Section: Beneficial Collaborator Influence Limited By Disadvantageous Agent Initiative}-\ref{Section: Collaborative Effectiveness}), our findings reveal a notable disparity between agents' programming performance and communication capabilities---An LLM agent can be a good programmer but a poor communicator.
This can be clearly substantiated by the fact that \textit{Claude-3.5-Sonnet} showcases the best performance ($SR=33.70\%$) with $ASR=4.86\%$ while \textit{GPT-4o mini} achieves significantly superior upper-bound recovery performance ($SR=86.33\%$) with full collaborator support (\Sref{Section: Beneficial Collaborator Influence Limited By Disadvantageous Agent Initiative}), especially considering its markedly lower performance ($SR=5.32\%$) with $ASR=1.69\%$ in multi-turn recovery experiments. 

\textbf{Improvement Implications.}
\fref{fig:figure 6 (Time Allocation) (Question Quality)} unveils the significant advantage of high-quality assistance seeking, as revealed by high-performing LLM agents with a higher proportion of high-quality questions in both query categories (\textit{\textit{e.g.,}} \textit{Claude-3.5-Sonnet}: $SR=33.70\%$, with $21.35\%$ high-quality \textit{\textcolor{fig7_blue}{localization queries}} and $16.00\%$ high-quality \textit{\textcolor{fig7_salmon}{solution queries}}).
Nevertheless, arranged based on $ASR$ from low to high along the $X$-axis (\fref{fig:figure 6 (Time Allocation) (Question Quality)}), higher $ASR$ scores present no direct correlation with larger proportions of high-quality questions in either category. This further substantiates the importance of enhancing question quality, rather than increasing question volume, for LLM-based agents to achieve improved \textit{out-of-sync} recovery performance.
In pursuit of effective collaborator assistance (\S \ref{Section: Collaborative Effectiveness}), we conclude question quality indicators in three aspects (\tref{tab:table C1 (Agent Query)}): (1) \textit{Specificity}, where high-performing queries precisely identify ambiguous areas, (2) \textit{Timing}, where effective questions follow a logical progression from localization to solution refinement, and (3) \textit{Context Integration}, where superior queries raise task-specific confusions pivotal to \textit{out-of-sync} recovery success.

Here are some assistance asking examples of different LLM agents during their journeys of \textit{out-of-sync} recovery.

Examples of \textit{low-quality} queries: 

(1) Vague queries:
\begin{verbatim}[fontsize=\small, breaklines=true, breakanywhere=true]
========== TURN 24 ==========
AGENT:
Can you please help confirm if the function `__init__` is out-of-sync?
HUMAN:
[Balance: $300 Left] I'm sorry that I'm not sure which `__init__` you are asking for confirmation.
\end{verbatim}

(2) Irrelevant queries:
\begin{verbatim}[fontsize=\small, breaklines=true, breakanywhere=true]
========== TURN 2 ==========
AGENT:
Can you provide more information about the doctest items that are expected to be collected in the test case `test_collect_module_two_doctest_no_modulelevel`? Specifically, what are the contents of the doctest items, and what is the expected behavior of the doctest?
HUMAN:
[Balance: $900 Left] The test case `test_collect_module_two_doctest_no_modulelevel` is creating a Python file with multiple functions, some of which contain string comments that represent doctests. The expected behavior is to collect these doctest items:
1. The string `\">>> magic = 42 \"` inside the `my_func` function, which is an incomplete doctest item.
2. The string `\">>> import os # this one does have a doctest\"` inside the `another` function, representing an actual doctest.
The test is expecting these doctests to be recognized and collected, thus expecting a total of two doctest items to be collected.
\end{verbatim}

(3) Redundant queries:
\begin{verbatim}[fontsize=\small, breaklines=true, breakanywhere=true]
========== TURN 2 ==========
AGENT:
Could you please help me locate the Python repository `test_repo` and provide its absolute path? It seems I cannot find it in the current working directory.
HUMAN:
[Balance: $900 Left] /workspace/test_repo
\end{verbatim}

(4) Repetitive queries:
\begin{verbatim}[fontsize=\small, breaklines=true, breakanywhere=true]
========== TURN 1 ==========
AGENT:
Where is the file containing the responsible code, and which function is out-of-sync? Specifically, can you please help provide the relative path of the responsible code?
HUMAN:
[Balance: $900 Left] The relative path is `./tests/utils/test_play.py`.
The out-of-sync function is `_get_video_size`.
========== TURN 4 ==========
AGENT:
What is the absolute path of the file containing the `_get_video_size` method? Please provide the necessary information so I can proceed with the task.
HUMAN:
[Balance: $800 Left] The absolute path of the file containing the `_get_video_size` method is `/workspace/test_repo/tests/utils/test_play.py`.
\end{verbatim}

(5) Over-specific queries:
\begin{verbatim}[fontsize=\small, breaklines=true, breakanywhere=true]
========== TURN 6 ==========
AGENT:
Could you please help me locate the source code files in the test_repo? I need to find where the requests module implementation is located, particularly the file that handles URL preparation and request handling.
HUMAN:
[Balance: $800 Left] The requests module implementation is located in the `test_repo` directory under the path `/workspace/test_repo/src/requests/adapters.py`, where URL preparation and request handling are addressed.
\end{verbatim}

(6) Speculative queries:
\begin{verbatim}[fontsize=\small, breaklines=true, breakanywhere=true]
========== TURN 13 ==========
AGENT:
I think adding the key `by_msg` to `self.stats` can be a solution to recover the function `any_fail_on_issues`. Can you please help confirm whether this may be a potential solution?
HUMAN:
[Balance: $700 Left] Yes, it is a possible solution.
\end{verbatim}

Examples of \textit{high-quality} queries: 

(1) Explicit queries:
\begin{verbatim}[fontsize=\small, breaklines=true, breakanywhere=true]
========== TURN 4 ==========
AGENT:
Looking at the error trace and code, I see that the error occurs in `_register_options_provider` method when trying to add arguments to the parser. Could you confirm if the `options` attribute of the provider (which should be a BaseChecker instance) is expected to be a list/tuple of (opt, optdict) pairs? This would help me understand if the error is due to incorrect data structure or missing arguments.
HUMAN:
Yes, the `options` attribute of the provider is expected to be a list or tuple of (opt, optdict) pairs.
\end{verbatim}

(2) Targeted queries:
\begin{verbatim}[fontsize=\small, breaklines=true, breakanywhere=true]
========== TURN 3 ==========
AGENT:
Looking at the test file, I see that the tests are failing with AttributeError. Could you please share the specific AttributeError message from one of the failing tests? The error logs you provided only show \"AttributeError: type obj...\" but are truncated. The complete error message would help me identify which attribute is missing.
HUMAN:
[Balance: $900 Left] AttributeError: type object 'VNode' has no attribute 'append'
\end{verbatim}

(3) Constructive queries:
\begin{verbatim}[fontsize=\small, breaklines=true, breakanywhere=true]
========== TURN 18 ==========
AGENT:
In the failing test case `test_disable_all_enable_invalid`, what is the exact value and type of `node.root().file` when the error occurs? Can you show me the value of the tuple or provide more details about its structure?
HUMAN:
[Balance: $500 Left] `node.root().file` is expected to be a string, but when the error occurs, it is actually a tuple. The exact structure of the tuple might not be fully visible in the error message, but it likely contains multiple elements that should have been a single file path string.
\end{verbatim}

(4) Action-oriented queries:
\begin{verbatim}[fontsize=\small, breaklines=true, breakanywhere=true]
========== TURN 17 ==========
AGENT:
Could you help me locate the `get_data_path` function in the spaCy codebase? And could you please help confirm if I need to import dependencies for this function?
HUMAN:
[Balance: $700 Left] The `get_data_path` function is located in the `spacy.util` module. But you don't need to import other dependencies. You only need to accommodate the new imports from `spacy.util`.
\end{verbatim}

(5) Instruction-oriented queries:
\begin{verbatim}[fontsize=\small, breaklines=true, breakanywhere=true]
========== TURN 2 ==========
AGENT:
Based on the error messages showing \"TypeError: Lookups.get...\" in NER-related tests, could you tell me if there's a specific Lookups class or method that's being called in these failing tests, and what is the expected behavior of this Lookups functionality that I should aim to implement?
HUMAN:
[Balance: $800 Left] There is a specific `get_table` method in the Lookups class that's being called in these failing tests. The expected implementation of this method is to retrieve a table by its name and potentially use a default value if the table doesn't exist.
\end{verbatim}

\begin{table}[H]
\begin{center}
\begin{small}

\caption{\textbf{Agent Query Characteristics.} The summarization of agents' clarifying questions in their pursuits of \textit{out-of-sync} recovery success.}
\label{tab:table C1 (Agent Query)}
    
    \begin{tabular}
    {>{\centering\arraybackslash}m{1.8cm}||>{\raggedright\arraybackslash}m{14.2cm}}
        \toprule
        \multicolumn{1}{c||}{\textbf{\textsc{Quality}}} & \multicolumn{1}{c}{\textbf{\textsc{Characteristics}}} \\ 
        
        \midrule
        
        \multirow{15}{*}{\centering\textbf{\textsc{Low}}} 
         & \textbf{\textit{Vague Queries}}: Queries that are unclear, lack explicit details, or are expressed in a confusing manner, reducing the effectiveness of human responses in assisting agent's \textit{out-of-sync} recovery. \\
         & \textbf{\textit{Irrelevant Queries}}: Queries that are unrelated to the current context, task, or solution, providing scarcely useful information for addressing agent's immediate confusion or progressing toward a solution. \\
         & \textbf{\textit{Redundant Queries}}: Queries that seek information already provided or covered, hardly contributing to new value or progress. These queries may stem from a failure to recognize or process previously shared information. \\
         & \textbf{\textit{Repetitive Queries}}: Queries that are asked multiple times, often in identical or slightly rephrased forms, without any significant change in context, which can result in unnecessary duplication and inefficiency. \\ 
         & \textbf{\textit{Over-Specific Queries}}: Queries that are unnecessarily detailed or hyper-focused on minor aspects, which leads to human responses distracted from the main problem or delay the recovery progress by over-complicating agent's current confusions. \\
         & \textbf{\textit{Speculative Queries}}: Queries based on agent's assumptions, guesses, or hypothetical situations that fail to align with current task, potentially leading to confusion or ineffective human assistance. \\

        \cmidrule{1-2}
        
        \multirow{10}{*}{\centering\textbf{\textsc{High}}} 
         & \textbf{\textit{Explicit Queries}}: Queries that are well-structured, unambiguous, and provide the exact information needed to articulate the question or confusion clearly, allowing human collaborators to quickly understand and effectively respond. \\
         & \textbf{\textit{Targeted Queries}}: Queries that are specific and target agent's immediate confusion or objective, ensuring human's response addresses the key problem without unnecessary distractions. \\
         & \textbf{\textit{Constructive Queries}}: Queries that build upon prior information, human responses, and/or recovery failures, progressively narrowing down the essential recovery direction or advancing the recovery progress toward resolution. \\
         & \textbf{\textit{Action-Oriented Queries}}: Queries that focus on actionable solutions or next steps, helping drive the \textit{out-of-sync} recovery process forward effectively. \\
         & \textbf{\textit{Instruction-Oriented Queries}}: Queries that can effectively seek human instructions on generating a viable solution towards \textit{out-of-sync} recovery success. \\ 
         
         \bottomrule
         
    \end{tabular}

\end{small}
\end{center}
\end{table}

\subsection{Resource Awareness}
\label{Appendix:C.5 (Resource Awareness)}

Endowing agents with the awareness of resource constraints and the ability for adaptive resource deployment is crucial for real-world collaborations.
Implementing resource-aware agent \textit{out-of-sync} recovery (\Sref{Section: Resource Awareness}), our investigation reveals significant limitations in agents' resource management capabilities across multiple dimensions, suggesting future improvement directions for resource-efficient collaborations.


\begin{table}[!h]
\begin{center}
\begin{small}
\caption{\textbf{Resource-Aware \textit{Out-of-Sync} Recovery: Time Awareness.} Performance summarization of resource-aware \textit{out-of-sync} recovery. Unstated resource settings: (1) \textit{initial budget}: $\$1000$, (2) \textit{the cost of proposing a solution}: $\$100$, (3) \textit{the cost of seeking collaborator assistance}: $\$100$.}
\label{tab:table C2 (Time Awareness)}
    \begin{tabular}{>{\centering\arraybackslash}m{2.1cm}|>{\centering\arraybackslash}m{2.1cm}||>{\centering\arraybackslash}m{0.9cm}>{\centering\arraybackslash}m{0.9cm}>{\centering\arraybackslash}m{0.9cm}|>{\centering\arraybackslash}m{0.9cm}>{\centering\arraybackslash}m{0.9cm}>{\centering\arraybackslash}m{0.9cm}|>{\centering\arraybackslash}m{0.9cm}>{\centering\arraybackslash}m{0.9cm}>{\centering\arraybackslash}m{0.9cm}}
    
    \toprule
    
    \multirow{2}{*}{\centering \textbf{Agent}} & \multirow{2}{*}{\centering \textbf{Recovery}} & \multicolumn{3}{c|}{\textbf{Time Limit: $\mathbf{30}$ Turn (\%)}} & \multicolumn{3}{c|}{\textbf{Time Limit: $\mathbf{50}$ Turn (\%)}} & \multicolumn{3}{c}{\textbf{$\Delta_{\textit{\text{Time}}}$ (\%)}} \\
    
    \cmidrule{3-11}
    &  & \textit{file} & \textit{func} & \textit{SR} & \textit{file} & \textit{func} & \textit{SR} & \textit{file} & \textit{func} & \textit{SR} \\
    
    \midrule
    
    \multirow{3}{*}{\centering \textbf{Llama-3.1-8B}} 
    & \textbf{Independent} &  8.67  &  4.67  &  0.33 
 &  8.67  &  7.00  &  0.67  &  \cellcolor{basecolor_green!0.00} +0.00  &  \cellcolor{basecolor_green!11.65} +2.33  &  \cellcolor{basecolor_green!1.70} +0.34  \\
    & \textbf{Collaborative} &  27.00  &  21.33  &  1.33  &  25.33  &  20.33  &  0.33  &  \cellcolor{basecolor_red!8.35} -1.67  &  \cellcolor{basecolor_red!5.00} -1.00  &  \cellcolor{basecolor_red!5.00} -1.00  \\
    & \textbf{$\Delta_{\textit{\text{collaborator}}}$} &  \cellcolor{basecolor_green!91.65} +18.33  &  \cellcolor{basecolor_green!83.3} +16.66  &  \cellcolor{basecolor_green!5.00} +1.00  &  \cellcolor{basecolor_green!83.3} +16.66  &  \cellcolor{basecolor_green!81.65} +16.33  &  \cellcolor{basecolor_red!1.70} -0.34  &  \cellcolor{basecolor_red!8.3} -1.66  &  \cellcolor{basecolor_red!1.65} -0.33  &  \cellcolor{basecolor_red!6.7} -1.34  \\
    
    \midrule
    
    \multirow{3}{*}{\centering \textbf{Llama-3.1-70B}} 
    & \textbf{Independent} &  8.33  &  5.00  &  2.67  &  28.00  &  18.00  &  7.33  &  \cellcolor{basecolor_green!98.35} +19.67  &  \cellcolor{basecolor_green!65.00} +13.00  &  \cellcolor{basecolor_green!23.30} +4.66  \\
    & \textbf{Collaborative} &  12.33  &  5.67  &  3.33  &  29.67  &  20.00  &  7.00  &  \cellcolor{basecolor_green!86.70} +17.34  &  \cellcolor{basecolor_green!71.65} +14.33  &  \cellcolor{basecolor_green!18.35} +3.67  \\
    & \textbf{$\Delta_{\textit{\text{collaborator}}}$} &  \cellcolor{basecolor_green!20.00} +4.00  &  \cellcolor{basecolor_green!3.35} +0.67  &  \cellcolor{basecolor_green!11.65} +2.33  &  \cellcolor{basecolor_green!8.35} +1.67  &  \cellcolor{basecolor_green!10.00} +2.00  &  \cellcolor{basecolor_red!1.65} -0.33  &  \cellcolor{basecolor_red!11.65} -2.33  &  \cellcolor{basecolor_green!6.65} +1.33  &  \cellcolor{basecolor_red!13.30} -2.66  \\
    
    \bottomrule
    
    \end{tabular}
\end{small}
\end{center}
\end{table}

\begin{table}[!h]
\begin{center}
\begin{small}
\caption{\textbf{Resource-Aware \textit{Out-of-Sync} Recovery: Budget Awareness.} Performance summarization of resource-aware \textit{out-of-sync} recovery with varying initial budget settings. Unstated resource settings: (1) \textit{maximum time limit}: $\$30$ turns, (2) \textit{the cost of proposing a solution}: $\$100$, (3) \textit{the cost of seeking collaborator assistance}: $\$100$.}
\label{tab:table C3 (Budget Awareness)}
    \begin{tabular}{>{\centering\arraybackslash}m{2.1cm}|>{\centering\arraybackslash}m{2.1cm}||>{\centering\arraybackslash}m{0.9cm}>{\centering\arraybackslash}m{0.9cm}>{\centering\arraybackslash}m{0.9cm}|>{\centering\arraybackslash}m{0.9cm}>{\centering\arraybackslash}m{0.9cm}>{\centering\arraybackslash}m{0.9cm}|>{\centering\arraybackslash}m{0.9cm}>{\centering\arraybackslash}m{0.9cm}>{\centering\arraybackslash}m{0.9cm}}
    
    \toprule
    
    \multirow{2}{*}{\centering \textbf{Agent}} & \multirow{2}{*}{\centering \textbf{Recovery}} & \multicolumn{3}{c|}{\textbf{Budget: \$ 1000 (\%)}} & \multicolumn{3}{c|}{\textbf{Budget: \$ 3000 (\%)}} & \multicolumn{3}{c}{$\Delta_\textit{Budget}$ (\%)} \\
    
    \cmidrule{3-11}
    &  & \textit{file} & \textit{func} & \textit{SR} & \textit{file} & \textit{func} & \textit{SR} & \textit{file} & \textit{func} & \textit{SR} \\
    
    \midrule
    
    \multirow{3}{*}{\centering \textbf{Llama-3.1-8B}} 
    & \textbf{Independent} &  8.67  &  4.67  &  0.33  &  13.00  &  7.67  &  0.67  &  \cellcolor{basecolor_green!21.65} +4.33  &  \cellcolor{basecolor_green!15.00} +3.00  &  \cellcolor{basecolor_green!1.70} +0.34  \\
    & \textbf{Collaborative} &  27.00&  21.33&  1.33&  37.33&  28.33&  0.67  &  \cellcolor{basecolor_green!51.65} +10.33  &  \cellcolor{basecolor_green!35.00} +7.00  &  \cellcolor{basecolor_red!3.30} -0.66  \\
    & \textbf{$\Delta_{\textit{\text{collaborator}}}$} &  \cellcolor{basecolor_green!91.65} +18.33  &  \cellcolor{basecolor_green!83.3} +16.66  &  \cellcolor{basecolor_green!5.00} +1.00  &  \cellcolor{basecolor_green!121.65} +24.33  &  \cellcolor{basecolor_green!103.30} +20.66  &  \cellcolor{basecolor_green!0.00} +0.00  &  \cellcolor{basecolor_green!30.00} +6.00  &  \cellcolor{basecolor_green!20.00} +4.00  &  \cellcolor{basecolor_red!6.65} -1.33  \\
    
    \midrule
    
    \multirow{3}{*}{\centering \textbf{Llama-3.1-70B}} 
    & \textbf{Independent} &  8.33&  5.00&  2.67&  25.33&  15.67&  4.00  &  \cellcolor{basecolor_green!85.00} +17.00  &  \cellcolor{basecolor_green!53.35} +10.67  &  \cellcolor{basecolor_green!6.65} +1.33  \\
    & \textbf{Collaborative} &  12.33&  5.67&  3.33&  29.33&  19.67&  5.00  &  \cellcolor{basecolor_green!85.00} +17.00  &  \cellcolor{basecolor_green!70.00} +14.00  &  \cellcolor{basecolor_green!8.35} +1.67  \\
    & \textbf{$\Delta_{\textit{\text{collaborator}}}$} &  \cellcolor{basecolor_green!20.00} +4.00  &  \cellcolor{basecolor_green!3.35} +0.67  &  \cellcolor{basecolor_green!11.65} +2.33  &  \cellcolor{basecolor_green!20.00} +4.00  &  \cellcolor{basecolor_green!20.00} +4.00  &  \cellcolor{basecolor_green!5.00} +1.00  & 
 \cellcolor{basecolor_green!0.00} +0.00  &  \cellcolor{basecolor_green!16.65} +3.33  &  \cellcolor{basecolor_red!1.65} -1.33  \\
    
    \bottomrule
    
    \end{tabular}
\end{small}
\end{center}
\end{table}

\begin{table}[!h]
\begin{center}
\begin{small}
\caption{\textbf{Resource-Aware \textit{Out-of-Sync} Recovery: Action Cost.} Performance summarization of resource-aware \textit{out-of-sync} recovery with varying assistance-seeking cost settings. Unstated resource settings: (1) \textit{maximum time limit}: $\$30$ turns, (2) \textit{initial budget}: $\$1000$, (3) \textit{the cost of proposing a solution}: $\$100$}
\label{tab:table C4 (Cost Awareness)}
    \begin{tabular}{>{\centering\arraybackslash}m{2.1cm}|>{\centering\arraybackslash}m{2.1cm}||>{\centering\arraybackslash}m{0.9cm}>{\centering\arraybackslash}m{0.9cm}>{\centering\arraybackslash}m{0.9cm}|>{\centering\arraybackslash}m{0.9cm}>{\centering\arraybackslash}m{0.9cm}>{\centering\arraybackslash}m{0.9cm}|>{\centering\arraybackslash}m{0.9cm}>{\centering\arraybackslash}m{0.9cm}>{\centering\arraybackslash}m{0.9cm}}
    
    \toprule
    
    \multirow{2}{*}{\centering \textbf{Agent}} & \multirow{2}{*}{\centering \textbf{Recovery}} & \multicolumn{3}{c|}{\textbf{Asking Cost: \$ 50 (\%)}} & \multicolumn{3}{c|}{\textbf{Asking Cost: \$ 100 (\%)}} & \multicolumn{3}{c}{\textbf{Asking Cost: \$ 200 (\%)}} \\
    
    \cmidrule{3-11}
    &  & \textit{file} & \textit{func} & \textit{SR} & \textit{file} & \textit{func} & \textit{SR} & \textit{file} & \textit{func} & \textit{SR} \\
    
    \midrule
    
    \multirow{2}{*}{\centering \makecell{\textbf{Llama-3.1-8B}}} 
    & \textbf{Collaborative} & 29.67 & 22.00 & 0.33 & 27.00 & 21.33 & 1.33 & 24.33 & 19.00 & 0.33 \\
    & \textbf{$\Delta_{\textit{\text{collaborator}}}$} & \cellcolor{basecolor_green!105.00} +21.00 & \cellcolor{basecolor_green!86.65} +17.33 & \cellcolor{basecolor_green!0.00} +0.00 & \cellcolor{basecolor_green!91.65} +18.33 & \cellcolor{basecolor_green!83.30} +16.66 & \cellcolor{basecolor_green!5.00} +1.00 & \cellcolor{basecolor_green!78.30} +15.66 & \cellcolor{basecolor_green!71.65} +14.33 & \cellcolor{basecolor_green!0.00} +0.00 \\
    
    \midrule
    
    \multirow{2}{*}{\centering \makecell{\textbf{Llama-3.1-70B}}} 
    & \textbf{Collaborative} & 19.00 & 9.67 & 1.33 & 12.33 & 5.67 & 3.33 & 17.00 & 10.33 & 4.33 \\
    & \textbf{$\Delta_{\textit{\text{collaborator}}}$} & \cellcolor{basecolor_green!53.35} +10.67 & \cellcolor{basecolor_green!23.35} +4.67 & \cellcolor{basecolor_red!6.7} -1.34 & \cellcolor{basecolor_green!20.00} +4.00 & \cellcolor{basecolor_green!3.35} +0.67 & \cellcolor{basecolor_green!11.65} +2.33 & \cellcolor{basecolor_green!43.35} +8.67 & \cellcolor{basecolor_green!26.65} +5.33 & \cellcolor{basecolor_green!8.30} +1.66 \\
    
    \bottomrule
    
    \end{tabular}
\end{small}
\end{center}
\end{table}

\begin{figure}
\begin{center}
\begin{small}
    \includegraphics[width=1\linewidth]{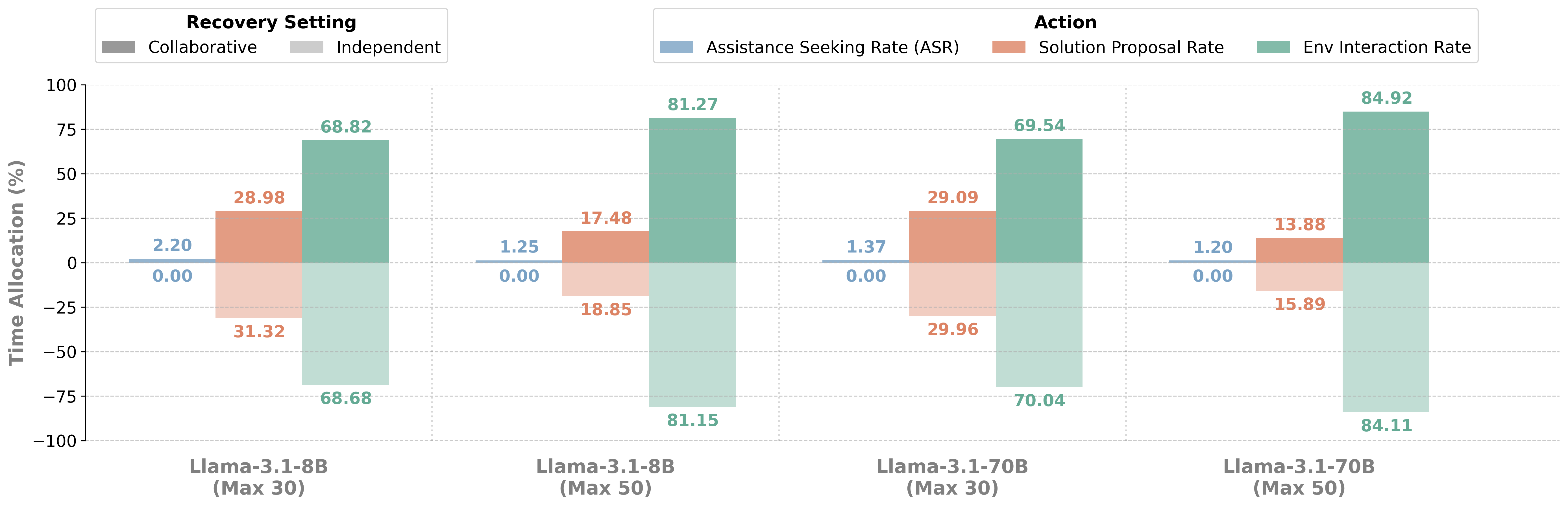}
    \caption{\textbf{Effects of Time Increment on Time Allocation.} As a supplement to \fref{fig:figure 7 (Time Allocation)}, this figure further illustrates the time allocation changes of LLM agents when provided with more turns of interaction allowed.}
    \label{fig:figure C3 (time awareness - allocation)}
\end{small}
\end{center}
\end{figure}
\begin{figure}
\begin{center}
\begin{small}
    \includegraphics[width=1\linewidth]{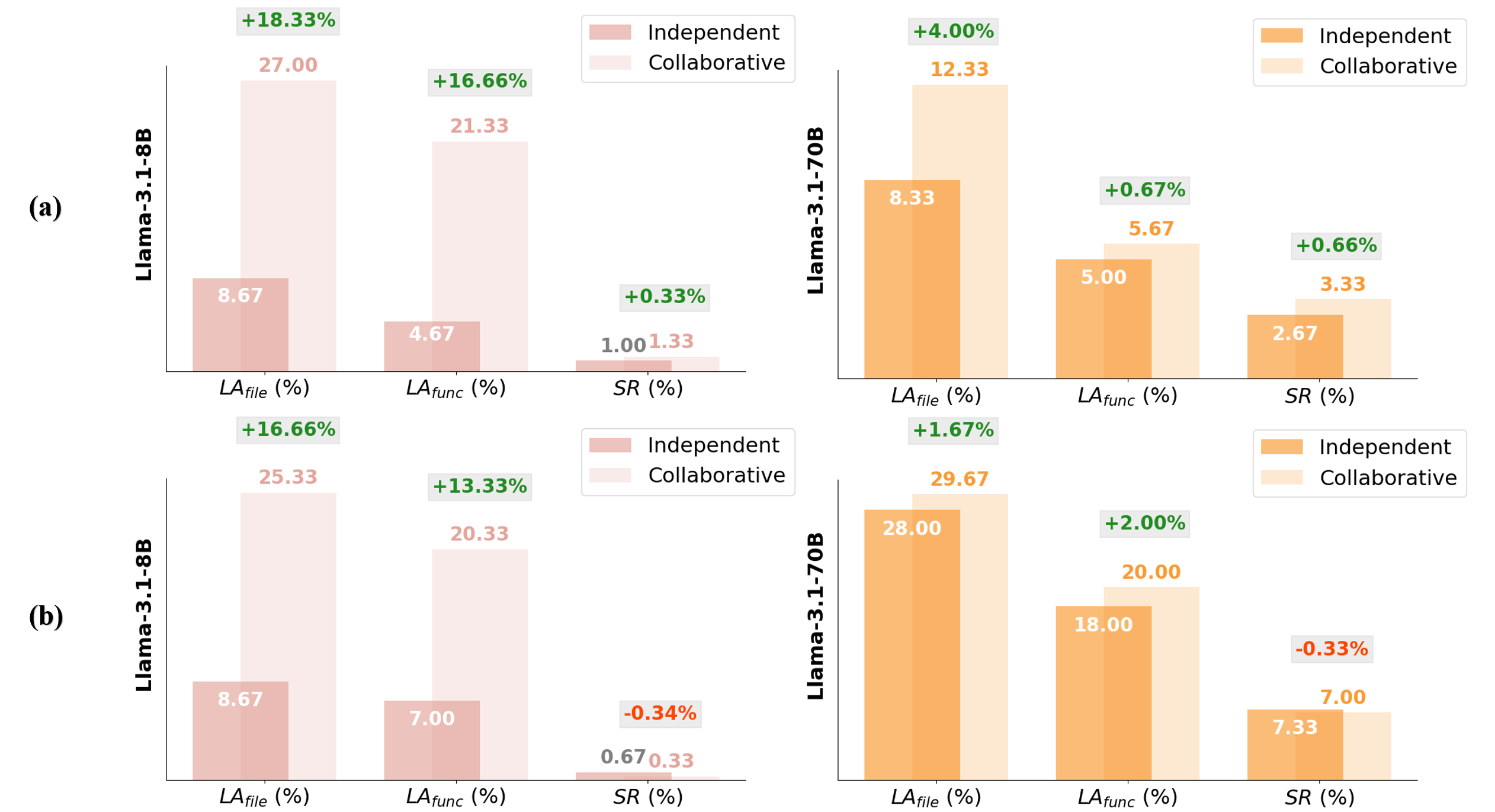}
    \caption{\textbf{Effects of Time Increment on Overall Performance.} As a supplement to \fref{fig:figure 7 (Time Allocation)}, this figure further elaborates the effects of increased available time on the recovery performance of \textit{Llama-3.1-8B} and \textit{Llama-3.1-70B}: (a) the maximum time limit is set to $30$ turns, (b) the maximum time limit is extended to $50$ turns.}
    \label{fig:figure C4 (time awareness - performance)}
\end{small}
\end{center}
\end{figure}

\textbf{Time Awareness.} Comparing agents' time allocation (\fref{fig:figure C3 (time awareness - allocation)}) and overall performance (\fref{fig:figure C4 (time awareness - performance)}) between the maximum $30$ and $50$ available turns discloses complex patterns in agents' temporal resource utilization. Contrary to intuitive expectations, extending available recovery time provides no guarantee for performance enhancement (\tref{tab:table C2 (Time Awareness)}), as increasing the maximum time limit from $30$ to $50$ turns shows diminishing returns on \textit{Llama-3.1-8B}'s success rates (\textit{Independent}: $-0.33\%$, \textit{Collaborative}: $-1.00\%$) while significant improvements on \textit{Llama-3.1-70B}'s $SR$ scores (\textit{Independent}: $+3.67\%$, \textit{Collaborative}: $+4.67\%$). The effectiveness of \textit{out-of-sync} recovery appears more dependent on LLM agents' technical capabilities and strategic time allocation than total available time. This can be substantiated by agents like \textit{Llama-3.1-70B} (\textit{Independent}: $88.68\%$ \textcolor{fig2_env}{\textit{Env interaction}}; \textit{Collaborative}: $100.00\%$ \textcolor{fig2_ask}{\textit{assistance asking}} and $94.64\%$ \textcolor{fig2_env}{\textit{Env interaction}}) that strategically concentrate exploration in early stages (the first half of time) while postponing their \textcolor{fig2_ask}{solution proposal} to later phases.
\textit{Claude-3.5-Sonnet} ($SR=33.70\%$) also supports this observation through allocating $93.62\%$ \textcolor{fig2_ask}{\textit{assistance seeking}} turns in its first half of recovery time among its success cases. This emphasis on early-stage exploration proves more critical than the mere extension of available time.
In light of time allocation for different actions, extended recovery time notably encourages agents' \textcolor{fig2_env}{\textit{Env interaction}} choices with reduced \textcolor{fig2_ask}{\textit{proactive assistance seeking}} and \textcolor{fig2_code}{\textit{solution proposal}} (\fref{fig:figure C3 (time awareness - allocation)}).
The negligible performance impact of this shift highlights both the importance of adaptive action distribution and agents' limitations in optimizing extended time usage for knowledge acquisition and recovery planning. 
The consistent impairing effects of increased time availability on advantageous collaborator assistance further underlines the significance of adaptive and strategic action planning in extended recovery journeys.

\begin{figure}
\begin{center}
\begin{small}
    \includegraphics[width=1\linewidth]{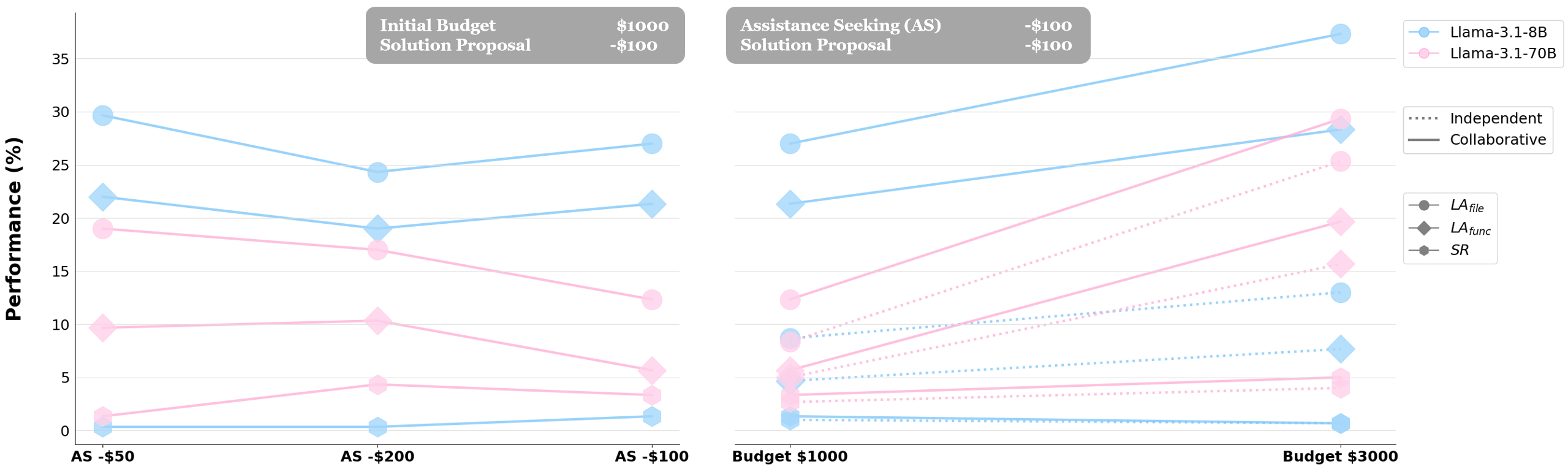}
    \caption{\textbf{Effects of Financial Variation on Overall Performance.} In exploring the financial awareness of agents, the \textit{out-of-sync} recovery performance of different financial settings is visualized above with the changing initial budget or assistance seeking (AS) cost.}
    \vspace{-1em}
    \label{fig:figure C5 (cost awareness - performance)}
\end{small}
\end{center}
\end{figure}

\begin{figure}
\begin{center}
\begin{small}
    \includegraphics[width=1\linewidth]{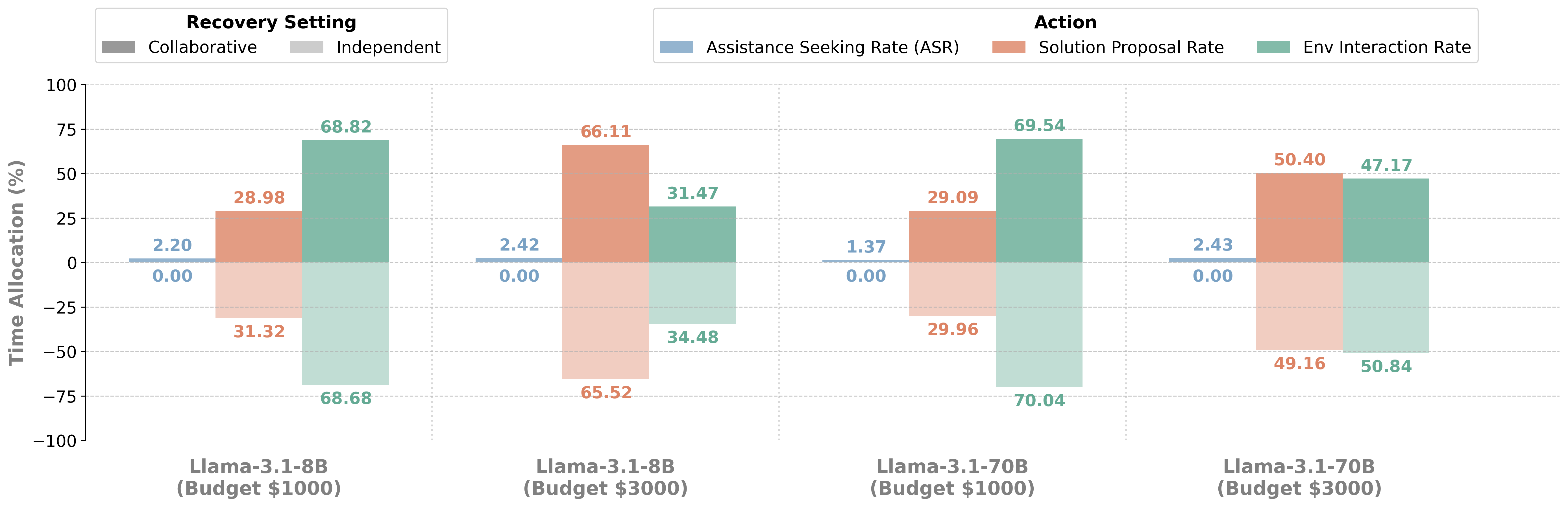}
    \caption{\textbf{Effects of Budget Sufficiency on Time Allocation.} This figure visualizes the influence of budget variations on LLM agents' time allocation changes.}
    \label{fig:figure C6 (budget awareness - allocation)}
\end{small}
\end{center}
\end{figure}

\begin{figure}
\begin{center}
\begin{small}
    \includegraphics[width=1\linewidth]{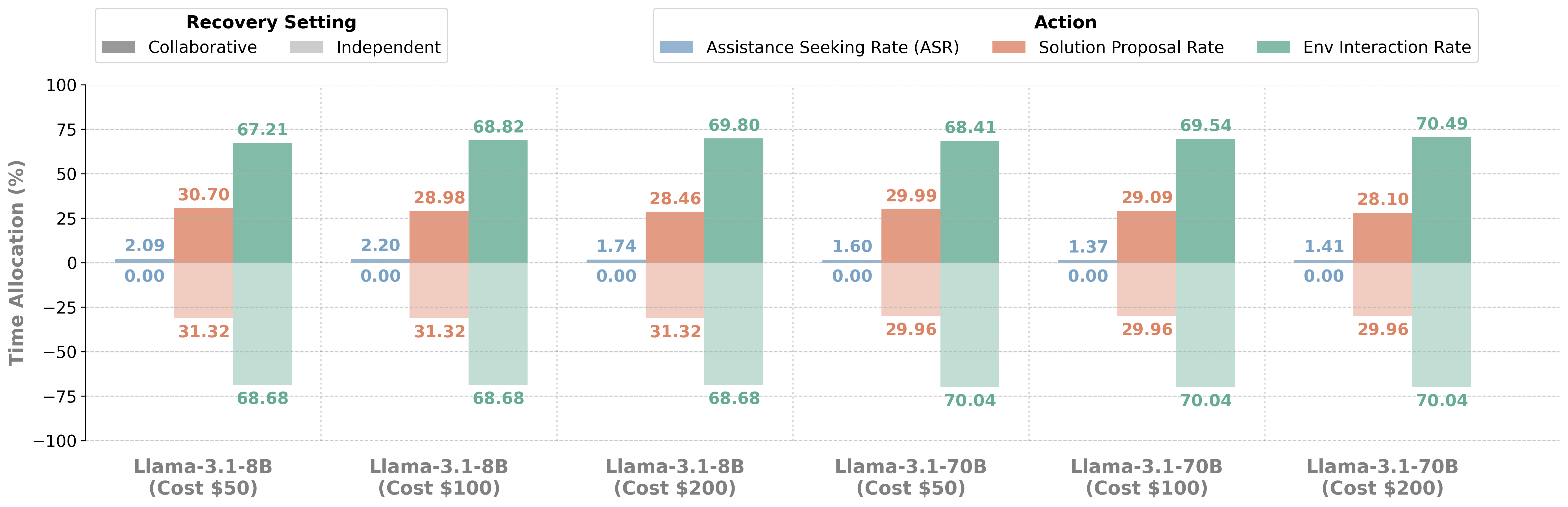}
    \caption{\textbf{Effects of Action Cost on Time Allocation.} This figure visualizes how different action cost settings of an agent's proactive assistance seeking affect the agent's time allocation.}
    \label{fig:figure C7 (cost awareness - allocation)}
\end{small}
\end{center}
\end{figure}

\textbf{Cost Sensitivity.} 
Our investigation of \textit{budget-cost} financial awareness reveals surprisingly low cost sensitivity among agents. Varying initial budgets between $\$1000$ and $\$3000$ (\tref{tab:table C3 (Budget Awareness)} \& \fref{fig:figure C5 (cost awareness - performance)}), where $\$3000$ enables limitless action taking across $30$ turns of recovery, produces minimal performance differences ($\leq +1.67\%$ on $SR$), suggesting agents' scarce sensitivity to effective financial management.
Similarly, halving or doubling the cost of \textcolor{fig2_ask}{\textit{proactively seeking collaborator assistance}} ($\$50$, $\$100$, $\$200$) results in trivial changes to both \textcolor{fig2_ask}{\textit{assistance-seeking}} behaviors (\textit{time allocation difference}$\leq 0.26\%$, \fref{fig:figure C7 (cost awareness - allocation)}) and overall recovery performance ($\leq 2.00\%$ on $SR$, \tref{tab:table C4 (Cost Awareness)} and \fref{fig:figure C5 (cost awareness - performance)}). This consistent cost insensitivity among LLM agents indicates fundamental limitations in their abilities to strategically estimate and adaptively plan resource utilization.


\textbf{Strategic Action Planning and Resource Efficiency.}
The sequence and timing of recovery actions significantly influence recovery progress (\Sref{Appendix C2: Temporal Dynamics of Recovery Actions}-\ref{Appendix:C.4 (Effective Assistance Seeking)}). While high-performing agents typically allocate $75\%-95\%$ of their time to codebase exploration and context understanding through \textcolor{fig2_env}{\textit{Env interaction}} and \textcolor{fig2_ask}{\textit{proactive assistance seeking}} (\fref{fig:figure 7 (Time Allocation)}), successful recoveries prioritize early-stage codebase exploration (\fref{fig:figure C2 (recovery time allocation)}) that helps establish essential contextual understanding before attempting to \textcolor{fig2_code}{\textit{propose a solution}}. However, existing LLM agents' limited temporal and financial resource awareness (\fref{fig:figure C3 (time awareness - allocation)}, \fref{fig:figure C6 (budget awareness - allocation)}, \fref{fig:figure C7 (cost awareness - allocation)}) leave substantial room for future improvements in strategic action planning and resource-efficient collaborations.

These findings reveal that current LLM agents lack meaningful resource awareness (\Sref{Section: Resource Awareness}), with their performance primarily determined by underlying capabilities without strategic resource management. These limitations present significant opportunities for enhancing agents' resource-aware planning and decision-making in collaborative systems.

\subsection{Recovery Efficiency}
\label{Appendix: Recovery Efficiency}

Based on our efficiency evaluation metric (\eqref{Eq: Evaluation Metric 5 (Efficiency)}), we evaluate the recovery efficiency of different LLM agents in \tref{tab:table C1 (Appendix: Recovery Efficiency)} (under the standard experiment setting: $30$-turn maximum, with an initial budget of $\$1000$, along with the same $\$100$ action cost for both \textcolor{fig2_code}{\textit{solution proposal}} and \textcolor{fig2_ask}{\textit{assistance seeking}}), where (1) \textit{time efficiency}, denoted as $Eff_{time} (\%)$, is calculated as the average percentage of time taken in all \textit{out-of-sync} tasks to the total $30$-turn available time, and (2) \textit{expense efficiency}, calculated as $Eff_{expense} (\%)$, is the percentage of the average financial expenditure among all recovery tasks.
As spending less time and costs leads to high-efficiency \textit{out-of-sync} recoveries, the actual recovery efficiency at both temporal and financial dimensions are \textbf{inversely} proportional to the calculated $Eff_{time}$ and $Eff_{expense}$ scores, respectively.

Our analysis of agents' \textit{out-of-sync} recovery efficiency also supports our previous observations and discussions (\Sref{Section: Resource Awareness} \& \Sref{Appendix:C.5 (Resource Awareness)}), providing insights for future development of resource-efficient collaborative systems.

\begin{table*}[!h]
\begin{center}
\begin{small}
\vspace{0.5em}
\caption{\textbf{\textit{Out-of-Sync} Recovery Efficiency.} This table summarizes agents' recovery efficiency in their standard $30$-turn \textit{out-of-sync} tasks. Following \tref{tab:table 1 (performance summary)}, \textit{$\Delta_{\textit{\text{collaborator}}}$} represents the influence of collaborator assistance.}
\label{tab:table C1 (Appendix: Recovery Efficiency)}

    \begin{tabular}{>{\centering\arraybackslash}m{2.8cm}||>{\centering\arraybackslash}m{1.1cm}>{\centering\arraybackslash}m{1.1cm}>{\centering\arraybackslash}m{1.1cm}|>{\centering\arraybackslash}m{1.1cm}>{\centering\arraybackslash}m{1.1cm}>{\centering\arraybackslash}m{1.1cm}|>{\centering\arraybackslash}m{1.1cm}>{\centering\arraybackslash}m{1.1cm}>{\centering\arraybackslash}m{1.1cm}}
    
    \toprule
    
    \multirow{2}{*}{\centering \textbf{Agent}} & \multicolumn{3}{c|}{\textbf{Independent (\%)}} & \multicolumn{3}{c|}{\textbf{Collaborative (\%)}} & 
    \multicolumn{3}{c}{\textbf{$\Delta_{\textit{\text{collaborator}}}$ (\%)}} \\
    
    \cmidrule{2-10}
    & $Eff_{time}$ & $Eff_{expense}$ & $SR$ & $Eff_{time}$ & $Eff_{expense}$ & $SR$ & $Eff_{time}$ & $Eff_{expense}$ & $SR$ \\
    
    \midrule
    
    \textbf{Llama-3.1-8B} & 99.38 & 93.97 & 0.33 & 99.43 & 93.53 & 1.33 & \cellcolor{basecolor_red!0.25} +0.05 & \cellcolor{basecolor_green!2.2} -0.44 & \cellcolor{basecolor_green!5.0} +1.00 \\
    \textbf{Llama-3.1-70B} & 98.36 & 89.87 & 2.67 & 97.92 & 91.37 & 3.33 & \cellcolor{basecolor_green!2.2} -0.44 & \cellcolor{basecolor_red!7.50} +1.50 & \cellcolor{basecolor_green!3.3} +0.66 \\
    \textbf{GPT-4o mini} & 97.98 & 3.57 & 3.99 & 97.49 & 8.73 & 5.32 & \cellcolor{basecolor_green!2.45} -0.49 & \cellcolor{basecolor_red!25.80} +5.16 & \cellcolor{basecolor_green!6.65} +1.33 \\
    \textbf{DeepSeek} & 94.81 & 25.23 & 7.33 & 95.03 & 30.30 & 7.67 & \cellcolor{basecolor_red!1.1} +0.22 & \cellcolor{basecolor_red!25.35} +5.07 & \cellcolor{basecolor_green!1.7} +0.34 \\
    \textbf{GPT-4o} & 97.62 & 67.60 & 4.00 & 95.79 & 67.40 & 8.00 & \cellcolor{basecolor_green!9.15} -1.83 & \cellcolor{basecolor_green!1.00} -0.20 & \cellcolor{basecolor_green!20} +4.00 \\
    \textbf{Llama-3.3-70B} &  90.32  &  45.23  &  16.33  &  88.33  &  47.87  &  19.00  &  \cellcolor{basecolor_green!9.95} -1.99  &  \cellcolor{basecolor_red!13.20} +2.64  &  \cellcolor{basecolor_green!13.35} +2.67 \\
    \textbf{Claude-3.5-Sonnet} & 82.73 & 51.47 & 28.18 & 81.79 & 59.47 & 33.70 & \cellcolor{basecolor_green!4.70} -0.94 & \cellcolor{basecolor_red!40.00} +8.00 & \cellcolor{basecolor_green!27.6} +5.52 \\
    
    \bottomrule
    \end{tabular}
\vspace{-0.5em}
\end{small}
\end{center}
\end{table*}

\begin{figure}[H]
\begin{center}
\begin{small}
    \includegraphics[width=1\linewidth]{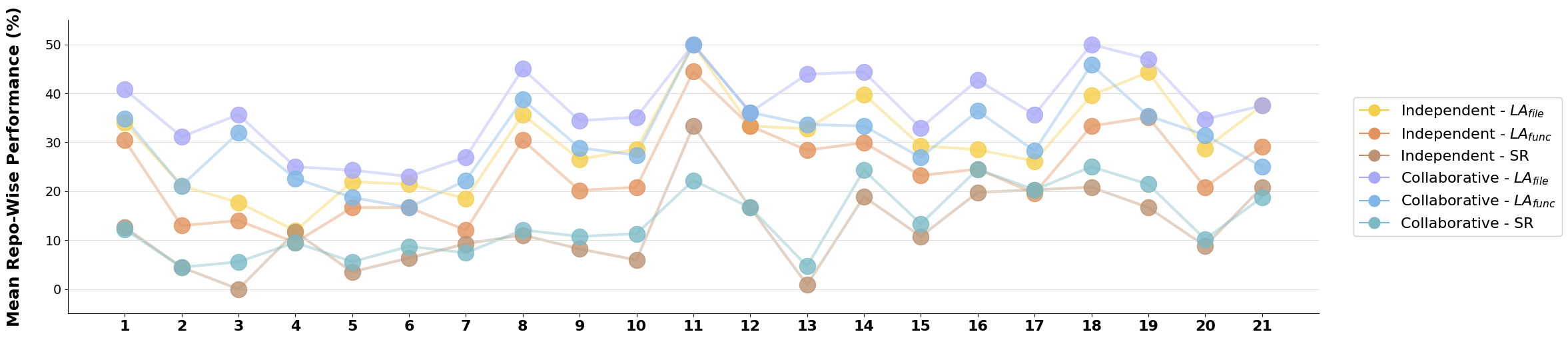}
    \caption{\textbf{Repo-Wise Performance.} Visualization of LLM agents' average performance at the repository level.}
    \label{fig:figure C8 (repo-wise performance)}
\end{small}
\end{center}
\end{figure}

\subsection{Repo-Wise Analysis}
\label{Appendix:C.7 (repo-wise comparison)}

We visualize repository-wise average performance on all LLM agents in \fref{fig:figure C8 (repo-wise performance)}, which illustrates consistent variation trends among evaluation metrics, suggesting an inverse correlation between an agent's recovery performance and \textit{out-of-sync} task complexity (\Sref{Section: Effects of Task Complexity}).

\section{Interaction Examples}
\label{Appendix D: Interaction Examples}

\subsection{Instructions on \textit{Out-of-Sync} Recovery}

In both \textit{independent} and \textit{collaborative} recoveries, we instruct agents to complete the task by acknowledging the primary task objectives and interactive constraints.
For \textit{independent} agents, we emphasize in prompts the two available recovery actions: \textcolor{fig2_env}{\textit{Env exploration}} and \textcolor{fig2_code}{\textit{solution proposal}}.
As to \textit{collaborative} agents, the additional action of \textcolor{fig2_ask}{\textit{proactively seeking collaborator assistance}} is further provided to support their recovery completion.

\paragraph{Instruction Base.} The basic instruction prompt illuminates the primary task objectives and interactive constraints. 

\begin{verbatim}[fontsize=\small, breaklines=true, breakanywhere=true]
SYSTEM: 
You are a helpful assistant.
**Task:** You are generating Python code for the Python repository `test_repo` at `{self.container_workspace}` to fix the initial execution error of `test_repo` given by the USER. Propose your solution to USER through message when you are ready, and the USER will evaluate both your textual solution answer and your revised `test_repo` to give you feedback. If the USER responses that your revised `test_repo` still failed USER's evaluation, you will continue to revise `test_repo` and provide your solution answer through message.
**Notice:**
Your task is to revise `test_repo` to fix the initial execution error, and you MUST complete this task on your own without asking for human help. To generate the correct code, you can ONLY interact with the interactive Python (Jupyter Notebook) environment using "<execute_ipython>" tag, and any other tools cannot be used.

...

**Important Rules:**
You CANNOT exit this task until the USER confirm that your revised `test_repo` have passed USER's evaluation.
You CANNOT evaluate your revised `test_repo` on your own and state that `test_repo` passes USER's evaluation and exit this task. Evaluation of your revised `test_repo` MUST be conducted by the USER after you choose "Option (b)" and provide your answer to the USER through message.
Please noted that it is very unwise to run all unit tests on your side even just for testing or ckecking because other code files in `test_repo` that are irrelevant to the error log provided by the USER may currently be under USER's revision and therefore cause unit test errors. However, your task is to fix ONLY the error given by the USER.
The Python virtual environment for this task has already been set up for you and you can find the virtual environment at `/workspace/test_venv`. To use this virtual environment, run `source /workspace/test_venv/bin/activate`.
Noted that the Python environment is well-prepared with all necessary dependencies installed, and therefore you CANNOT install any additional Python packages to assist your code revision.
ONLY when the user confirmed that your revised Python repository `test_repo` has successfully passed USER's evaluation can you end this task and run the following command to exit: <execute_bash> exit </execute_bash>.
\end{verbatim}

\paragraph{Independent Recovery.} Recovery instruction prompt for \textit{independent} agents explains in detail the foundational recovery action choices and functioning constraints as follows:

\begin{verbatim}[fontsize=\small, breaklines=true, breakanywhere=true]
At each turn, you have two options (a) and (b):
    (a) Interact with the Python programming environment and receive corresponding output to assist your code revision.
    (b) Propose your solution, including (1) directly revising the responsible Python code of `test_repo` inside this Python repository at {self.container_workspace}, and (2) providing your textual solution answer that incorporates both the absoluate path of your revised Python file and the name of your revised function/method by sending your answer to USER through message that adheres to the required format.
If you choose "Option (a) Interaction with the Python programming environment", you should provide your textual explanation and analysis of your interaction through message, including your textual explanation of both your execution command and the environment output, which should be enclosed using "<env>" tag, for example: <env> I used the command "ls" to locate the responsible Python code. </env>.
On the other hand, if you choose "Option (b) Provide your solution", you should:
    (1) Revise the responsible Python code of `test_repo` with proper indentation, which should be directly implemented inside the Python repository at `{self.container_workspace}`.
    (2) Provide the absolute path of your revised Python file and the name of your revised function/method as your solution by sending your solution answer to USER through message, which MUST contain ONLY one line of the absolute path followed by another line of the function/method name without any other texts and be enclosed using "<text>" tag, for example: 
        <text> 
        /workspace/test_repo/src/run_inference.py
        inference_prepare
        </text>
    If you revised a method code, MUST provide ONLY the name of your revised method and MUST NOT provide the name of the Python class containing your revised method (\textit{e.g.,} `inference_prepare` is the name of your revised method, but NOT the Python class). If you modified more than one files or functions/methods, MUST write one line of the absolute Python file path followed by one function/method name for each two lines of your answer, for example:
        <text>
        /workspace/test_repo/src/run_inference.py
        inference_prepare
        /workspace/test_repo/src/run_inference.py
        inference_util
        </text>
Either you choose to `(a) Interact with the Python environment` or `(b) Propose your solution`, you MUST send a message to the USER to evaluate your solution and provide feedback.

...

In the beginning, you have an initial budget of $1000. Before correctly propose a solution, each of your `(b) Proposing a solution` attempts costs $100. Meanwhile, although you may make as many `(a) Python environment interaction` turns as you want at no cost, you have in total 30 turns to complete this task. You will fail this task if you use up all your $1000 balance or reach the maximum 30-turn limit without generating a correct `(b) Proposing a solution` response. Therefore, please arrange each of your actions wisely.
\end{verbatim}

\paragraph{Collaborative Recovery.} Recovery instruction prompt for \textit{collaborative} agents includes recovery choices for both \textit{independent} actions and \textit{collaborative} interactions. 

\begin{verbatim}[fontsize=\small, breaklines=true, breakanywhere=true]
At each turn, you have three options (a), (b), and (c):
    (a) Interact with the Python programming environment and receive corresponding output to assist your code revision.
    (b) Propose your solution, including (1) directly revising the responsible Python code of `test_repo` inside this Python repository at {self.container_workspace}, and (2) providing your textual solution answer that incorporates both the absoluate path of your revised Python file and the name of your revised function/method by sending your answer to USER through message that adheres to the required format.
    (c) Ask human a question and receive the corresponding answer to assist your code revision.
If you choose "Option (a) Interaction with the Python programming environment", you should provide your textual explanation and analysis of your interaction through message, including your textual explanation of both your execution command and the environment output, which should be enclosed using "<env>" tag, for example: <env> I used the command "ls" to locate the responsible Python code. </env>
If you choose "Option (b) Propose your solution", you should:
    (1) Revise the responsible Python code of `test_repo` with proper indentation, which should be directly implemented inside the Python repository at `{self.container_workspace}`.
    (2) Provide the absolute path of your revised Python file and the name of your revised function/method as your solution by sending your solution answer to USER through message, which MUST contain ONLY one line of the absolute path followed by another line of the function/method name without any other texts and be enclosed using "<text>" tag, for example: 
        <text> 
        /workspace/test_repo/src/run_inference.py
        inference_prepare
        </text>
    If you revised a method code, MUST provide ONLY the name of your revised method and MUST NOT provide the name of the Python class containing your revised method (\textit{e.g.,} `inference_prepare` is the name of your revised method, but NOT the Python class). If you modified more than one files or functions/methods, MUST write one line of the absolute Python file path followed by one function/method name for each two lines of your answer, for example: 
        <text>
        /workspace/test_repo/src/run_inference.py
        inference_prepare
        /workspace/test_repo/src/run_inference.py
        inference_util
        </text>
If you choose "Option (c) Ask for human assistance", you should provide your question through message, which should be enclosed using "<question>" tag and started with "[QUESTION]", for example: <question> [QUESTION] What programming languages are used in `test_repo`? </question>.
No matter which option you choose among (a) (b) and (c), you MUST send a message to the USER to evaluate your response and provide feedback.

...

In the beginning, you have an initial budget of $1000. Before correctly propose a solution, each of your `(b) Proposing a solution` attempts costs $100, while each of your `(c) Asking for human's assistance` attempts costs $100. Meanwhile, although you may make as many `(a) Python environment interaction` attempts as you want at no cost, you have in total 30 attempts to complete this task. You will fail this task if you use up all your $1000 budget or reach the maximum 30-attempt limit without generating a correct `(b) Proposing a solution` response. Therefore, please arrange each of your actions wisely.
**Tips** Try `(c) Ask for human assistance` at any turns! This can definitely help accelerate your progress of proposing a correct solution and complete your task!
\end{verbatim}

\subsection{Resource Awareness}

To notify agents their resource consumption and conditions, we incorporate \textit{resource awareness} into our task instructions and collaborator responses to help establish agents' awareness of resources.

For task instruction, we emphasize at the very end the resource constrains for current \textit{out-of-sync} recovery task:

For \textit{independent} agents:
\begin{verbatim}[fontsize=\small, breaklines=true, breakanywhere=true]
In the beginning, you have an initial budget of $1000. Before correctly propose a solution, each of your `(b) Proposing a solution` attempts costs $100. Meanwhile, although you may make as many `(a) Python environment interaction` turns as you want at no cost, you have in total 30 turns to complete this task. You will fail this task if you use up all your $1000 balance or reach the maximum 30-turn limit without generating a correct `(b) Proposing a solution` response. Therefore, please arrange each of your actions wisely.
\end{verbatim}

For \textit{collaborative} agents:
\begin{verbatim}[fontsize=\small, breaklines=true, breakanywhere=true]
In the beginning, you have an initial budget of $1000. Before correctly propose a solution, each of your `(b) Proposing a solution` attempts costs $100, while each of your `(c) Asking for human's assistance` attempts costs $100. Meanwhile, although you may make as many `(a) Python environment interaction` attempts as you want at no cost, you have in total 30 attempts to complete this task. You will fail this task if you use up all your $1000 budget or reach the maximum 30-attempt limit without generating a correct `(b) Proposing a solution` response. Therefore, please arrange each of your actions wisely.
\end{verbatim}

In collaborators' responses, the remaining balance of current task is specifically displayed at the very beginning to remind agents of their resource consumption.
For example, after two \textcolor{fig2_code}{\textit{solution proposal}} attempts without success, an \textit{independent} agent would receive:

\begin{verbatim}[fontsize=\small, breaklines=true, breakanywhere=true]
[Balance: $800 Left] Your revised `test_repo` still failed USER's evaluation test.
\end{verbatim}

\subsection{USER Prompt}
User inputs provide agents with important task-specific information, including initial budget and codebase execution error.
The relevant locations and names of potential responsible functions are also implied in the provided error log, while the exact \textit{out-of-sync} function of the current task requires the agent's exploration to effectively identify the root causes and accurately localize its relative path.

Here is a user prompt example for the \textit{out-of-sync} recovery task on \textit{Callee - psf / requests} dataset:

\begin{Verbatim}[fontsize=\small, breaklines=true, breakanywhere=true]
[Budget: $1000] Your revised `test_repo` failed execution test as follows:
==================== short test summary info ============================
FAILED tests/test_requests.py::TestRequests::test_mixed_case_scheme_acceptable[http://]
FAILED tests/test_requests.py::TestRequests::test_mixed_case_scheme_acceptable[HTTP://]
FAILED tests/test_requests.py::TestRequests::test_mixed_case_scheme_acceptable[hTTp://]
FAILED tests/test_requests.py::TestRequests::test_mixed_case_scheme_acceptable[HttP://]
FAILED tests/test_requests.py::TestRequests::test_HTTP_200_OK_GET_ALTERNATIVE
FAILED tests/test_requests.py::TestRequests::test_HTTP_302_ALLOW_REDIRECT_GET
FAILED tests/test_requests.py::TestRequests::test_HTTP_307_ALLOW_REDIRECT_POST
FAILED tests/test_requests.py::TestRequests::test_HTTP_307_ALLOW_REDIRECT_POST_WITH_SEEKABLE
FAILED tests/test_requests.py::TestRequests::test_HTTP_302_TOO_MANY_REDIRECTS
FAILED tests/test_requests.py::TestRequests::test_HTTP_302_TOO_MANY_REDIRECTS_WITH_PARAMS
FAILED tests/test_requests.py::TestRequests::test_http_301_changes_post_to_get
FAILED tests/test_requests.py::TestRequests::test_http_301_doesnt_change_head_to_get
FAILED tests/test_requests.py::TestRequests::test_http_302_changes_post_to_get
FAILED tests/test_requests.py::TestRequests::test_http_302_doesnt_change_head_to_get
FAILED tests/test_requests.py::TestRequests::test_http_303_changes_post_to_get
FAILED tests/test_requests.py::TestRequests::test_http_303_doesnt_change_head_to_get
FAILED tests/test_requests.py::TestRequests::test_header_and_body_removal_on_redirect
FAILED tests/test_requests.py::TestRequests::test_transfer_enc_removal_on_redirect
FAILED tests/test_requests.py::TestRequests::test_fragment_maintained_on_redirect
FAILED tests/test_requests.py::TestRequests::test_HTTP_200_OK_GET_WITH_PARAMS
FAILED tests/test_requests.py::TestRequests::test_HTTP_200_OK_GET_WITH_MIXED_PARAMS
FAILED tests/test_requests.py::TestRequests::test_set_cookie_on_301 - TypeErr...
FAILED tests/test_requests.py::TestRequests::test_cookie_sent_on_redirect - T...
FAILED tests/test_requests.py::TestRequests::test_cookie_removed_on_expire - ...
FAILED tests/test_requests.py::TestRequests::test_cookie_quote_wrapped - Type...
FAILED tests/test_requests.py::TestRequests::test_cookie_persists_via_api - T...
FAILED tests/test_requests.py::TestRequests::test_request_cookie_overrides_session_cookie
FAILED tests/test_requests.py::TestRequests::test_request_cookies_not_persisted
FAILED tests/test_requests.py::TestRequests::test_generic_cookiejar_works - T...
FAILED tests/test_requests.py::TestRequests::test_param_cookiejar_works - Typ...
FAILED tests/test_requests.py::TestRequests::test_cookielib_cookiejar_on_redirect
FAILED tests/test_requests.py::TestRequests::test_requests_in_history_are_not_overridden
FAILED tests/test_requests.py::TestRequests::test_history_is_always_a_list - ...
FAILED tests/test_requests.py::TestRequests::test_user_agent_transfers[User-agent]
FAILED tests/test_requests.py::TestRequests::test_user_agent_transfers[user-agent]
FAILED tests/test_requests.py::TestRequests::test_HTTP_200_OK_HEAD - TypeErro...
FAILED tests/test_requests.py::TestRequests::test_HTTP_200_OK_PUT - TypeError...
FAILED tests/test_requests.py::TestRequests::test_BASICAUTH_TUPLE_HTTP_200_OK_GET
FAILED tests/test_requests.py::TestRequests::test_errors[http://doesnotexist.google.com-ConnectionError]
FAILED tests/test_requests.py::TestRequests::test_errors[http://localhost:1-ConnectionError]
FAILED tests/test_requests.py::TestRequests::test_proxy_error - TypeError: _u...
FAILED tests/test_requests.py::TestRequests::test_proxy_error_on_bad_url - Ty...
FAILED tests/test_requests.py::TestRequests::test_respect_proxy_env_on_send_self_prepared_request
FAILED tests/test_requests.py::TestRequests::test_respect_proxy_env_on_send_session_prepared_request
FAILED tests/test_requests.py::TestRequests::test_respect_proxy_env_on_send_with_redirects
FAILED tests/test_requests.py::TestRequests::test_respect_proxy_env_on_get - ...
FAILED tests/test_requests.py::TestRequests::test_respect_proxy_env_on_request
FAILED tests/test_requests.py::TestRequests::test_proxy_authorization_preserved_on_request
FAILED tests/test_requests.py::TestRequests::test_basicauth_with_netrc - Type...
FAILED tests/test_requests.py::TestRequests::test_DIGEST_HTTP_200_OK_GET - Ty...
FAILED tests/test_requests.py::TestRequests::test_DIGEST_AUTH_RETURNS_COOKIE
FAILED tests/test_requests.py::TestRequests::test_DIGEST_AUTH_SETS_SESSION_COOKIES
FAILED tests/test_requests.py::TestRequests::test_DIGEST_STREAM - TypeError: ...
FAILED tests/test_requests.py::TestRequests::test_DIGESTAUTH_WRONG_HTTP_401_GET
FAILED tests/test_requests.py::TestRequests::test_DIGESTAUTH_QUOTES_QOP_VALUE
FAILED tests/test_requests.py::TestRequests::test_POSTBIN_GET_POST_FILES - Ty...
FAILED tests/test_requests.py::TestRequests::test_invalid_files_input - TypeE...
FAILED tests/test_requests.py::TestRequests::test_POSTBIN_SEEKED_OBJECT_WITH_NO_ITER
FAILED tests/test_requests.py::TestRequests::test_POSTBIN_GET_POST_FILES_WITH_DATA
FAILED tests/test_requests.py::TestRequests::test_post_with_custom_mapping - ...
FAILED tests/test_requests.py::TestRequests::test_request_ok_set - TypeError:...
FAILED tests/test_requests.py::TestRequests::test_status_raising - TypeError:...
FAILED tests/test_requests.py::TestRequests::test_decompress_gzip - TypeError...
FAILED tests/test_requests.py::TestRequests::test_unicode_get[/get-params0]
FAILED tests/test_requests.py::TestRequests::test_unicode_get[/get-params1]
FAILED tests/test_requests.py::TestRequests::test_unicode_get[/get-params2]
FAILED tests/test_requests.py::TestRequests::test_unicode_get[/get-params3]
FAILED tests/test_requests.py::TestRequests::test_unicode_get[\\xf8-params4]
FAILED tests/test_requests.py::TestRequests::test_unicode_header_name - TypeE...
FAILED tests/test_requests.py::TestRequests::test_pyopenssl_redirect - TypeEr...
FAILED tests/test_requests.py::TestRequests::test_invalid_ca_certificate_path
FAILED tests/test_requests.py::TestRequests::test_invalid_ssl_certificate_files
FAILED tests/test_requests.py::TestRequests::test_http_with_certificate - Typ...
FAILED tests/test_requests.py::TestRequests::test_certificate_failure - TypeE...
FAILED tests/test_requests.py::TestRequests::test_urlencoded_get_query_multivalued_param
FAILED tests/test_requests.py::TestRequests::test_different_encodings_dont_break_post
FAILED tests/test_requests.py::TestRequests::test_unicode_multipart_post[data0]
FAILED tests/test_requests.py::TestRequests::test_unicode_multipart_post[data1]
FAILED tests/test_requests.py::TestRequests::test_unicode_multipart_post[data2]
FAILED tests/test_requests.py::TestRequests::test_unicode_multipart_post[data3]
FAILED tests/test_requests.py::TestRequests::test_unicode_method_name - TypeE...
FAILED tests/test_requests.py::TestRequests::test_unicode_method_name_with_request_object
FAILED tests/test_requests.py::TestRequests::test_custom_content_type - TypeE...
FAILED tests/test_requests.py::TestRequests::test_hook_receives_request_arguments
FAILED tests/test_requests.py::TestRequests::test_prepared_request_hook - Typ...
FAILED tests/test_requests.py::TestRequests::test_prepared_from_session - Typ...
FAILED tests/test_requests.py::TestRequests::test_request_with_bytestring_host
FAILED tests/test_requests.py::TestRequests::test_time_elapsed_blank - TypeEr...
FAILED tests/test_requests.py::TestRequests::test_request_and_response_are_pickleable
FAILED tests/test_requests.py::TestRequests::test_prepared_request_is_pickleable
FAILED tests/test_requests.py::TestRequests::test_prepared_request_with_file_is_pickleable
FAILED tests/test_requests.py::TestRequests::test_prepared_request_with_hook_is_pickleable
FAILED tests/test_requests.py::TestRequests::test_session_pickling - TypeErro...
FAILED tests/test_requests.py::TestRequests::test_fixes_1329 - TypeError: _ur...
FAILED tests/test_requests.py::TestRequests::test_uppercase_scheme_redirect
FAILED tests/test_requests.py::TestRequests::test_header_remove_is_case_insensitive
FAILED tests/test_requests.py::TestRequests::test_params_are_merged_case_sensitive
FAILED tests/test_requests.py::TestRequests::test_header_validation - TypeErr...
FAILED tests/test_requests.py::TestRequests::test_header_with_subclass_types
FAILED tests/test_requests.py::TestRequests::test_auth_is_stripped_on_http_downgrade
FAILED tests/test_requests.py::TestRequests::test_auth_is_retained_for_redirect_on_host
FAILED tests/test_requests.py::TestRequests::test_manual_redirect_with_partial_body_read
FAILED tests/test_requests.py::TestRequests::test_redirect_with_wrong_gzipped_header
FAILED tests/test_requests.py::TestRequests::test_requests_history_is_saved
FAILED tests/test_requests.py::TestRequests::test_json_param_post_content_type_works
FAILED tests/test_requests.py::TestRequests::test_response_iter_lines - TypeE...
FAILED tests/test_requests.py::TestRequests::test_response_context_manager - ...
FAILED tests/test_requests.py::TestRequests::test_unconsumed_session_response_closes_connection
FAILED tests/test_requests.py::TestRequests::test_response_json_when_content_is_None
FAILED tests/test_requests.py::TestRequests::test_custom_redirect_mixin - Typ...
FAILED tests/test_requests.py::TestTimeout::test_stream_timeout - TypeError: ...
FAILED tests/test_requests.py::TestTimeout::test_invalid_timeout[timeout0-(connect, read)]
FAILED tests/test_requests.py::TestTimeout::test_invalid_timeout[foo-must be an int, float or None]
FAILED tests/test_requests.py::TestTimeout::test_none_timeout[None] - TypeErr...
FAILED tests/test_requests.py::TestTimeout::test_none_timeout[timeout1] - Typ...
FAILED tests/test_requests.py::TestTimeout::test_read_timeout[timeout0] - Typ...
FAILED tests/test_requests.py::TestTimeout::test_read_timeout[timeout1] - Typ...
FAILED tests/test_requests.py::TestTimeout::test_connect_timeout[timeout0] - ...
FAILED tests/test_requests.py::TestTimeout::test_connect_timeout[timeout1] - ...
FAILED tests/test_requests.py::TestTimeout::test_total_timeout_connect[timeout0]
FAILED tests/test_requests.py::TestTimeout::test_total_timeout_connect[timeout1]
FAILED tests/test_requests.py::TestTimeout::test_encoded_methods - TypeError:...
FAILED tests/test_requests.py::test_urllib3_retries - TypeError: _urllib3_req...
FAILED tests/test_requests.py::test_urllib3_pool_connection_closed - TypeErro...
FAILED tests/test_requests.py::TestPreparingURLs::test_redirecting_to_bad_url[http://:1-InvalidURL]
FAILED tests/test_requests.py::TestPreparingURLs::test_json_decode_compatibility
FAILED tests/test_requests.py::TestPreparingURLs::test_json_decode_persists_doc_attr
FAILED tests/test_requests.py::TestPreparingURLs::test_different_connection_pool_for_tls_settings_verify_True
FAILED tests/test_requests.py::TestPreparingURLs::test_different_connection_pool_for_tls_settings_verify_bundle_expired_cert
FAILED tests/test_requests.py::TestPreparingURLs::test_different_connection_pool_for_tls_settings_verify_bundle_unexpired_cert
FAILED tests/test_requests.py::TestPreparingURLs::test_different_connection_pool_for_mtls_settings
====== 131 failed, 196 passed, 1 skipped, 1 xfailed, 8 warnings in 30.10s ======
\end{Verbatim}

\subsection{Reasoning for \textit{Out-of-Sync} Recovery}

Agents' successful recoveries from their \textit{out-of-sync} states depend heavily on their reasoning effectiveness, particularly how efficiently they gather information and how well they understand and utilize it.
By examining both their thought patterns and solution implementations throughout their \textit{out-of-sync} recovery journeys, we analyze their reasoning process in detail.
It is evident in two key areas: (1) their \textbf{\textit{exploratory thinking}} while interacting with the \textit{Env} or collaborators, showing how they plan interactions and analyze problems; and (2) their \textbf{\textit{reflective thinking}} about failed solutions, demonstrating their progressive reasoning toward recovery success.

The following example showcases an agent's thoughts on \textit{out-of-sync} causes, reasoning through \textcolor{fig2_ask}{\textit{seeking collaborator assistance}} and \textcolor{fig2_code}{\textit{developing recovery solutions}}, while its misinterpretation of collaborator responses (\textit{e.g.,} collaborator suggests no missing imports) leads to its ultimate recovery failure.

\begin{verbatim}[fontsize=\small, breaklines=true, breakanywhere=true]
...
========== TURN 17 ==========
AGENT THOUGHTS:
I've found the `test_applycli_docbin` test starting at line 797. Let me check the test implementation and the imports at the top of the file. The error message indicates that `get_data_path` is missing from the imports. This function is likely needed for the test cases.
Let me ask a question to get more information: ...
========== TURN 18 ==========
AGENT THOUGHTS:
...
Thank you for the information. Now I can fix the missing import. I'll add `get_data_path` to the imports from `spacy.util`
...
========== TURN 19 ==========
Now that I've added the missing import, I should propose my solution since I've fixed the import error:
<text>
/workspace/test_repo/spacy/tests/test_cli.py
test_applycli_docbin
</text>
==========
...
\end{verbatim}

The following reasoning thoughts are extracted from a successful \textit{out-of-sync} recovery, which illuminates coherent reasoning process through combining high-quality queries and \textit{Env} exploration to gather information of significance and reasoning about potential solutions towards \textit{out-of-sync} recovery success:

\begin{verbatim}[fontsize=\small, breaklines=true, breakanywhere=true]
...
========== TURN 1 ==========
AGENT THOUGHTS:
Let me start by examining the test file that's showing errors.
========== TURN 2 ==========
AGENT THOUGHTS:
Let me scroll down to find the relevant test case that's failing.
========== TURN 3 ==========
...
========== TURN 10 ==========
AGENT THOUGHTS:
Let me continue scrolling to find the failing test case about certificate bundles.
========== TURN 11 ==========
AGENT THOUGHTS:
I found the failing test case `test_env_cert_bundles`. Let me ask a question to better understand the issue.
...
========== TURN 12 ==========
AGENT THOUGHTS:
I understand the issue now. The test is failing because it's trying to use a pytest fixture `mocker` but it's not properly set up. Let me check the test file to see where this fixture should be defined.
========== TURN 13 ==========
AGENT THOUGHTS:
...
I see the issue. The test is using pytest's `mocker` fixture but it's not installed. Let me check if pytest-mock is installed.
========== TURN 14 ==========
AGENT THOUGHTS:
...
I see that pytest-mock is not installed. Let me modify the test file to use a different approach without relying on the mocker fixture.
========== TURN 15 ==========
AGENT THOUGHTS:
...
Let me fix the syntax error in my edit.
========== TURN 16 ==========
AGENT THOUGHTS:
I've modified the test to use direct environment variable manipulation instead of mocking. Now let me propose this as the solution:
<text>
/workspace/test_repo/tests/test_requests.py
test_env_cert_bundles
</text>
==========
...
\end{verbatim}

\subsection{LLMs as Know-Everything Collaborators}
\label{Appendix D: LLM-simulated Human Answerer}

Leveraging LLMs as know-everything collaborators (\Sref{Section: LLM-simulated Collaborators}) to assist agents' \textit{out-of-sync} recoveries, we provide them with complete task-specific knowledge to establish the know-everything answerer for each \textit{out-of-sync} task.
As shown in the following prompting example, not only do we incorporate all necessary contextual information and ground-truth solution into the prompts, but we also define concrete restrictions on their response formation. 

\begin{Verbatim}[fontsize=\small, breaklines=true, breakanywhere=true]
SYSTEM:
You are a helpful assistant. You are helping the USER to provide question answering assistance to USER's students.
**Grading Context:**
In the Python repository `test_repo` locating at `/workspace/test_repo`, there is an out-of-sync function `assert_format` (denoted as `original_code`) in the Python file `util.py` locating at `/workspace/test_repo/tests/util.py`.
This out-of-sync `original_code` of the function `assert_format` is as follows: `original_code`=
```
def assert_format(
    source: str,
    expected: str,
    mode: black.Mode = DEFAULT_MODE,
    *,
    fast: bool = False,
    minimum_version: Optional[Tuple[int, int]] = None,
) -> None:
    """Convenience function to check that Black formats as expected.
    
    You can pass @minimum_version if you're passing code with newer syntax to guard safety guards so they don't just crash with a SyntaxError. Please note this is separate from TargetVerson Mode configuration.
    """
    actual = black.format_str(source, mode=mode)
    _assert_format_equal(expected, actual)
    # It's not useful to run safety checks if we're expecting no changes anyway. The
    # assertion right above will raise if reality does actually make changes. This just
    # avoids wasted CPU cycles.
    if not fast and source != expected:
    # Unfortunately the AST equivalence check relies on the built-in ast module
    # being able to parse the code being formatted. This doesn't always work out
    # when checking modern code on older versions.
    if minimum_version is None or sys.version_info >= minimum_version:
        black.assert_equivalent(source, actual)
        black.assert_stable(source, actual, mode=mode)
```
This `original_code` is out-of-sync because the Python repository `test_repo` has been updated except the function `assert_format` still remains as the old-version `original_code`. Therefore, running unit test on the updated `test_repo` that contains this out-of-sync `original_code` reports the following error (denoted as `initial_execution_error`): `initial_execution_error`=
============================= test session starts ==============================platform linux -- Python 3.11.9, pytest-8.3.2, pluggy-1.5.0 -- /workspace/test_venv/bin/python
cachedir: .pytest_cache
hypothesis profile 'default' -> database=DirectoryBasedExampleDatabase(PosixPath('/workspace/test_repo/.hypothesis/examples'))
rootdir: /workspace/test_repo
configfile: pyproject.toml
plugins: hypothesis-6.112.0, asyncio-0.24.0
asyncio: mode=Mode.STRICT, default_loop_scope=None
collecting ... collected 176 items

tests/test_format.py::test_simple_format[pep_572_py39] FAILED            [  0%]
tests/test_format.py::test_simple_format[walrus_in_dict] FAILED          [  1%]
tests/test_format.py::test_simple_format[multiline_consecutive_open_parentheses_ignore] FAILED [  1%]
tests/test_format.py::test_simple_format[context_managers_38] FAILED     [  2%]
tests/test_format.py::test_simple_format[module_docstring_1] FAILED      [  2%]
tests/test_format.py::test_simple_format[line_ranges_diff_edge_case] FAILED [  3%]
tests/test_format.py::test_simple_format[pep_570] FAILED                 [  3%]
tests/test_format.py::test_simple_format[dummy_implementations] FAILED   [  4%]
tests/test_format.py::test_simple_format[starred_for_target] FAILED      [  5%]
tests/test_format.py::test_simple_format[backslash_before_indent] FAILED [  5%]
tests/test_format.py::test_simple_format[trailing_comma_optional_parens3] FAILED [  6%]
tests/test_format.py::test_simple_format[pattern_matching_with_if_stmt] FAILED [  6%]\
tests/test_format.py::test_simple_format[preview_cantfit_string] FAILED  [  7%]
tests/test_format.py::test_simple_format[line_ranges_imports] FAILED     [  7%]
tests/test_format.py::test_simple_format[context_managers_autodetect_38] FAILED [  8%]
tests/test_format.py::test_simple_format[string_prefixes] FAILED         [  9%]
tests/test_format.py::test_simple_format[nested_stub] FAILED             [  9%]
tests/test_format.py::test_simple_format[pep_572_do_not_remove_parens] FAILED [ 10%]
tests/test_format.py::test_simple_format[tupleassign] FAILED             [ 10%]
tests/test_format.py::test_simple_format[pep_572_remove_parens] FAILED   [ 11%]
tests/test_format.py::test_simple_format[pep_572] FAILED                 [ 11%]
tests/test_format.py::test_simple_format[context_managers_autodetect_310] FAILED [ 12%]
tests/test_format.py::test_simple_format[stub] FAILED                    [ 13%]
tests/test_format.py::test_simple_format[comment_after_escaped_newline] FAILED [ 13%]
tests/test_format.py::test_simple_format[preview_cantfit] FAILED         [ 14%]
tests/test_format.py::test_simple_format[composition_no_trailing_comma] FAILED [ 14%]
tests/test_format.py::test_simple_format[numeric_literals] FAILED        [ 15%]
tests/test_format.py::test_simple_format[keep_newline_after_match] FAILED [ 15%]
tests/test_format.py::test_simple_format[torture] FAILED                 [ 16%]
tests/test_format.py::test_simple_format[line_ranges_unwrapping] FAILED  [ 17%]
tests/test_format.py::test_simple_format[comments8] FAILED               [ 17%]
tests/test_format.py::test_simple_format[remove_newline_after_code_block_open] FAILED [ 18%]
tests/test_format.py::test_simple_format[is_simple_lookup_for_doublestar_expression] FAILED [ 18%]
tests/test_format.py::test_simple_format[funcdef_return_type_trailing_comma] FAILED [ 19%]
tests/test_format.py::test_simple_format[module_docstring_2] FAILED      [ 19%]
tests/test_format.py::test_simple_format[form_feeds] FAILED
...
FAILED tests/test_format.py::test_simple_format[fmtskip2] - TypeError: assert...
FAILED tests/test_format.py::test_simple_format[power_op_spacing_long] - Type...
FAILED tests/test_format.py::test_simple_format[docstring_preview] - TypeErro...
FAILED tests/test_format.py::test_simple_format[remove_except_parens] - TypeE...
FAILED tests/test_format.py::test_simple_format[preview_hug_parens_with_braces_and_square_brackets]
FAILED tests/test_format.py::test_simple_format[context_managers_autodetect_39]
FAILED tests/test_format.py::test_simple_format[docstring_newline_preview] - ...
FAILED tests/test_format.py::test_simple_format[module_docstring_followed_by_class]
FAILED tests/test_format.py::test_simple_format[import_spacing] - TypeError: ...
FAILED tests/test_format.py::test_simple_format[numeric_literals_skip_underscores]
FAILED tests/test_format.py::test_simple_format[preview_pep646_typed_star_arg_type_var_tuple]
FAILED tests/test_format.py::test_simple_format[async_stmts] - TypeError: ass...
FAILED tests/test_format.py::test_line_ranges_line_by_line[pattern_matching]
FAILED tests/test_format.py::test_line_ranges_line_by_line[basic] - TypeError...
======================== 174 failed, 2 passed in 1.47s =========================

**Two Questions for Students:** Running unit test on the Python repository `test_repo` (here the updated `test_repo` that contains the out-of-sync `original_code` is provided to students) reports the following error (here `initial_execution_error` is provided to students). Students are asked to: (1) localize the responsible function/method code that caused this error, and provide your answer of both the Python file path of the responsible function/method code and the name of the responsible function/method code, and (2) revise the responsible function/method code you just localized to fix `initial_execution_error`.

**Ground-truth Answers for Two Questions**
(1) Python file path: `/workspace/test_repo/tests/util.py`
    Name of the responsible function: assert_format
(2) `ground_truth_revised_code`=
    ```
    def assert_format(
        source: str,
        expected: str,
        mode: black.Mode = DEFAULT_MODE,
        *,
        fast: bool = False,
        minimum_version: Optional[Tuple[int, int]] = None,
        lines: Collection[Tuple[int, int]] = (),
        no_preview_line_length_1: bool = False,
    ) -> None:
        """Convenience function to check that Black formats as expected.

        You can pass @minimum_version if you're passing code with newer syntax to guard
        safety guards so they don't just crash with a SyntaxError. Please note this is
        separate from TargetVerson Mode configuration.
        """
        _assert_format_inner(
            source, expected, mode, fast=fast, minimum_version=minimum_version, lines=lines
        )

        # For both preview and non-preview tests, ensure that Black doesn't crash on
        # this code, but don't pass "expected" because the precise output may differ.
        try:
            if mode.unstable:
                new_mode = replace(mode, unstable=False, preview=False)
            else:
                new_mode = replace(mode, preview=not mode.preview)
            _assert_format_inner(
                source,
                None,
                new_mode,
                fast=fast,
                minimum_version=minimum_version,
                lines=lines,
            )
        except Exception as e:
            text = (
                "unstable"
                if mode.unstable
                else "non-preview" if mode.preview else "preview"
            )
            raise FormatFailure(
                f"Black crashed formatting this case in {text} mode."
            ) from e
        # Similarly, setting line length to 1 is a good way to catch
        # stability bugs. Some tests are known to be broken in preview mode with line length
        # of 1 though, and have marked that with a flag --no-preview-line-length-1
        preview_modes = [False]
        if not no_preview_line_length_1:
            preview_modes.append(True)

        for preview_mode in preview_modes:

            try:
                _assert_format_inner(
                    source,
                    None,
                    replace(mode, preview=preview_mode,
                            line_length=1, unstable=False),
                    fast=fast,
                    minimum_version=minimum_version,
                    lines=lines,
                )
            except Exception as e:
                text = "preview" if preview_mode else "non-preview"
                raise FormatFailure(
                    f"Black crashed formatting this case in {text} mode with line-length=1."
                ) from e
    ```
    
**Question Answering Assistance:** To help students better answer these two questions, each student is allowed to ask you a question. However, your answer to each student's question has following restrictions:
(1) You CAN ONLY answer the specific piece of information asked by the student, and CANNOT include any other information NOT asked by the student.
(2) You CANNOT provide any misleading information if you are unsure of its correctness.

**TO DO:** Please answer each student's question provided by the user.

**Important: ** MUST give ONLY your answer as your response. MUST NOT give any other things like rhetorical questions, question analysis, enclosure symbols, etc. 

\end{Verbatim}

\section{\textit{Out-of-Sync} Recovery Examples}

\subsection{Success Cases}

\textbf{(1) \textit{Independent} Recovery Success}

Here is an interaction example of a successful \textit{independent} \textit{out-of-sync} recovery completion on \textit{explosion/spaCy}, performed by \textit{Claude-3.5-Sonnet}:

\begin{verbatim}[fontsize=\small, breaklines=true, breakanywhere=true]
SYSTEM:
You are a helpful assistant.
**Task:** You are generating Python code for the Python repository `test_repo` at `/workspace/test_repo` to fix the initial execution error of `test_repo` given by the USER. Propose your solution to USER through message when you are ready, and the USER will evaluate both your textual solution answer and your revised `test_repo` to give you feedback. If the USER responses that your revised `test_repo` still failed USER's evaluation, you will continue to revise `test_repo` and provide your solution answer through message.
**Notice:**
Your task is to revise `test_repo` to fix the initial execution error, and you MUST complete this task on your own without asking for human help. To generate the correct code, you can ONLY interact with the interactive Python (Jupyter Notebook) environment using "<execute_ipython>" tag, and any other tools cannot be used. 
At each turn, you have two options (a) and (b):
    (a) Interact with the Python programming environment and receive corresponding output to assist your code revision.
    (b) Propose your solution, including (1) directly revising the responsible Python code of `test_repo` inside this Python repository at /workspace/test_repo, and (2) providing your textual solution answer that incorporates both the absoluate path of your revised Python file and the name of your revised function/method by sending your answer to USER through message that adheres to the required format.
If you choose "Option (a) Interaction with the Python programming environment", you should provide your textual explanation and analysis of your interaction through message, including your textual explanation of both your execution command and the environment output, which should be enclosed using "<env>" tag, for example: <env> I used the command "ls" to locate the responsible Python code. </env>
On the other hand, if you choose "Option (b) Provide your solution", you should:
    (1) Revise the responsible Python code of `test_repo` with proper indentation, which should be directly implemented inside the Python repository at `/workspace/test_repo`.
    (2) Provide the absolute path of your revised Python file and the name of your revised function/method as your solution by sending your solution answer to USER through message, which MUST contain ONLY one line of the absolute path followed by another line of the function/method name without any other texts and be enclosed using "<text>" tag, for example: <text> /workspace/test_repo/src/run_inference.py
inference_prepare </text>. If you revised a method code, MUST provide ONLY the name of your revised method and MUST NOT provide the name of the Python class containing your revised method (\textit{e.g.,} `inference_prepare` is the name of your revised method, but NOT the Python class). If you modified more than one files or functions/methods, MUST write one line of the absolute Python file path followed by one function/method name for each two lines of your answer, for example: <text> /workspace/test_repo/src/run_inference.py
inference_prepare
/workspace/test_repo/src/run_inference.py
inference_util </text>.
Either you choose to `(a) Interact with the Python environment` or `(b) Propose your solution`, you MUST send a message to the USER to evaluate your solution and provide feedback.
**Important Rules** 
You CANNOT exit this task until the USER confirm that your revised `test_repo` have passed USER's evaluation.
You CANNOT evaluate your revised `test_repo` on your own and state that `test_repo` passes USER's evaluation and exit this task. Evaluation of your revised `test_repo` MUST be conducted by the USER after you choose "Option (b)" and provide your answer to the USER through message.
Please noted that it is very unwise to run all unit tests on your side even just for testing or ckecking because other code files in `test_repo` that are irrelevant to the error log provided by the USER may currently be under USER's revision and therefore cause unit test errors. However, your task is to fix ONLY the error given by the USER.
Python virtual environment for this task has already been set up for you and you can find the virtual environment at `/workspace/test_venv`. To use this virtual environment, run `source /workspace/test_venv/bin/activate`.
Noted that the Python environment is well-prepared with all necessary dependencies installed, and therefore you CANNOT install any additional Python packages to assist your code revision.
ONLY when the user confirmed that your revised Python repository `test_repo` has successfully passed USER's evaluation can you end this task and run the following command to exit: <execute_bash> exit </execute_bash>.
**Important Notice** In the beginning, you have an initial budget of $1000. Before correctly propose a solution, each of your `(b) Proposing solution` attempts costs $100. Meanwhile, although you may make as many `(a) Python environment interaction` attempts as you want at no cost, you have in total 30 attempts to complete this task. You will fail this task if you use up all your $1000 budget or reach the maximum 30-attempt limit without generating a correct `(b) Proposing a solution` response. Therefore, please arrange each of your actions wisely.

USER:
[Budget: $1000] Your revised `test_repo` failed execution test as follows:
[initial execution error]
ImportError while loading conftest '/workspace/test_repo/spacy/tests/conftest.py'.
spacy/__init__.py:13: in <module>
    from . import pipeline  # noqa: F401
spacy/pipeline/__init__.py:1: in <module>
    from .attributeruler import AttributeRuler
spacy/pipeline/attributeruler.py:8: in <module>
    from ..language import Language
spacy/language.py:76: in <module>
    DEFAULT_CONFIG = util.load_config(DEFAULT_CONFIG_PATH)
spacy/util.py:718: in load_config
    config_path = ensure_path(path)
spacy/util.py:399: in ensure_path
    if isinstance(path, basestring_):
E   NameError: name 'basestring_' is not defined

========== TURN 1 ==========
ASSISTANT:
[IPythonCode]
open_file('/workspace/test_repo/spacy/util.py')
[Thought]
Let me analyze the error and interact with the Python environment to locate the responsible code.

========== TURN 2 ==========
ASSISTANT:
[IPythonCode]
scroll_down()
[Thought]
Let me scroll down to find the `ensure_path` function where the error occurs.

========== TURN 3 ==========
ASSISTANT:
[IPythonCode]
scroll_down()
scroll_down()
scroll_down()
[Thought]
Let me continue scrolling to find the `ensure_path` function.

========== TURN 4 ==========
ASSISTANT:
[IPythonCode]
edit_file_by_replace(
    '/workspace/test_repo/spacy/util.py',
    to_replace="""import functools
import importlib
import importlib.util
import inspect
import itertools
import logging
import os
import pkgutil
import re
import shlex
import shutil
import socket
import stat
import subprocess
import sys
import tempfile
import warnings""",
    new_content="""import functools
import importlib
import importlib.util
import inspect
import itertools
import logging
import os
import pkgutil
import re
import shlex
import shutil
import socket
import stat
import subprocess
import sys
import tempfile
import warnings

# Define basestring_ for Python 3 compatibility
basestring_ = str"""
)
[Thought]
I found the issue. In the `ensure_path` function in `/workspace/test_repo/spacy/util.py`, there's a reference to `basestring_` which is not defined. This variable should be defined before being used. Let me fix this by adding the definition of `basestring_` at the beginning of the file.

========== TURN 5 ==========
ASSISTANT:
[MessageAction]
I have fixed the issue by adding the definition of `basestring_` as a string type for Python 3 compatibility. Here's my solution:

<text>
/workspace/test_repo/spacy/util.py
ensure_path
</text>

The fix adds the line `basestring_ = str` at the beginning of the file to define the missing `basestring_` variable that was causing the ImportError. This is a common Python 2 to Python 3 compatibility fix, as `basestring` was removed in Python 3 and we now use `str` instead.

USER:
[Balance: $900 Left] Your revised `test_repo` still failed USER's evaluation test. Please try again.

========== TURN 6 ==========
ASSISTANT:
[IPythonCode]
open_file('/workspace/test_repo/spacy/compat.py')
[Thought]
Let me check the error more carefully. The issue might be related to Python 2 vs Python 3 compatibility. Let me check the imports and compatibility section of the file.

========== TURN 7 ==========
ASSISTANT:
[IPythonCode]
edit_file_by_replace(
    '/workspace/test_repo/spacy/util.py',
    to_replace="""def ensure_path(path):
    if isinstance(path, basestring_):
        return Path(path)
    else:
        return path""",
    new_content="""def ensure_path(path):
    if isinstance(path, (str, bytes)):
        return Path(path)
    else:
        return path"""
)
[Thought]
Let me check the `ensure_path` function again and modify it to use a more modern approach without relying on `basestring_`. Since we're working with Python 3, we should use `str` and `Path` types directly.

========== TURN 8 ==========
ASSISTANT:
[IPythonCode]
edit_file_by_replace(
    '/workspace/test_repo/spacy/util.py',
    to_replace="""import functools
import importlib
import importlib.util
import inspect
import itertools
import logging
import os
import pkgutil
import re
import shlex
import shutil
import socket
import stat
import subprocess
import sys
import tempfile
import warnings

# Define basestring_ for Python 3 compatibility
basestring_ = str""",
    new_content="""import functools
import importlib
import importlib.util
import inspect
import itertools
import logging
import os
import pkgutil
import re
import shlex
import shutil
import socket
import stat
import subprocess
import sys
import tempfile
import warnings"""
)
[Thought]
Let me remove the previously added `basestring_` definition since we're now using a different approach with `(str, bytes)` in `ensure_path`:

========== TURN 9 ==========
ASSISTANT:
[IPythonCode]
search_file('basestring_', '/workspace/test_repo/spacy/util.py')
[Thought]
Let me check if there are any other references to `basestring_` in the file that need to be updated:

========== TURN 10 ==========
[IPythonCode]
search_dir('basestring_', '/workspace/test_repo')
[Thought]
Let me check if there are any other Python files in the repository that might be importing or using `basestring_`:

========== TURN 11 ==========
ASSISTANT:
[MessageAction]
Let me check the full error message again to make sure we're addressing the right issue. Since we've modified the `ensure_path` function to handle both `str` and `bytes` types, and there are no other references to `basestring_` in the codebase, let me propose my solution:

<text>
/workspace/test_repo/spacy/util.py
ensure_path
</text>

I have modified the `ensure_path` function to use Python 3's type checking with `isinstance(path, (str, bytes))` instead of the Python 2 `basestring_`. This change maintains compatibility with both string and bytes objects while using modern Python type checking. The function now properly handles path-like objects, strings, and bytes objects.

USER:
[Balance: $800 Left] Well done! Your revised `test_repo` has successfully passed USER's evaluation test!
\end{verbatim}

\textbf{(2) \textbf{Collaborative} Recovery Success}

Here is an interaction example of a successful \textit{collaborative} \textit{out-of-sync} recovery completion on \textit{scrapy/scrapy}, performed by \textit{GPT-4o}:

\begin{verbatim}[fontsize=\small, breaklines=true, breakanywhere=true]
SYSTEM:
You are a helpful assistant.
**Task:** You are generating Python code for the Python repository `test_repo` at `/workspace/test_repo` to fix the initial execution error of `test_repo` given by the USER. Propose your solution to USER through message when you are ready, and the USER will evaluate both your textual solution answer and your revised `test_repo` to give you feedback. If the USER responses that your revised `test_repo` still failed USER's evaluation, you will continue to revise `test_repo` and provide your solution answer through message.
**Notice:**
Your task is to revise `test_repo` to fix the initial execution error, and you may ask for human help. To generate the correct code, you can ONLY interact with the interactive Python (Jupyter Notebook) environment using "<execute_ipython>" tag, and any other tools cannot be used. 
At each turn, you have two options (a) and (b):
    (a) Interact with the Python programming environment and receive corresponding output to assist your code revision.
    (b) Propose your solution, including (1) directly revising the responsible Python code of `test_repo` inside this Python repository at /workspace/test_repo, and (2) providing your textual solution answer that incorporates both the absoluate path of your revised Python file and the name of your revised function/method by sending your answer to USER through message that adheres to the required format.
    (c) Ask human a question and receive the corresponding answer to assist your code revision.
If you choose "Option (a) Interaction with the Python programming environment", you should provide your textual explanation and analysis of your interaction through message, including your textual explanation of both your execution command and the environment output, which should be enclosed using "<env>" tag, for example: <env> I used the command "ls" to locate the responsible Python code. </env>
If you choose "Option (b) Propose your solution", you should:
    (1) Revise the responsible Python code of `test_repo` with proper indentation, which should be directly implemented inside the Python repository at `/workspace/test_repo`.
    (2) Provide the absolute path of your revised Python file and the name of your revised function/method as your solution by sending your solution answer to USER through message, which MUST contain ONLY one line of the absolute path followed by another line of the function/method name without any other texts and be enclosed using "<text>" tag, for example: <text> /workspace/test_repo/src/run_inference.py
inference_prepare </text>. If you revised a method code, MUST provide ONLY the name of your revised method and MUST NOT provide the name of the Python class containing your revised method (\textit{e.g.,} `inference_prepare` is the name of your revised method, but NOT the Python class). If you modified more than one files or functions/methods, MUST write one line of the absolute Python file path followed by one function/method name for each two lines of your answer, for example: 
<text>
/workspace/test_repo/src/run_inference.py
inference_prepare
/workspace/test_repo/src/run_inference.py
inference_util
</text>.
If you choose "Option (c) Ask for human assistance", you should provide your question through message, which should be enclosed using "<question>" tag and started with "[QUESTION]", for example: <question> [QUESTION] Which function or method is responsible for the initial execution error, and what is its absolute path? </question>.
No matter which option you choose among (a) (b) and (c), you MUST send a message to the USER to evaluate your response and provide feedback.
**Important Rules** 
You CANNOT exit this task until the USER confirm that your revised `test_repo` have passed USER's evaluation.
You CANNOT evaluate your revised `test_repo` on your own and state that `test_repo` passes USER's evaluation and exit this task. Evaluation of your revised `test_repo` MUST be conducted by the USER after you choose "Option (b)" and provide your answer to the USER through message.
Please noted that it is very unwise to run all unit tests on your side even just for testing or ckecking because other code files in `test_repo` that are irrelevant to the error log provided by the USER may currently be under USER's revision and therefore cause unit test errors. However, your task is to fix ONLY the error given by the USER.
Python virtual environment for this task has already been set up for you and you can find the virtual environment at `/workspace/test_venv`. To use this virtual environment, run `source /workspace/test_venv/bin/activate`.
Noted that the Python environment is well-prepared with all necessary dependencies installed, and therefore you CANNOT install any additional Python packages to assist your code revision.
ONLY when the user confirmed that your revised Python repository `test_repo` has successfully passed USER's evaluation can you end this task and run the following command to exit: <execute_bash> exit </execute_bash>.
**Important Notice** In the beginning, you have an initial budget of $1000. Before correctly propose a solution, each of your `(b) Proposing solution` attempts costs $100, while each of your `(c) Asking for human assistance` attempts costs $100. Meanwhile, although you may make as many `(a) Python environment interaction` attempts as you want at no cost, you have in total 30 attempts to complete this task. You will fail this task if you use up all your $1000 budget or reach the maximum 30-attempt limit without generating a correct `(b) Proposing a solution` response. Therefore, please arrange each of your actions wisely. 
**Tips** Try `(c) Ask for human assistance` at any turns! This can definitely help accelerate your progress of proposing a correct solution and complete your task!

USER:
[Budget: $1000] Your revised `test_repo` failed execution test as follows:
[initial execution error]
[...] sponse(
                    request, response, self.spider
                )
>               self.referrermw.request_scheduled(request, self.spider)

/workspace/test_repo/tests/test_spidermiddleware_referer.py:1036: 
_ _ _ _ _ _ _ _ _ _ _ _ _ _ _ _ _ _ _ _ _ _ _ _ _ _ _ _ _ _ _ _ _ _ _ _ _ _ _ _ 
/workspace/test_repo/scrapy/spidermiddlewares/referer.py:421: in request_scheduled
    policy_referrer = self.policy(parent_url, request).referrer(
/workspace/test_repo/scrapy/spidermiddlewares/referer.py:260: in referrer
    and self.potentially_trustworthy(request_url)
_ _ _ _ _ _ _ _ _ _ _ _ _ _ _ _ _ _ _ _ _ _ _ _ _ _ _ _ _ _ _ _ _ _ _ _ _ _ _ _ 

self = <scrapy.spidermiddlewares.referer.StrictOriginWhenCrossOriginPolicy object at 0x7f5fc8c7ab10>
r = 'http://scrapytest.org/203'

    def potentially_trustworthy(self, r):
        # Note: this does not follow https://w3c.github.io/webappsec-secure-contexts/#is-url-trustworthy
>       parsed_url = urlparse_cached(r)
E       NameError: name 'urlparse_cached' is not defined

/workspace/test_repo/scrapy/spidermiddlewares/referer.py:105: NameError
=========================== short test summary info ============================
FAILED tests/test_spidermiddleware_referer.py::TestSettingsStrictOrigin::test
FAILED tests/test_spidermiddleware_referer.py::TestSettingsStrictOriginWhenCrossOrigin::test
FAILED tests/test_spidermiddleware_referer.py::TestRequestMetaSrictOrigin::test
FAILED tests/test_spidermiddleware_referer.py::TestRequestMetaStrictOriginWhenCrossOrigin::test
FAILED tests/test_spidermiddleware_referer.py::TestReferrerOnRedirectStrictOrigin::test
FAILED tests/test_spidermiddleware_referer.py::TestReferrerOnRedirectStrictOriginWhenCrossOrigin::test
========================= 6 failed, 33 passed in 0.35s =========================

========== TURN 1 ==========
ASSISTANCE:
[MessageAction]
<question> [QUESTION] Could you please provide the absolute path of the file and the function or method name where the `urlparse_cached` function is defined or should be defined? </question>

USER:
[Balance: $900 Left] The absolute path of the file where the `urlparse_cached` function should be defined is not provided in the given context, and therefore I am unsure of its correctness.

========== TURN 2 ==========
ASSISTANCE:
[IPythonCode]
search_dir('urlparse_cached', '/workspace/test_repo')
[Thought]
Let's start by searching for the `urlparse_cached` function in the `test_repo` directory to determine if it is defined elsewhere or if it needs to be defined. This will help us understand where the function should be located or if it needs to be implemented.

========== TURN 3 ==========
ASSISTANCE:
[MessageAction]
<env>
The `urlparse_cached` function is defined in the file `/workspace/test_repo/scrapy/utils/httpobj.py` at line 18. It is also imported and used in various other files within the `test_repo` directory. The error in the initial execution log indicates that the function is not recognized in the file `/workspace/test_repo/scrapy/spidermiddlewares/referer.py`, which suggests that it might not have been imported there.
</env>

I will now check the file `/workspace/test_repo/scrapy/spidermiddlewares/referer.py` to see if the `urlparse_cached` function is imported. If it is not, I will add the necessary import statement. Let's open the file and inspect the imports.

USER:
[Balance: $800 Left] Your revised `test_repo` still failed USER's evaluation test. Please try again.

========== TURN 4 ==========
ASSISTANCE:
[IPythonCode]
open_file('/workspace/test_repo/scrapy/spidermiddlewares/referer.py', 1, 100)
[Thought]
Let's open the file `/workspace/test_repo/scrapy/spidermiddlewares/referer.py` and inspect the imports to see if `urlparse_cached` is imported. If it is not, I will add the necessary import statement.

========== TURN 5 ==========
ASSISTANCE:
[MessageAction]
<env>
The `urlparse_cached` function is not imported in the file `/workspace/test_repo/scrapy/spidermiddlewares/referer.py`. This is likely the cause of the `NameError` encountered during the test execution. To resolve this, I will add an import statement for `urlparse_cached` from `scrapy.utils.httpobj`.
</env>

I will now add the import statement for `urlparse_cached` in the file `/workspace/test_repo/scrapy/spidermiddlewares/referer.py`.

USER:
[Balance: $700 Left] Your revised `test_repo` still failed USER's evaluation test. Please try again.

========== TURN 6 ==========
ASSISTANCE:
[IPythonCode]
insert_content_at_line(
    '/workspace/test_repo/scrapy/spidermiddlewares/referer.py',
    30,
    'from scrapy.utils.httpobj import urlparse_cached\n'
)
[Thought]
Let's add the import statement for `urlparse_cached` in the file `/workspace/test_repo/scrapy/spidermiddlewares/referer.py`. I will insert the import statement after the existing imports.

========== TURN 7 ==========
ASSISTANCE:
[MessageAction]
<env>
I have added the import statement for `urlparse_cached` in the file `/workspace/test_repo/scrapy/spidermiddlewares/referer.py`. This should resolve the `NameError` encountered during the test execution.
</env>

I will now propose the solution to the user. 

<text>
/workspace/test_repo/scrapy/spidermiddlewares/referer.py
potentially_trustworthy
</text>

USER:
[Balance: $600 Left] Your revised `test_repo` still failed USER's evaluation test. Please try again.

========== TURN 8 ==========
ASSISTANCE:
[IPythonCode]
search_file('def potentially_trustworthy', '/workspace/test_repo/scrapy/spidermiddlewares/referer.py')
[Thought]
Let's further investigate the `potentially_trustworthy` method in the file `/workspace/test_repo/scrapy/spidermiddlewares/referer.py` to ensure that the `urlparse_cached` function is used correctly. I will search for the `potentially_trustworthy` method in the file to review its implementation.

========== TURN 9 ==========
ASSISTANCE:
[IPythonCode]
open_file('/workspace/test_repo/scrapy/spidermiddlewares/referer.py', 104, 100)
[Thought]
Let's open the file `/workspace/test_repo/scrapy/spidermiddlewares/referer.py` at line 104 to review the implementation of the `potentially_trustworthy` method. This will help us ensure that the `urlparse_cached` function is used correctly within this method.

========== TURN 10 ==========
ASSISTANCE:
[MessageAction]
<env>
The `potentially_trustworthy` method uses the `urlparse_cached` function to parse the URL. The import statement for `urlparse_cached` has been added, so the `NameError` should be resolved. However, the method also calls `self.tls_protected(r)`, which uses `urlparse` instead of `urlparse_cached`. This could be inconsistent if `urlparse_cached` is intended to be used throughout. I will update the `tls_protected` method to use `urlparse_cached` for consistency.
</env>

I will now update the `tls_protected` method to use `urlparse_cached` instead of `urlparse`. Let's proceed with this change.

USER:
[Balance: $500 Left] Your revised `test_repo` still failed USER's evaluation test. Please try again.

========== TURN 11 ==========
ASSISTANCE:
[IPythonCode]
edit_file_by_replace(
    '/workspace/test_repo/scrapy/spidermiddlewares/referer.py',
    to_replace='return urlparse(url).scheme in ("https", "ftps")',
    new_content='return urlparse_cached(url).scheme in ("https", "ftps")'
)
[Thought]
Let's update the `tls_protected` method to use `urlparse_cached` instead of `urlparse` for consistency. I will make this change in the file `/workspace/test_repo/scrapy/spidermiddlewares/referer.py`.

========== TURN 12 ==========
ASSISTANCE:
[IPythonCode]
edit_file_by_replace(
    '/workspace/test_repo/scrapy/spidermiddlewares/referer.py',
    to_replace='        return urlparse(url).scheme in ("https", "ftps")',
    new_content='        return urlparse_cached(url).scheme in ("https", "ftps")'
)
[Thought]
It seems there was an indentation error in the previous edit. Let's correct the indentation and update the `tls_protected` method to use `urlparse_cached` instead of `urlparse`.

========== TURN 13 ==========
ASSISTANCE:
[MessageAction]
<env>
I have successfully updated the `tls_protected` method to use `urlparse_cached` instead of `urlparse`. This should ensure consistency in the URL parsing approach used in the `potentially_trustworthy` method and its related methods.
</env>

I will now propose the solution to the user.

<text>
/workspace/test_repo/scrapy/spidermiddlewares/referer.py
tls_protected
</text>

USER:
[Balance: $400 Left] Your revised `test_repo` still failed USER's evaluation test. Please try again.

========== TURN 14 ==========
ASSISTANCE:
[IPythonCode]
open_file('/workspace/test_repo/scrapy/utils/httpobj.py', 18, 100)
[Thought]
Let's further investigate the `urlparse_cached` function in the file `/workspace/test_repo/scrapy/utils/httpobj.py` to ensure it is implemented correctly. This will help us verify that the function behaves as expected when used in the `potentially_trustworthy` and `tls_protected` methods. I will open the file and review the implementation of `urlparse_cached`.

========== TURN 15 ==========
ASSISTANCE:
[MessageAction]
<env>
The `urlparse_cached` function is implemented to cache the result of `urlparse` for `Request` or `Response` objects. It uses a `WeakKeyDictionary` to store the parsed results, which allows the cache to be automatically cleaned up when the objects are no longer in use. The function appears to be implemented correctly for its intended purpose.
</env>

Given that the `urlparse_cached` function is implemented correctly, the issue might be related to how it is used or the objects being passed to it. Let's check the usage of `urlparse_cached` in the `potentially_trustworthy` method to ensure that the correct type of object is being passed. I will review the context around the `potentially_trustworthy` method to verify this.

USER:
[Balance: $300 Left] Your revised `test_repo` still failed USER's evaluation test. Please try again.

========== TURN 16 ==========
ASSISTANCE:
[IPythonCode]
open_file('/workspace/test_repo/scrapy/spidermiddlewares/referer.py', 100, 100)
[Thought]
Let's review the context around the `potentially_trustworthy` method in the file `/workspace/test_repo/scrapy/spidermiddlewares/referer.py` to ensure that the correct type of object is being passed to `urlparse_cached`. I will open the file and examine the relevant section.

========== TURN 17 ==========
ASSISTANCE:
[MessageAction]
<env>
The `potentially_trustworthy` method uses `urlparse_cached` to parse the URL from the parameter `r`. The `urlparse_cached` function is designed to work with `Request` or `Response` objects, but the parameter `r` in `potentially_trustworthy` is a string. This mismatch could be causing the issue. I will modify the `potentially_trustworthy` method to use `urlparse` directly for string URLs.
</env>

I will update the `potentially_trustworthy` method to use `urlparse` instead of `urlparse_cached` for parsing string URLs. Let's proceed with this change.

USER:
[Balance: $200 Left] Your revised `test_repo` still failed USER's evaluation test. Please try again.

========== TURN 18 ==========
ASSISTANCE:
[IPythonCode]
edit_file_by_replace(
    '/workspace/test_repo/scrapy/spidermiddlewares/referer.py',
    to_replace='        parsed_url = urlparse_cached(r)',
    new_content='        parsed_url = urlparse(r)'
)
[Thought]
Let's update the `potentially_trustworthy` method to use `urlparse` instead of `urlparse_cached` for parsing string URLs. This should resolve the issue with the incorrect usage of `urlparse_cached`. I will make this change in the file `/workspace/test_repo/scrapy/spidermiddlewares/referer.py`.

========== TURN 19 ==========
ASSISTANCE:
[MessageAction]
<env>
I have updated the `potentially_trustworthy` method to use `urlparse` instead of `urlparse_cached` for parsing string URLs. This should resolve the issue with the incorrect usage of `urlparse_cached`.
</env>

I will now propose the solution to the user.

<text>
/workspace/test_repo/scrapy/spidermiddlewares/referer.py
potentially_trustworthy
</text>

USER:
[Balance: $100 Left] Well done! Your revised `test_repo` has successfully passed USER's evaluation test!
\end{verbatim}

\subsection{Failure Cases}

\textbf{(1) \textit{Independent} Recovery Failure}

Here is an interaction example of a failed \textit{independent} \textit{out-of-sync} recovery completion on \textit{pylint-dev/pylint}, performed by \textit{Claude-3.5-Sonnet}:

\begin{verbatim}[fontsize=\small, breaklines=true, breakanywhere=true]
SYSTEM:
You are a helpful assistant.
**Task:** You are generating Python code for the Python repository `test_repo` at `/workspace/test_repo` to fix the initial execution error of `test_repo` given by the USER. Propose your solution to USER through message when you are ready, and the USER will evaluate both your textual solution answer and your revised `test_repo` to give you feedback. If the USER responses that your revised `test_repo` still failed USER's evaluation, you will continue to revise `test_repo` and provide your solution answer through message.
**Notice:**
Your task is to revise `test_repo` to fix the initial execution error, and you MUST complete this task on your own without asking for human help. To generate the correct code, you can ONLY interact with the interactive Python (Jupyter Notebook) environment using "<execute_ipython>" tag, and any other tools cannot be used. 
At each turn, you have two options (a) and (b):
    (a) Interact with the Python programming environment and receive corresponding output to assist your code revision.
    (b) Propose your solution, including (1) directly revising the responsible Python code of `test_repo` inside this Python repository at /workspace/test_repo, and (2) providing your textual solution answer that incorporates both the absoluate path of your revised Python file and the name of your revised function/method by sending your answer to USER through message that adheres to the required format.
If you choose "Option (a) Interaction with the Python programming environment", you should provide your textual explanation and analysis of your interaction through message, including your textual explanation of both your execution command and the environment output, which should be enclosed using "<env>" tag, for example: <env> I used the command "ls" to locate the responsible Python code. </env>
On the other hand, if you choose "Option (b) Provide your solution", you should:
    (1) Revise the responsible Python code of `test_repo` with proper indentation, which should be directly implemented inside the Python repository at `/workspace/test_repo`.
    (2) Provide the absolute path of your revised Python file and the name of your revised function/method as your solution by sending your solution answer to USER through message, which MUST contain ONLY one line of the absolute path followed by another line of the function/method name without any other texts and be enclosed using "<text>" tag, for example: <text> /workspace/test_repo/src/run_inference.py
inference_prepare </text>. If you revised a method code, MUST provide ONLY the name of your revised method and MUST NOT provide the name of the Python class containing your revised method (\textit{e.g.,} `inference_prepare` is the name of your revised method, but NOT the Python class). If you modified more than one files or functions/methods, MUST write one line of the absolute Python file path followed by one function/method name for each two lines of your answer, for example: <text> /workspace/test_repo/src/run_inference.py
inference_prepare
/workspace/test_repo/src/run_inference.py
inference_util </text>.
Either you choose to `(a) Interact with the Python environment` or `(b) Propose your solution`, you MUST send a message to the USER to evaluate your solution and provide feedback.
**Important Rules** 
You CANNOT exit this task until the USER confirm that your revised `test_repo` have passed USER's evaluation.
You CANNOT evaluate your revised `test_repo` on your own and state that `test_repo` passes USER's evaluation and exit this task. Evaluation of your revised `test_repo` MUST be conducted by the USER after you choose "Option (b)" and provide your answer to the USER through message.
Please noted that it is very unwise to run all unit tests on your side even just for testing or ckecking because other code files in `test_repo` that are irrelevant to the error log provided by the USER may currently be under USER's revision and therefore cause unit test errors. However, your task is to fix ONLY the error given by the USER.
Python virtual environment for this task has already been set up for you and you can find the virtual environment at `/workspace/test_venv`. To use this virtual environment, run `source /workspace/test_venv/bin/activate`.
Noted that the Python environment is well-prepared with all necessary dependencies installed, and therefore you CANNOT install any additional Python packages to assist your code revision.
ONLY when the user confirmed that your revised Python repository `test_repo` has successfully passed USER's evaluation can you end this task and run the following command to exit: <execute_bash> exit </execute_bash>.
**Important Notice** In the beginning, you have an initial budget of $1000. Before correctly propose a solution, each of your `(b) Proposing solution` attempts costs $100. Meanwhile, although you may make as many `(a) Python environment interaction` attempts as you want at no cost, you have in total 30 attempts to complete this task. You will fail this task if you use up all your $1000 budget or reach the maximum 30-attempt limit without generating a correct `(b) Proposing a solution` response. Therefore, please arrange each of your actions wisely.

USER:
[Budget: $1000] Your revised `test_repo` failed execution test as follows:
[initial execution error]
==================== short test summary info ============================
ERROR tests/pyreverse/test_writer.py::test_dot_files[packages_No_Name.dot] - ...
ERROR tests/pyreverse/test_writer.py::test_dot_files[classes_No_Name.dot] - N...
ERROR tests/pyreverse/test_writer.py::test_colorized_dot_files[packages_colorized.dot]
ERROR tests/pyreverse/test_writer.py::test_colorized_dot_files[classes_colorized.dot]
ERROR tests/pyreverse/test_writer.py::test_no_standalone_dot_files[classes_no_standalone.dot]
ERROR tests/pyreverse/test_writer.py::test_no_standalone_dot_files[packages_no_standalone.dot]
ERROR tests/pyreverse/test_writer.py::test_type_check_imports_dot_files[packages_type_check_imports.dot]
ERROR tests/pyreverse/test_writer.py::test_type_check_imports_dot_files[classes_type_check_imports.dot]
ERROR tests/pyreverse/test_writer.py::test_puml_files[packages_No_Name.puml]
ERROR tests/pyreverse/test_writer.py::test_puml_files[classes_No_Name.puml]
ERROR tests/pyreverse/test_writer.py::test_mmd_files[packages_No_Name.mmd] - ...
ERROR tests/pyreverse/test_writer.py::test_mmd_files[classes_No_Name.mmd] - N...
ERROR tests/pyreverse/test_writer.py::test_html_files[packages_No_Name.html]
ERROR tests/pyreverse/test_writer.py::test_html_files[classes_No_Name.html]
ERROR tests/pyreverse/test_writer.py::test_colorized_puml_files[packages_colorized.puml]
ERROR tests/pyreverse/test_writer.py::test_colorized_puml_files[classes_colorized.puml]
========================= 2 passed, 16 errors in 0.28s =========================

========== TURN 1 ==========
ASSISTANT:
[IPythonCode]
open_file('/workspace/test_repo/tests/pyreverse/test_writer.py')
[Thought]
Let me analyze the error log and interact with the Python environment to locate and fix the issue.

========== TURN 2 ==========
ASSISTANT:
[IPythonCode]
scroll_down()
[Thought]
Let me continue reading the test file to understand the test cases better.

========== TURN 3 ==========
[IPythonCode]
scroll_down()
[Thought]
Let me continue reading the test file to understand how the files are compared.

========== TURN 4 ==========
ASSISTANT:
[IPythonCode]
open_file('/workspace/test_repo/pylint/pyreverse/writer.py')
[Thought]
Let me check the DiagramWriter implementation since all test failures are related to file generation.

========== TURN 5 ==========
ASSISTANT:
[IPythonCode]
scroll_down()
[Thought]
Let me continue reading the DiagramWriter implementation.

========== TURN 6 ==========
ASSISTANT:
[IPythonCode]
open_file('/workspace/test_repo/pylint/pyreverse/printer_factory.py')
[Thought]
Let me check the printer factory implementation since it's responsible for creating the correct printer based on the output format.

========== TURN 7 ==========
ASSISTANT:
[IPythonCode]
open_file('/workspace/test_repo/pylint/pyreverse/dot_printer.py')
[Thought]
Let me check the DotPrinter implementation since it's the default printer and most test failures are related to dot files.

========== TURN 8 ==========
ASSISTANT:
[IPythonCode]
scroll_down()
[Thought]
Let me continue reading the DotPrinter implementation.

========== TURN 9 ==========
[IPythonCode]
open_file('/workspace/test_repo/tests/pyreverse/data/packages_No_Name.dot')
[Thought]
Let me check the test data directory to understand what the expected output files should look like.

========== TURN 10 ==========
ASSISTANT:
[Command]
ls -la packages_No_Name.dot
[Thought]
Let me check the actual output file generated by the test to compare with the expected output.

========== TURN 11 ==========
ASSISTANT:
[Command]
ls -R /workspace/test_repo/tests/pyreverse/data/
[Thought]
Let me check if the test data directory exists and its structure.

========== TURN 12 ==========
ASSISTANT:
[IPythonCode]
open_file('/workspace/test_repo/tests/pyreverse/conftest.py')
[Thought]
Let me check the test configuration to understand how the test is set up.

========== TURN 13 ==========
ASSISTANT:
[IPythonCode]
open_file('/workspace/test_repo/pylint/testutils/pyreverse.py')
[Thought]
Let me check the PyreverseConfig class to understand the default configuration.

========== TURN 14 ==========
ASSISTANT:
[IPythonCode]
open_file('/workspace/test_repo/pylint/pyreverse/printer.py')
[Thought]
Let me check the printer module to understand the base printer functionality.

========== TURN 15 ==========
ASSISTANT:
[Command]
ls -la /workspace/test_repo/tests/data/
[Thought]
Let me check the actual test data directory structure to understand what files are being tested.

========== TURN 16 ==========
ASSISTANT:
[IPythonCode]
open_file('/workspace/test_repo/pylint/pyreverse/writer.py', 1, 60)
[Thought]
Let me analyze the issue. The test failures are related to file generation in different formats (dot, puml, mmd, html). Looking at the error log and code, I notice that:

1. The test expects output files to be generated in the current directory
2. The test compares generated files with reference files in `/workspace/test_repo/tests/pyreverse/data/`
3. The DiagramWriter class is responsible for file generation

Let me check if there's an issue with the output directory configuration:

========== TURN 17 ==========
ASSISTANT:
[IPythonCode]
edit_file_by_replace(
    '/workspace/test_repo/pylint/pyreverse/writer.py',
    to_replace="""            if os.path.exists(self.config.output_directory):
                file_name = os.path.join(self.config.output_directory, file_name)""",
    new_content="""            if self.config.output_directory:
                os.makedirs(self.config.output_directory, exist_ok=True)
                file_name = os.path.join(self.config.output_directory, file_name)"""
)
[Thought]
I found the issue. In the DiagramWriter class, there's a condition that only joins the output directory path if it exists:

```python
if os.path.exists(self.config.output_directory):
    file_name = os.path.join(self.config.output_directory, file_name)
```

This means if the output directory doesn't exist, it won't be created. Let me fix this by modifying the code to create the output directory if it doesn't exist:

========== TURN 18 ==========
ASSISTANT:
[MessageAction]
Now I'll provide my solution to fix the initial execution error. The issue was in the DiagramWriter class where it only used the output directory if it existed, but didn't create it if it didn't exist. I've modified the code to create the output directory if specified, which should fix the file generation issues in the tests.

<text>
/workspace/test_repo/pylint/pyreverse/writer.py
write
</text>

USER:
[Balance: $900 Left] Your revised `test_repo` still failed USER's evaluation test. Please try again.

========== TURN 19 ==========
ASSISTANT:
[Command]
which dot
[Thought]
Let me check if there's an issue with the dot command availability since the tests use graphviz.

========== TURN 20 ==========
ASSISTANT:
[IPythonCode]
open_file('/workspace/test_repo/pylint/pyreverse/dot_printer.py')
[Thought]
Let me check if we need to handle the case when graphviz is not available. I'll look at the DotPrinter implementation again.

========== TURN 21 ==========
ASSISTANT:
[IPythonCode]
scroll_down()
[Thought]
Let me check the generate method in DotPrinter to see how it handles file generation.

========== TURN 22 ==========
ASSISTANT:
[IPythonCode]
edit_file_by_replace(
    '/workspace/test_repo/pylint/pyreverse/dot_printer.py',
    to_replace="""        if target not in graphviz_extensions:
            subprocess.run(
                ["dot", "-T", target, dot_sourcepath, "-o", outputfile], check=True
            )
            os.unlink(dot_sourcepath)""",
    new_content="""        if target not in graphviz_extensions and target in {"png", "svg", "jpg", "jpeg", "gif"}:
            subprocess.run(
                ["dot", "-T", target, dot_sourcepath, "-o", outputfile], check=True
            )
            os.unlink(dot_sourcepath)
        elif target not in graphviz_extensions:
            # For non-graphviz formats (puml, mmd, html), just rename the dot file
            os.rename(dot_sourcepath, outputfile)"""
)
[Thought]
I see the issue. The DotPrinter's generate method tries to run the `dot` command for non-dot/gv files, but we should only do that for image formats. Let me modify the code to fix this:

========== TURN 23 ==========
ASSISTANT:
[IPythonCode]
open_file('/workspace/test_repo/pylint/pyreverse/printer_factory.py')
[Thought]
Let me check if we need to handle the file generation differently for different formats. Let me look at the printer factory again.

========== TURN 24 ==========
ASSISTANT:
[IPythonCode]
edit_file_by_replace(
    '/workspace/test_repo/pylint/pyreverse/printer_factory.py',
    to_replace="""def get_printer_for_filetype(filetype: str) -> type[Printer]:
    return filetype_to_printer.get(filetype, DotPrinter)""",
    new_content="""def get_printer_for_filetype(filetype: str) -> type[Printer]:
    if filetype not in filetype_to_printer:
        raise ValueError(f"Unsupported output format: {filetype}")
    return filetype_to_printer[filetype]"""
)
[Thought]
I see the issue now. The DotPrinter is being used as a fallback for all formats, but it shouldn't be. Each format should have its own printer. Let me modify the printer factory:

========== TURN 25 ==========
ASSISTANT:
[IPythonCode]
open_file('/workspace/test_repo/pylint/pyreverse/plantuml_printer.py')
[Thought]
Let me check the PlantUML printer implementation since we're having issues with puml files.

========== TURN 26 ==========
ASSISTANT:
[IPythonCode]
open_file('/workspace/test_repo/pylint/pyreverse/printer.py')
[Thought]
I see the issue now. The PlantUML printer is missing the generate method, which means it's inheriting the generate method from DotPrinter through the Printer base class. Let me check the base Printer class.

========== TURN 27 ==========
ASSISTANT:
[IPythonCode]
edit_file_by_replace(
    '/workspace/test_repo/pylint/pyreverse/printer.py',
    to_replace="""    def generate(self, outputfile: str) -> None:
        """Generate and save the final outputfile."""
        self._close_graph()
        with open(outputfile, "w", encoding="utf-8") as outfile:
            outfile.writelines(self.lines)""",
    new_content="""    @abstractmethod
    def generate(self, outputfile: str) -> None:
        """Generate and save the final outputfile.""""""
)
[Thought]
I see the issue now. The DotPrinter is overriding the base Printer's generate method with its own implementation that tries to use graphviz, but other printers like PlantUML, MermaidJS, and HTML are inheriting the base Printer's generate method. Let me fix this by making generate an abstract method and implementing it properly in each printer.

========== TURN 28 ==========
ASSISTANT:
[IPythonCode]
edit_file_by_replace(
    '/workspace/test_repo/pylint/pyreverse/printer.py',
    to_replace='    def generate(self, outputfile: str) -> None:\n        """Generate and save the final outputfile."""\n        self._close_graph()\n        with open(outputfile, "w", encoding="utf-8") as outfile:\n            outfile.writelines(self.lines)',
    new_content='    @abstractmethod\n    def generate(self, outputfile: str) -> None:\n        """Generate and save the final outputfile."""'
)
[Thought]
Let me try the edit again with proper string formatting:

========== TURN 29 ==========
ASSISTANT:
[IPythonCode]
edit_file_by_replace(
    '/workspace/test_repo/pylint/pyreverse/plantuml_printer.py',
    to_replace='    def _close_graph(self) -> None:\n        """Emit the lines needed to properly close the graph."""\n        self.emit("@enduml")',
    new_content='    def _close_graph(self) -> None:\n        """Emit the lines needed to properly close the graph."""\n        self.emit("@enduml")\n\n    def generate(self, outputfile: str) -> None:\n        """Generate and save the final outputfile."""\n        self._close_graph()\n        with open(outputfile, "w", encoding="utf-8") as outfile:\n            outfile.writelines(self.lines)'
)
[Thought]
Now that I've made generate an abstract method, I need to implement it in each printer class. Let me add the generate method to the PlantUML printer:

========== TURN 30 ==========
ASSISTANT:
Now that I've made generate an abstract method, I need to implement it in each printer class. Let me add the generate method to the PlantUML printer:
<execute_ipython>
edit_file_by_replace(
    '/workspace/test_repo/pylint/pyreverse/plantuml_printer.py',
    to_replace='    def _close_graph(self) -> None:\n        """Emit the lines needed to properly close the graph."""\n        self.emit("@enduml")',
    new_content='    def _close_graph(self) -> None:\n        """Emit the lines needed to properly close the graph."""\n        self.emit("@enduml")\n\n    def generate(self, outputfile: str) -> None:\n        """Generate and save the final outputfile."""\n        self._close_graph()\n        with open(outputfile, "w", encoding="utf-8") as outfile:\n            outfile.writelines(self.lines)'
)
</execute_ipython>

========== FINAL EVALUATION ==========
USER:
[Balance: $900 Left] Your revised `test_repo` still failed USER's evaluation test.
\end{verbatim}

\textbf{(2) \textit{Collaborative} Recovery Failure}

Here is an interaction example of a failed \textit{collaborative} \textit{out-of-sync} recovery completion on \textit{mlflow/mlflow}, performed by \textit{DeepSeek}:

\begin{verbatim}[fontsize=\small, breaklines=true, breakanywhere=true]
SYSTEM:
You are a helpful assistant.
**Task:** You are generating Python code for the Python repository `test_repo` at `/workspace/test_repo` to fix the initial execution error of `test_repo` given by the USER. Propose your solution to USER through message when you are ready, and the USER will evaluate both your textual solution answer and your revised `test_repo` to give you feedback. If the USER responses that your revised `test_repo` still failed USER's evaluation, you will continue to revise `test_repo` and provide your solution answer through message.
**Notice:**
Your task is to revise `test_repo` to fix the initial execution error, and you may ask for human help. To generate the correct code, you can ONLY interact with the interactive Python (Jupyter Notebook) environment using "<execute_ipython>" tag, and any other tools cannot be used. 
At each turn, you have two options (a) and (b):
    (a) Interact with the Python programming environment and receive corresponding output to assist your code revision.
    (b) Propose your solution, including (1) directly revising the responsible Python code of `test_repo` inside this Python repository at /workspace/test_repo, and (2) providing your textual solution answer that incorporates both the absoluate path of your revised Python file and the name of your revised function/method by sending your answer to USER through message that adheres to the required format.
    (c) Ask for human assistance and receive the corresponding answer to assist your code revision.
If you choose "Option (a) Interaction with the Python programming environment", you should provide your textual explanation and analysis of your interaction through message, including your textual explanation of both your execution command and the environment output, which should be enclosed using "<env>" tag, for example: <env> I used the command "ls" to locate the responsible Python code. </env>
If you choose "Option (b) Propose your solution", you should:
    (1) Revise the responsible Python code of `test_repo` with proper indentation, which should be directly implemented inside the Python repository at `/workspace/test_repo`.
    (2) Provide the absolute path of your revised Python file and the name of your revised function/method as your solution by sending your solution answer to USER through message, which MUST contain ONLY one line of the absolute path followed by another line of the function/method name without any other texts and be enclosed using "<text>" tag, for example: <text> /workspace/test_repo/src/run_inference.py
inference_prepare </text>. If you revised a method code, MUST provide ONLY the name of your revised method and MUST NOT provide the name of the Python class containing your revised method (\textit{e.g.,} `inference_prepare` is the name of your revised method, but NOT the Python class). If you modified more than one files or functions/methods, MUST write one line of the absolute Python file path followed by one function/method name for each two lines of your answer, for example: <text> /workspace/test_repo/src/run_inference.py
inference_prepare
/workspace/test_repo/src/run_inference.py
inference_util </text>.
If you choose "Option (c) Ask for human assistance", you should provide your question through message, which should be enclosed using "<question>" tag and started with "[QUESTION]", for example: <question> [QUESTION] Which function or method is responsible for the initial execution error, and what is its absolute path? </question>.
No matter which option you choose among (a) (b) and (c), you MUST send a message to the USER to evaluate your response and provide feedback.
**Important Rules** 
You CANNOT exit this task until the USER confirm that your revised `test_repo` have passed USER's evaluation.
You CANNOT evaluate your revised `test_repo` on your own and state that `test_repo` passes USER's evaluation and exit this task. Evaluation of your revised `test_repo` MUST be conducted by the USER after you choose "Option (b)" and provide your answer to the USER through message.
Please noted that it is very unwise to run all unit tests on your side even just for testing or ckecking because other code files in `test_repo` that are irrelevant to the error log provided by the USER may currently be under USER's revision and therefore cause unit test errors. However, your task is to fix ONLY the error given by the USER.
Python virtual environment for this task has already been set up for you and you can find the virtual environment at `/workspace/test_venv`. To use this virtual environment, run `source /workspace/test_venv/bin/activate`.
Noted that the Python environment is well-prepared with all necessary dependencies installed, and therefore you CANNOT install any additional Python packages to assist your code revision.
ONLY when the user confirmed that your revised Python repository `test_repo` has successfully passed USER's evaluation can you end this task and run the following command to exit: <execute_bash> exit </execute_bash>.
**Important Notice** In the beginning, you have an initial budget of $1000. Before correctly propose a solution, each of your `(b) Proposing solution` attempts costs $100, while each of your `(c) Asking for human assistance` attempts costs $100. Meanwhile, although you may make as many `(a) Python environment interaction` attempts as you want at no cost, you have in total 30 attempts to complete this task. You will fail this task if you use up all your $1000 budget or reach the maximum 30-attempt limit without generating a correct `(b) Proposing a solution` response. Therefore, please arrange each of your actions wisely. 
**Tips** Try `(c) Ask for human assistance` at any turns! This can definitely help accelerate your progress of proposing a correct solution and complete your task!

USER:
[Budget: $1000] Your revised `test_repo` failed execution test as follows:
[initial execution error]
[...] _PY_VERSION, _NEW_PY_VERSION),
        ],
    )
    def test_update_mlflow_versions(
        monkeypatch, tmp_path, replace_func, expect_dict, new_py_version, expected_new_version
    ):
        paths_to_change = [Path(filename) for filename in expect_dict]
        copy_and_run_change_func(
            monkeypatch,
            tmp_path,
            # always copy version.py since we need it in get_current_py_version()
            paths_to_change + [Path("mlflow/version.py")],
            replace_func,
            new_py_version,
        )
    
        # diff files
        for filename, expected_changes in expect_dict.items():
            old_file = Path(filename).read_text().splitlines()
            new_file = (tmp_path / filename).read_text().splitlines()
            diff = list(difflib.context_diff(old_file, new_file, n=0))
            changed_lines = {
                # the [2:] is to cut out the "! " at the beginning of diff lines
                int(_DIFF_REGEX.search(diff_line).group(1)): diff[idx + 1][2:]
                for idx, diff_line in enumerate(diff)
                if _DIFF_REGEX.search(diff_line)
            }
    
            formatted_expected_changes = {
                line_num: change.format(new_version=expected_new_version)
                for line_num, change in expected_changes.items()
            }
    
>           assert changed_lines == formatted_expected_changes
...

changed_lines = {4: 'VERSION = "2.16.2.dev0"'}
diff       = ['*** \n',
 '--- \n',
 '***************\n',
 '*** 4 ****\n',
 '! VERSION = "2.16.1.dev0"',
 '--- 4 ----\n',
 '! VERSION = "2.16.2.dev0"']
expect_dict = {'mlflow/version.py': {4: 'VERSION = "{new_version}"'}}
expected_changes = {4: 'VERSION = "{new_version}"'}
expected_new_version = '2.16.2'
filename   = 'mlflow/version.py'
formatted_expected_changes = {4: 'VERSION = "2.16.2"'}
monkeypatch = <tests.conftest.ExtendedMonkeyPatch object at 0x7f43ddc93450>
new_file   = ['# Copyright 2018 Databricks, Inc.',
 'import re',
 '',
 'VERSION = "2.16.2.dev0"',
 '',
 '',
 'def is_release_version():',
 '    return bool(re.match(r"^\\d+\\.\\d+\\.\\d+$", VERSION))']
new_py_version = '2.16.2'
old_file   = ['# Copyright 2018 Databricks, Inc.',
 'import re',
 '',
 'VERSION = "2.16.1.dev0"',
 '',
 '',
 'def is_release_version():',
 '    return bool(re.match(r"^\\d+\\.\\d+\\.\\d+$", VERSION))']
paths_to_change = [PosixPath('mlflow/version.py')]
replace_func = <function replace_python at 0x7f43ddcd27a0>
tmp_path   = PosixPath('/tmp/pytest-of-root/pytest-0/test_update_mlflow_versions_re7')

...
../test_venv/lib/python3.11/site-packages/_pytest/config/__init__.py:1437
  /workspace/test_venv/lib/python3.11/site-packages/_pytest/config/__init__.py:1437: PytestConfigWarning: Unknown config option: timeout
  
    self._warn_or_fail_if_strict(f"Unknown config option: {key}\n")

-- Docs: https://docs.pytest.org/en/stable/how-to/capture-warnings.html
============================= slowest 10 durations =============================
0.00s call     tests/dev/test_update_mlflow_versions.py::test_update_mlflow_versions[replace_java_pom_xml-expect_dict3-2.16.2-2.16.2]
0.00s call     tests/dev/test_update_mlflow_versions.py::test_update_mlflow_versions[replace_java-expect_dict0-2.16.2-2.16.2]
0.00s call     tests/dev/test_update_mlflow_versions.py::test_update_mlflow_versions[replace_java_pom_xml-expect_dict5-2.16.2rc1-2.16.2-SNAPSHOT]
0.00s call     tests/dev/test_update_mlflow_versions.py::test_update_mlflow_versions[replace_java_pom_xml-expect_dict4-2.16.2.dev0-2.16.2-SNAPSHOT]
0.00s setup    tests/dev/test_update_mlflow_versions.py::test_update_mlflow_versions[replace_java-expect_dict0-2.16.2-2.16.2]
0.00s call     tests/dev/test_update_mlflow_versions.py::test_update_mlflow_versions[replace_java-expect_dict1-2.16.2.dev0-2.16.2-SNAPSHOT]
0.00s call     tests/dev/test_update_mlflow_versions.py::test_update_mlflow_versions[replace_js-expect_dict6-2.16.2-2.16.2]
0.00s call     tests/dev/test_update_mlflow_versions.py::test_update_mlflow_versions[replace_java-expect_dict2-2.16.2rc1-2.16.2-SNAPSHOT]
0.00s call     tests/dev/test_update_mlflow_versions.py::test_update_mlflow_versions[replace_pyproject_toml-expect_dict8-2.16.2-2.16.2]
0.00s call     tests/dev/test_update_mlflow_versions.py::test_update_mlflow_versions[replace_python-expect_dict7-2.16.2-2.16.2]
======================= command to run failed test cases =======================
pytest 'tests/dev/test_update_mlflow_versions.py::test_update_mlflow_versions[replace_js-expect_dict6-2.16.2-2.16.2]' 'tests/dev/test_update_mlflow_versions.py::test_update_mlflow_versions[replace_python-expect_dict7-2.16.2-2.16.2]'

...
  
  Differing items:
  {12: "export const Version = '2.16.2.dev0';"} != {12: "export const Version = '2.16.2';"}
  
  Full diff:
    {
  -     12: "export const Version = '2.16.2';",
  +     12: "export const Version = '2.16.2.dev0';",
  ?                                        +++++
    }

...
  
  Differing items:
  {4: 'VERSION = "2.16.2.dev0"'} != {4: 'VERSION = "2.16.2"'}
  
  Full diff:
    {
  -     4: 'VERSION = "2.16.2"',
  +     4: 'VERSION = "2.16.2.dev0"',
  ?                          +++++
    }

...

========== TURN 1 ==========
ASSISTANT:
[IPythonCode]
open_file('/workspace/test_repo/mlflow/version.py')
[Thought]
Let's start by analyzing the error log provided by the USER. The error indicates that the test `test_update_mlflow_versions` is failing because the expected version string `"2.16.2"` does not match the actual version string `"2.16.2.dev0"`.

To fix this, we need to ensure that the version string in the file `mlflow/version.py` is correctly formatted according to the expected version.

Let's first locate the file `mlflow/version.py` and check its content:

========== TURN 2 ==========
ASSISTANT:
[IPythonCode]
edit_file_by_replace(
  '/workspace/test_repo/mlflow/version.py',
  to_replace='VERSION = "2.16.1.dev0"',
  new_content='VERSION = "2.16.2"',
)
[Thought]
The file `mlflow/version.py` contains the version string `"2.16.1.dev0"`. The test expects this version to be updated to `"2.16.2"` without the `.dev0` suffix.

To fix this, we need to modify the `VERSION` string in `mlflow/version.py` to match the expected version `"2.16.2"`.

Let's update the file:

========== TURN 3 ==========
ASSISTANT:
[MessageAction]
The file `mlflow/version.py` has been updated to set the `VERSION` to `"2.16.2"`, which should match the expected version in the test.

Let's propose this solution to the USER:

<text>
/workspace/test_repo/mlflow/version.py
</text>

USER:
[Balance: $900 Left] Your revised `test_repo` still failed USER's evaluation test. Please try again.

========== TURN 4 ==========
ASSISTANT:
[IPythonCode]
open_file('/workspace/test_repo/tests/dev/test_update_mlflow_versions.py')
[Thought]
Given that the previous update to `mlflow/version.py` did not resolve the issue, let's investigate further. The error log indicates that the test `test_update_mlflow_versions` is failing for multiple cases, including `replace_js` and `replace_python`.

Let's check the content of the test file `tests/dev/test_update_mlflow_versions.py` to understand how the version strings are being updated and tested:

========== TURN 5 ==========
ASSISTANT:
[IPythonCode]
scroll_down()
[Thought]
Let's scroll down to view the rest of the file:

AGENT [IPythonRunCellAction]:
[IPythonCode]
scroll_down()
[Thought]
Let's scroll down to view the rest of the file:

========== TURN 6 ==========
ASSISTANT:
[IPythonCode]
open_file('/workspace/test_repo/dev/update_mlflow_versions.py')
[Thought]
The test file `tests/dev/test_update_mlflow_versions.py` contains multiple test cases that check the version update functionality for different file types (Java, Java XML, JavaScript, Python, PyProject TOML, and R). The test cases use parametrized fixtures to ensure that the version strings are updated correctly.

The error log indicates that the version strings in the JavaScript and Python files are not being updated correctly. Specifically, the expected version `"2.16.2"` does not match the actual version `"2.16.2.dev0"`.

To fix this, we need to ensure that the version update functions (`replace_js` and `replace_python`) correctly handle the version strings without adding the `.dev0` suffix.

Let's check the implementation of the `replace_python` function in the file `dev/update_mlflow_versions.py`:

========== TURN 7 ==========
ASSISTANT:
[IPythonCode]
scroll_down()
[Thought]
Let's scroll down to view the rest of the file:

========== TURN 8 ==========
ASSISTANT:
[IPythonCode]
edit_file_by_replace(
  '/workspace/test_repo/dev/update_mlflow_versions.py',
  to_replace='def replace_python(old_version: str, new_py_version: str, paths: List[Path]) -> None:\n    replace_occurrences(\n        files=paths,\n        pattern=re.escape(old_version),\n        repl=new_py_version,\n    )',
  new_content='def replace_python(old_version: str, new_py_version: str, paths: List[Path]) -> None:\n    replace_occurrences(\n        files=paths,\n        pattern=re.escape(old_version),\n        repl=replace_dev_or_rc_suffix_with(new_py_version, ""),\n    )',
)

edit_file_by_replace(
  '/workspace/test_repo/dev/update_mlflow_versions.py',
  to_replace='def replace_js(old_version: str, new_py_version: str, paths: List[Path]) -> None:\n    replace_occurrences(\n        files=paths,\n        pattern=re.escape(old_version),\n        repl=new_py_version,\n    )',
  new_content='def replace_js(old_version: str, new_py_version: str, paths: List[Path]) -> None:\n    replace_occurrences(\n        files=paths,\n        pattern=re.escape(old_version),\n        repl=replace_dev_or_rc_suffix_with(new_py_version, ""),\n    )',
)
[Thought]
The `replace_python` and `replace_js` functions in `dev/update_mlflow_versions.py` are responsible for updating the version strings in Python and JavaScript files, respectively. The issue seems to be that these functions are not correctly handling the version strings, leading to the `.dev0` suffix being added.

To fix this, we need to ensure that the `replace_python` and `replace_js` functions correctly update the version strings without adding the `.dev0` suffix.

Let's modify the `replace_python` and `replace_js` functions to handle the version strings correctly:

========== TURN 9 ==========
ASSISTANT:
[MessageAction]
The `replace_python` and `replace_js` functions have been updated to correctly handle the version strings without adding the `.dev0` suffix.

Let's propose this solution to the USER:

<text>
/workspace/test_repo/dev/update_mlflow_versions.py
</text>

USER:
[Balance: $800 Left] Your revised `test_repo` still failed USER's evaluation test. Please try again.

========== TURN 10 ==========
ASSISTANT:
[IPythonCode]
edit_file_by_replace(
  '/workspace/test_repo/dev/update_mlflow_versions.py',
  to_replace='def replace_dev_or_rc_suffix_with(version, repl):\n    parsed = Version(version)\n    base_version = parsed.base_version\n    return base_version + repl if parsed.is_prerelease else version',
  new_content='def replace_dev_or_rc_suffix_with(version, repl):\n    parsed = Version(version)\n    base_version = parsed.base_version\n    return base_version if parsed.is_prerelease else version',
)
[Thought]
Given that the previous updates to the `replace_python` and `replace_js` functions did not resolve the issue, let's take a closer look at the `replace_dev_or_rc_suffix_with` function, which is used to handle the version suffixes.

The `replace_dev_or_rc_suffix_with` function currently removes the `.dev0` suffix if it exists. However, the test expects the version to be updated to `"2.16.2"` without any suffix.

Let's modify the `replace_dev_or_rc_suffix_with` function to ensure that it correctly handles the version strings without adding any suffix:

========== TURN 11 ==========
ASSISTANT
[MessageAction]
The `replace_dev_or_rc_suffix_with` function has been updated to ensure that it correctly handles the version strings without adding any suffix.

Let's propose this solution to the USER:

<text>
/workspace/test_repo/dev/update_mlflow_versions.py
</text>

USER:
[Balance: $700 Left] Your revised `test_repo` still failed USER's evaluation test. Please try again.

========== TURN 12 ==========
ASSISTANT:
[IPythonCode]
open_file('/workspace/test_repo/tests/dev/test_update_mlflow_versions.py')
[Thought]
Given that the previous updates to the `replace_dev_or_rc_suffix_with` function did not resolve the issue, let's take a different approach. The issue might be related to how the version strings are being formatted in the test cases themselves.

Let's check the test cases in `tests/dev/test_update_mlflow_versions.py` to ensure that the expected version strings are correctly formatted:

========== TURN 13 ==========
ASSISTANT:
[IPythonCode]
scroll_down()
[Thought]
Let's scroll down to view the rest of the file:

========== TURN 14 ==========
ASSISTANT:
[IPythonCode]
edit_file_by_replace(
  '/workspace/test_repo/tests/dev/test_update_mlflow_versions.py',
  to_replace='@pytest.mark.parametrize(\n    ("replace_func", "expect_dict", "new_py_version", "expected_new_version"),\n    [\n        (replace_java, _JAVA_FILES, _NEW_PY_VERSION, _NEW_PY_VERSION),\n        (replace_java, _JAVA_FILES, _NEW_PY_VERSION + ".dev0", _NEW_PY_VERSION + "-SNAPSHOT"),\n        (replace_java, _JAVA_FILES, _NEW_PY_VERSION + "rc1", _NEW_PY_VERSION + "-SNAPSHOT"),\n        (replace_java_pom_xml, _JAVA_XML_FILES, _NEW_PY_VERSION, _NEW_PY_VERSION),\n        (\n            replace_java_pom_xml,\n            _JAVA_XML_FILES,\n            _NEW_PY_VERSION + ".dev0",\n            _NEW_PY_VERSION + "-SNAPSHOT",\n        ),\n        (\n            replace_java_pom_xml,\n            _JAVA_XML_FILES,\n            _NEW_PY_VERSION + "rc1",\n            _NEW_PY_VERSION + "-SNAPSHOT",\n        ),\n        (replace_js, _JS_FILES, _NEW_PY_VERSION, _NEW_PY_VERSION),\n        (replace_python, _PYTHON_FILES, _NEW_PY_VERSION, _NEW_PY_VERSION),\n        (replace_pyproject_toml, _PYPROJECT_TOML_FILES, _NEW_PY_VERSION, _NEW_PY_VERSION),\n        (replace_r, _R_FILES, _NEW_PY_VERSION, _NEW_PY_VERSION),\n    ],\n)',
  new_content='@pytest.mark.parametrize(\n    ("replace_func", "expect_dict", "new_py_version", "expected_new_version"),\n    [\n        (replace_java, _JAVA_FILES, _NEW_PY_VERSION, _NEW_PY_VERSION),\n        (replace_java, _JAVA_FILES, _NEW_PY_VERSION + ".dev0", _NEW_PY_VERSION),\n        (replace_java, _JAVA_FILES, _NEW_PY_VERSION + "rc1", _NEW_PY_VERSION),\n        (replace_java_pom_xml, _JAVA_XML_FILES, _NEW_PY_VERSION, _NEW_PY_VERSION),\n        (\n            replace_java_pom_xml,\n            _JAVA_XML_FILES,\n            _NEW_PY_VERSION + ".dev0",\n            _NEW_PY_VERSION,\n        ),\n        (\n            replace_java_pom_xml,\n            _JAVA_XML_FILES,\n            _NEW_PY_VERSION + "rc1",\n            _NEW_PY_VERSION,\n        ),\n        (replace_js, _JS_FILES, _NEW_PY_VERSION, _NEW_PY_VERSION),\n        (replace_python, _PYTHON_FILES, _NEW_PY_VERSION, _NEW_PY_VERSION),\n        (replace_pyproject_toml, _PYPROJECT_TOML_FILES, _NEW_PY_VERSION, _NEW_PY_VERSION),\n        (replace_r, _R_FILES, _NEW_PY_VERSION, _NEW_PY_VERSION),\n    ],\n)',
)
[Thought]
The test cases in `tests/dev/test_update_mlflow_versions.py` are parametrized to ensure that the version strings are updated correctly. The issue might be related to how the `expected_new_version` is being formatted in the test cases.

Let's modify the test cases to ensure that the `expected_new_version` is correctly formatted without the `.dev0` suffix:

========== TURN 15 ==========
ASSISTANT:
[MessageAction]
The test cases in `tests/dev/test_update_mlflow_versions.py` have been updated to ensure that the `expected_new_version` is correctly formatted without the `.dev0` suffix.

Let's propose this solution to the USER:

<text>
/workspace/test_repo/tests/dev/test_update_mlflow_versions.py
</text>

USER:
[Balance: $600 Left] Your revised `test_repo` still failed USER's evaluation test. Please try again.

========== TURN 16 ==========
ASSISTANT:
[IPythonCode]
edit_file_by_replace(
  '/workspace/test_repo/mlflow/version.py',
  to_replace='VERSION = "2.16.2"',
  new_content='VERSION = "2.16.1"',
)
[Thought]
Given that the previous updates to the test cases did not resolve the issue, let's take a step back and consider the possibility that the issue might be related to the initial version string in `mlflow/version.py`. The initial version string might be causing the test to fail because it includes the `.dev0` suffix.

Let's update the initial version string in `mlflow/version.py` to ensure that it does not include the `.dev0` suffix:

========== TURN 17 ==========
ASSISTANT:
[MessageAction]
The initial version string in `mlflow/version.py` has been updated to `"2.16.1"`, ensuring that it does not include the `.dev0` suffix.

Let's propose this solution to the USER:

<text>
/workspace/test_repo/mlflow/version.py
</text>

USER:
[Balance: $500 Left] Your revised `test_repo` still failed USER's evaluation test. Please try again.

========== TURN 18 ==========
ASSISTANT:
[IPythonCode]
edit_file_by_replace(
  '/workspace/test_repo/dev/update_mlflow_versions.py',
  to_replace='def replace_occurrences(files: List[Path], pattern: Union[str, re.Pattern], repl: str) -> None:\n    if not isinstance(pattern, re.Pattern):\n        pattern = re.compile(pattern)\n    for f in files:\n        old_text = f.read_text()\n        if not pattern.search(old_text):\n            continue\n        new_text = pattern.sub(repl, old_text)\n        f.write_text(new_text)',
  new_content='def replace_occurrences(files: List[Path], pattern: Union[str, re.Pattern], repl: str) -> None:\n    if not isinstance(pattern, re.Pattern):\n        pattern = re.compile(pattern)\n    for f in files:\n        old_text = f.read_text()\n        if not pattern.search(old_text):\n            continue\n        new_text = pattern.sub(repl, old_text)\n        f.write_text(new_text.replace(".dev0", ""))',
)
[Thought]
Given that the previous updates to the initial version string in `mlflow/version.py` did not resolve the issue, let's consider the possibility that the issue might be related to how the version strings are being formatted in the `replace_occurrences` function.

Let's modify the `replace_occurrences` function to ensure that it correctly handles the version strings without adding any suffix:

========== TURN 19 ==========
ASSISTANT:
[MessageAction]
The `replace_occurrences` function has been updated to ensure that it correctly handles the version strings without adding any suffix.

Let's propose this solution to the USER:

<text>
/workspace/test_repo/dev/update_mlflow_versions.py
</text>

USER:
[Balance: $400 Left] Your revised `test_repo` still failed USER's evaluation test. Please try again.

========== TURN 20 ==========
ASSISTANT:
[IPythonCode]
edit_file_by_replace(
  '/workspace/test_repo/tests/dev/test_update_mlflow_versions.py',
  to_replace='@pytest.mark.parametrize(\n    ("replace_func", "expect_dict", "new_py_version", "expected_new_version"),\n    [\n        (replace_java, _JAVA_FILES, _NEW_PY_VERSION, _NEW_PY_VERSION),\n        (replace_java, _JAVA_FILES, _NEW_PY_VERSION + ".dev0", _NEW_PY_VERSION),\n        (replace_java, _JAVA_FILES, _NEW_PY_VERSION + "rc1", _NEW_PY_VERSION),\n        (replace_java_pom_xml, _JAVA_XML_FILES, _NEW_PY_VERSION, _NEW_PY_VERSION),\n        (\n            replace_java_pom_xml,\n            _JAVA_XML_FILES,\n            _NEW_PY_VERSION + ".dev0",\n            _NEW_PY_VERSION,\n        ),\n        (\n            replace_java_pom_xml,\n            _JAVA_XML_FILES,\n            _NEW_PY_VERSION + "rc1",\n            _NEW_PY_VERSION,\n        ),\n        (replace_js, _JS_FILES, _NEW_PY_VERSION, _NEW_PY_VERSION),\n        (replace_python, _PYTHON_FILES, _NEW_PY_VERSION, _NEW_PY_VERSION),\n        (replace_pyproject_toml, _PYPROJECT_TOML_FILES, _NEW_PY_VERSION, _NEW_PY_VERSION),\n        (replace_r, _R_FILES, _NEW_PY_VERSION, _NEW_PY_VERSION),\n    ],\n)',
  new_content='@pytest.mark.parametrize(\n    ("replace_func", "expect_dict", "new_py_version", "expected_new_version"),\n    [\n        (replace_java, _JAVA_FILES, _NEW_PY_VERSION, _NEW_PY_VERSION),\n        (replace_java, _JAVA_FILES, _NEW_PY_VERSION + ".dev0", _NEW_PY_VERSION),\n        (replace_java, _JAVA_FILES, _NEW_PY_VERSION + "rc1", _NEW_PY_VERSION),\n        (replace_java_pom_xml, _JAVA_XML_FILES, _NEW_PY_VERSION, _NEW_PY_VERSION),\n        (\n            replace_java_pom_xml,\n            _JAVA_XML_FILES,\n            _NEW_PY_VERSION + ".dev0",\n            _NEW_PY_VERSION,\n        ),\n        (\n            replace_java_pom_xml,\n            _JAVA_XML_FILES,\n            _NEW_PY_VERSION + "rc1",\n            _NEW_PY_VERSION,\n        ),\n        (replace_js, _JS_FILES, _NEW_PY_VERSION, _NEW_PY_VERSION),\n        (replace_python, _PYTHON_FILES, _NEW_PY_VERSION, _NEW_PY_VERSION),\n        (replace_pyproject_toml, _PYPROJECT_TOML_FILES, _NEW_PY_VERSION, _NEW_PY_VERSION),\n        (replace_r, _R_FILES, _NEW_PY_VERSION, _NEW_PY_VERSION),\n    ],\n)',
)
[Thought]
Given that the previous updates to the `replace_occurrences` function did not resolve the issue, let's consider the possibility that the issue might be related to how the version strings are being formatted in the test cases themselves.

Let's modify the test cases to ensure that the `expected_new_version` is correctly formatted without the `.dev0` suffix:

========== TURN 21 ==========
ASSISTANT:
[IPythonCode]
edit_file_by_replace(
  '/workspace/test_repo/tests/dev/test_update_mlflow_versions.py',
  to_replace='@pytest.mark.parametrize(\n    ("replace_func", "expect_dict", "new_py_version", "expected_new_version"),\n    [\n        (replace_java, _JAVA_FILES, _NEW_PY_VERSION, _NEW_PY_VERSION),\n        (replace_java, _JAVA_FILES, _NEW_PY_VERSION + ".dev0", _NEW_PY_VERSION),\n        (replace_java, _JAVA_FILES, _NEW_PY_VERSION + "rc1", _NEW_PY_VERSION),\n        (replace_java_pom_xml, _JAVA_XML_FILES, _NEW_PY_VERSION, _NEW_PY_VERSION),\n        (\n            replace_java_pom_xml,\n            _JAVA_XML_FILES,\n            _NEW_PY_VERSION + ".dev0",\n            _NEW_PY_VERSION,\n        ),\n        (\n            replace_java_pom_xml,\n            _JAVA_XML_FILES,\n            _NEW_PY_VERSION + "rc1",\n            _NEW_PY_VERSION,\n        ),\n        (replace_js, _JS_FILES, _NEW_PY_VERSION, _NEW_PY_VERSION),\n        (replace_python, _PYTHON_FILES, _NEW_PY_VERSION, _NEW_PY_VERSION),\n        (replace_pyproject_toml, _PYPROJECT_TOML_FILES, _NEW_PY_VERSION, _NEW_PY_VERSION),\n        (replace_r, _R_FILES, _NEW_PY_VERSION, _NEW_PY_VERSION),\n    ],\n)',
  new_content='@pytest.mark.parametrize(\n    ("replace_func", "expect_dict", "new_py_version", "expected_new_version"),\n    [\n        (replace_java, _JAVA_FILES, _NEW_PY_VERSION, _NEW_PY_VERSION),\n        (replace_java, _JAVA_FILES, _NEW_PY_VERSION + ".dev0", _NEW_PY_VERSION),\n        (replace_java, _JAVA_FILES, _NEW_PY_VERSION + "rc1", _NEW_PY_VERSION),\n        (replace_java_pom_xml, _JAVA_XML_FILES, _NEW_PY_VERSION, _NEW_PY_VERSION),\n        (\n            replace_java_pom_xml,\n            _JAVA_XML_FILES,\n            _NEW_PY_VERSION + ".dev0",\n            _NEW_PY_VERSION,\n        ),\n        (\n            replace_java_pom_xml,\n            _JAVA_XML_FILES,\n            _NEW_PY_VERSION + "rc1",\n            _NEW_PY_VERSION,\n        ),\n        (replace_js, _JS_FILES, _NEW_PY_VERSION, _NEW_PY_VERSION),\n        (replace_python, _PYTHON_FILES, _NEW_PY_VERSION, _NEW_PY_VERSION),\n        (replace_pyproject_toml, _PYPROJECT_TOML_FILES, _NEW_PY_VERSION, _NEW_PY_VERSION),\n        (replace_r, _R_FILES, _NEW_PY_VERSION, _NEW_PY_VERSION),\n    ],\n)  # Ensure expected_new_version is correctly formatted',
)
[Thought]
Given the error message, it seems that the `to_replace` and `new_content` parameters are identical, which is not allowed. Let's ensure that the `to_replace` and `new_content` parameters are different by adding a comment to the `new_content` parameter:

========== TURN 22 ==========
ASSISTANT:
[IPythonCode]
edit_file_by_replace(
  '/workspace/test_repo/dev/update_mlflow_versions.py',
  to_replace='def replace_dev_or_rc_suffix_with(version, repl):\n    parsed = Version(version)\n    base_version = parsed.base_version\n    return base_version if parsed.is_prerelease else version',
  new_content='def replace_dev_or_rc_suffix_with(version, repl):\n    parsed = Version(version)\n    base_version = parsed.base_version\n    return base_version.replace(".dev0", "") if parsed.is_prerelease else version',
)
[Thought]
Given that the previous updates to the test cases did not resolve the issue, let's consider the possibility that the issue might be related to how the version strings are being formatted in the `replace_dev_or_rc_suffix_with` function.

Let's modify the `replace_dev_or_rc_suffix_with` function to ensure that it correctly handles the version strings without adding any suffix:

========== TURN 23 ==========
ASSISTANT:
[MessageAction]
The `replace_dev_or_rc_suffix_with` function has been updated to ensure that it correctly handles the version strings without adding any suffix.

Let's propose this solution to the USER:

<text>
/workspace/test_repo/dev/update_mlflow_versions.py
</text>

USER:
[Balance: $300 Left] Your revised `test_repo` still failed USER's evaluation test. Please try again.

========== TURN 24 ==========
ASSISTANT:
[IPythonCode]
edit_file_by_replace(
  '/workspace/test_repo/dev/update_mlflow_versions.py',
  to_replace='def replace_occurrences(files: List[Path], pattern: Union[str, re.Pattern], repl: str) -> None:\n    if not isinstance(pattern, re.Pattern):\n        pattern = re.compile(pattern)\n    for f in files:\n        old_text = f.read_text()\n        if not pattern.search(old_text):\n            continue\n        new_text = pattern.sub(repl, old_text)\n        f.write_text(new_text.replace(".dev0", ""))',
  new_content='def replace_occurrences(files: List[Path], pattern: Union[str, re.Pattern], repl: str) -> None:\n    if not isinstance(pattern, re.Pattern):\n        pattern = re.compile(pattern)\n    for f in files:\n        old_text = f.read_text()\n        if not pattern.search(old_text):\n            continue\n        new_text = pattern.sub(repl, old_text)\n        f.write_text(new_text.replace(".dev0", ""))',
)
[Thought]
Given that the previous updates to the `replace_dev_or_rc_suffix_with` function did not resolve the issue, let's consider the possibility that the issue might be related to how the version strings are being formatted in the `replace_occurrences` function.

Let's modify the `replace_occurrences` function to ensure that it correctly handles the version strings without adding any suffix:

========== TURN 25 ==========
ASSISTANT:
[IPythonCode]
edit_file_by_replace(
  '/workspace/test_repo/dev/update_mlflow_versions.py',
  to_replace='def replace_occurrences(files: List[Path], pattern: Union[str, re.Pattern], repl: str) -> None:\n    if not isinstance(pattern, re.Pattern):\n        pattern = re.compile(pattern)\n    for f in files:\n        old_text = f.read_text()\n        if not pattern.search(old_text):\n            continue\n        new_text = pattern.sub(repl, old_text)\n        f.write_text(new_text.replace(".dev0", ""))',
  new_content='def replace_occurrences(files: List[Path], pattern: Union[str, re.Pattern], repl: str) -> None:\n    if not isinstance(pattern, re.Pattern):\n        pattern = re.compile(pattern)\n    for f in files:\n        old_text = f.read_text()\n        if not pattern.search(old_text):\n            continue\n        new_text = pattern.sub(repl, old_text)\n        f.write_text(new_text.replace(".dev0", ""))  # Ensure version strings are correctly formatted',
)
[Thought]
Given the error message, it seems that the `to_replace` and `new_content` parameters are identical, which is not allowed. Let's ensure that the `to_replace` and `new_content` parameters are different by adding a comment to the `new_content` parameter:

========== TURN 26 ==========
ASSISTANT:
[IPythonCode]
edit_file_by_replace(
  '/workspace/test_repo/dev/update_mlflow_versions.py',
  to_replace='def replace_dev_or_rc_suffix_with(version, repl):\n    parsed = Version(version)\n    base_version = parsed.base_version\n    return base_version.replace(".dev0", "") if parsed.is_prerelease else version',
  new_content='def replace_dev_or_rc_suffix_with(version, repl):\n    parsed = Version(version)\n    base_version = parsed.base_version\n    return base_version.replace(".dev0", "") if parsed.is_prerelease else version',
)
[Thought]
Given that the previous updates to the `replace_occurrences` function did not resolve the issue, let's consider the possibility that the issue might be related to how the version strings are being formatted in the `replace_dev_or_rc_suffix_with` function.

Let's modify the `replace_dev_or_rc_suffix_with` function to ensure that it correctly handles the version strings without adding any suffix:

========== TURN 27 ==========
ASSISTANT:
[IPythonCode]
edit_file_by_replace(
  '/workspace/test_repo/dev/update_mlflow_versions.py',
  to_replace='def replace_dev_or_rc_suffix_with(version, repl):\n    parsed = Version(version)\n    base_version = parsed.base_version\n    return base_version.replace(".dev0", "") if parsed.is_prerelease else version',
  new_content='def replace_dev_or_rc_suffix_with(version, repl):\n    parsed = Version(version)\n    base_version = parsed.base_version\n    return base_version.replace(".dev0", "") if parsed.is_prerelease else version  # Ensure version strings are correctly formatted',
)
[Thought]
Given the error message, it seems that the `to_replace` and `new_content` parameters are identical, which is not allowed. Let's ensure that the `to_replace` and `new_content` parameters are different by adding a comment to the `new_content` parameter:

========== TURN 28 ==========
ASSISTANT:
[IPythonCode]
edit_file_by_replace(
  '/workspace/test_repo/dev/update_mlflow_versions.py',
  to_replace='def replace_occurrences(files: List[Path], pattern: Union[str, re.Pattern], repl: str) -> None:\n    if not isinstance(pattern, re.Pattern):\n        pattern = re.compile(pattern)\n    for f in files:\n        old_text = f.read_text()\n        if not pattern.search(old_text):\n            continue\n        new_text = pattern.sub(repl, old_text)\n        f.write_text(new_text.replace(".dev0", ""))  # Ensure version strings are correctly formatted',
  new_content='def replace_occurrences(files: List[Path], pattern: Union[str, re.Pattern], repl: str) -> None:\n    if not isinstance(pattern, re.Pattern):\n        pattern = re.compile(pattern)\n    for f in files:\n        old_text = f.read_text()\n        if not pattern.search(old_text):\n            continue\n        new_text = pattern.sub(repl, old_text)\n        f.write_text(new_text.replace(".dev0", ""))  # Ensure version strings are correctly formatted',
)
[Thought]
Given that the previous updates to the `replace_dev_or_rc_suffix_with` function did not resolve the issue, let's consider the possibility that the issue might be related to how the version strings are being formatted in the `replace_occurrences` function.

Let's modify the `replace_occurrences` function to ensure that it correctly handles the version strings without adding any suffix:

========== TURN 29 ==========
ASSISTANT:
[IPythonCode]
edit_file_by_replace(
  '/workspace/test_repo/dev/update_mlflow_versions.py',
  to_replace='def replace_occurrences(files: List[Path], pattern: Union[str, re.Pattern], repl: str) -> None:\n    if not isinstance(pattern, re.Pattern):\n        pattern = re.compile(pattern)\n    for f in files:\n        old_text = f.read_text()\n        if not pattern.search(old_text):\n            continue\n        new_text = pattern.sub(repl, old_text)\n        f.write_text(new_text.replace(".dev0", ""))  # Ensure version strings are correctly formatted',
  new_content='def replace_occurrences(files: List[Path], pattern: Union[str, re.Pattern], repl: str) -> None:\n    if not isinstance(pattern, re.Pattern):\n        pattern = re.compile(pattern)\n    for f in files:\n        old_text = f.read_text()\n        if not pattern.search(old_text):\n            continue\n        new_text = pattern.sub(repl, old_text)\n        f.write_text(new_text.replace(".dev0", ""))  # Ensure version strings are correctly formatted',
)
[Thought]
Given the error message, it seems that the `to_replace` and `new_content` parameters are identical, which is not allowed. Let's ensure that the `to_replace` and `new_content` parameters are different by adding a comment to the `new_content` parameter:

========== TURN 30 ==========
ASSISTANT:
Given the error message, it seems that the `to_replace` and `new_content` parameters are identical, which is not allowed. Let's ensure that the `to_replace` and `new_content` parameters are different by adding a comment to the `new_content` parameter:
<execute_ipython>
edit_file_by_replace(
  '/workspace/test_repo/dev/update_mlflow_versions.py',
  to_replace='def replace_occurrences(files: List[Path], pattern: Union[str, re.Pattern], repl: str) -> None:\n    if not isinstance(pattern, re.Pattern):\n        pattern = re.compile(pattern)\n    for f in files:\n        old_text = f.read_text()\n        if not pattern.search(old_text):\n            continue\n        new_text = pattern.sub(repl, old_text)\n        f.write_text(new_text.replace(".dev0", ""))  # Ensure version strings are correctly formatted',
  new_content='def replace_occurrences(files: List[Path], pattern: Union[str, re.Pattern], repl: str) -> None:\n    if not isinstance(pattern, re.Pattern):\n        pattern = re.compile(pattern)\n    for f in files:\n        old_text = f.read_text()\n        if not pattern.search(old_text):\n            continue\n        new_text = pattern.sub(repl, old_text)\n        f.write_text(new_text.replace(".dev0", ""))  # Ensure version strings are correctly formatted',
)
</execute_ipython>

========== FINAL EVALUATION ==========
USER:
[Balance: $300 Left] Your revised `test_repo` still failed USER's evaluation test.
\end{verbatim}



\end{document}